\documentclass[twocolumn, times]{aastex63}

\usepackage{amsmath,verbatim, nicefrac,color, soul}
\usepackage{gensymb}
\usepackage{natbib,graphicx}
\usepackage{array}

\usepackage[caption=false]{subfig}

\makeatletter
\newcommand\notsotiny{\@setfontsize\notsotiny\@vipt\@viipt}
\makeatother

\newcommand\Hl[1]{\colorbox{yellow}}

\DeclareCaptionFormat{cont}{#1 (cont.)#2#3\par}

\shorttitle{Evaluating the Prospects of Cyclic Deconvolution}
\shortauthors{J. E. Turner}

\begin{document}

\title{Evaluating the Prospects of Cyclic Deconvolution Across 312 Pulsars} 

\author[0000-0002-2451-7288]{Jacob E. Turner}
\affiliation{Green Bank Observatory, P.O. Box 2, Green Bank, WV 24944, USA}

\begin{abstract}
We use the cyclic figure of merit to determine the likelihood of achieving cyclic deconvolution for 312 pulsars with sub-40 ms spin periods across 15 different telescope-observing frequency combinations. We find that the optimal frequency range for achieving cyclic deconvolution for most pulsars is between $\sim$80$-$300 MHz, making low frequency observatories such as uGMRT, LOFAR, and MWA the best-suited instruments for the technique. Moreover, we find that, as quantified by the total number sources with sufficient cyclic merits that are observed within the full deconvolution regime, uGMRT is likely the best current instrument for cyclic spectroscopy among the ten telescopes we considered, with LOFAR being the second best, although our simulations predict that the DSA may become the top instrument once a greater fraction of galactic millisecond pulsars are discovered. The relatively high cyclic merit of the Crab Pulsar in the frequency ranges considered for GBT, MWA, LOFAR, and uGMRT suggests that some faster-spinning canonical pulsars may be able to achieve cyclic deconvolution, and we discuss potential follow-up analyses on other non-recycled pulsars. We conclude by advocating for near real-time cyclic spectroscopy backends to be considered for current and upcoming low frequency telescopes to increase the accessibility of this technique.
\end{abstract}

\section{Introduction}\label{intro}
\par Cyclic spectroscopy \citep{cyc_spec} is a signal processing technique that can utilize the periodicity and harmonic information of pulsar emission to extract more information from data than can be accomplished through conventional phase-resolved spectroscopy. The namesake cyclic spectrum for pulsar emission interacting with an ionized medium is given by
\begin{equation}
    S_{E}(\nu,\alpha_k) = \langle H(\nu+\alpha_k/2)H^*(\nu-\alpha_k/2)\rangle S_x(\nu, \alpha_k),
\label{pulse_cs}
\end{equation}
where $\nu$, is the observing frequency, $\alpha_k$ is the cyclic frequency, or harmonic, in which $\alpha_k=k/P$, where $k$ is an integer and $P$ is the pulse period, $S_x(\nu, \alpha)$ is the intrinsic pulse profile's Fourier transform, and $H(\nu)$ is the frequency response, or transfer function, of the interstellar medium \citep[ISM,][]{wdv13}. The cyclic spectrum is valid up to a pulsar's diffractive timescale, $\Delta t_{\rm d}$, also referred to as the scintillation or decorrelation timescale, after which it must be resampled \citep{dsj+20}. 
\par One can use the phase information within the cyclic spectrum to deconvolve the intrinsic pulse profile from the pulse broadening function, which is the Fourier transform of the transfer function. This serves a few benefits: First, the pulse broadening function can be used to directly measure pulse time-of-arrival delays caused by interstellar scattering, allowing for improvements in pulsar timing array sensitivity to gravitational waves \citep{Levin_Scat,turner_scat,epta_scint}. Scattering delays recovered in this way have been shown in simulations to be more precise, accurate, and better correlated with the true delay of the signal, even at moderate levels of signal-to-noise, when compared to indirect inference through taking autocorrelations of pulsar dynamic spectra \citep{turner_cyc}, which are time and frequency-evolving diffraction patterns that form due to pulsar emission undergoing multipath propagation through an ionized medium. Second, if there is sufficient S/N that one can resample the cyclic spectrum many times within a scintillation timescale, one can stack recovered transfer functions temporally to create an observation's dynamic wavefield, which constitutes a full phase-recovered reconstruction of the time and frequency-evolving signal the ISM imparts onto pulsar emission as it propagates through the medium. Dynamic spectra consist of the amplitude information of this signal at the 0$^{\rm th}$ harmonic.

\par In addition to the benefits gained from cyclic deconvolution, by using cyclic spectroscopy to resample data modulo the pulsar spin frequency, one can circumvent the time-frequency sampling uncertainty relation described by the Gabor limit to maintain high pulse phase resolution while achieving frequency resolutions down to $1/P$. This is incredibly valuable for studies of the ISM, since dynamic spectra require sufficient channelization to resolve scintles, which are bright patches of constructive interference within the spectra. Understanding the scintillation pattern  within dynamic spectra is key to studying the structure of the ISM along a given line of sight. Additionally, one can use the characteristic widths of scintillation structures within these spectra, referred to as the diffractive, decorrelation, or scintillation bandwidth, given by $\Delta \nu_{\rm d}$, to infer scattering delays, which is useful if the pulse broadening function cannot be obtained \citep{Levin_Scat, turner_scat, Liu_2022}. $\Delta \nu_{\rm d}$ can be related to the scattering delay, $\tau_{\rm d}$, via

\begin{equation}
    \label{tau_nu_relation}
    2\pi\tau_{\rm d}\Delta \nu_{\rm d} = C_1,
\end{equation}
where $C_1$ is a constant that depends on the geometry of the scattering screen and the electron density wavenumber spectrum along a given line of sight \citep{Cordes_1998}. \par In practice, unless one has sufficiently high pulse phase and frequency resolution, as would be the case with cyclic spectroscopy-processed data or baseband data processed twice to achieve one data product with high pulse phase resolution and another with high frequency resolution, $C_1$ will be unknown for a given line of sight and must be assumed to convert between the two quantities. This means that a direct measurement of either $\tau_{\rm d}$ or $\Delta \nu_{\rm d}$ in an observation without sufficient resolution in both domains would result in a potentially incorrect estimation of the other quantity. Improperly estimated scattering delays inferred from scintillation bandwidths in this manor could potentially reduce the sensitivity of pulsar timing arrays to gravitational wave signals, particularly when pulsars in those arrays have scattering delays similar to or greater than their median pulse time-of-arrival uncertainties \citep{Turner_1937}.    
\par Additionally, if  scintillation features are sufficiently resolved, one can often observe parabolic features, known as scintillation arcs \citep{OG_arcs}, in the Fourier counterparts of dynamic spectra, known as secondary spectra. These arcs are crucial to studying and localizing the distance to and studying the features within structures along a line of sight that are responsible for the majority of scattering experienced by pulsar emission. Furthermore, scintillation arcs can be used to track  these structures with sub-astronomical unit precision as they transit our line of sight \citep{Brisken_2010}.
\par While all millisecond pulsars should be able to achieve simultaneous high pulse phase and frequency resolution with cyclic spectroscopy, there is no such guarantee that a given pulsar will achieve complete phase retrieval of the transfer function, and consequently, full cyclic deconvolution. The likelihood of this retrieval's success is quantified by the cyclic figure of merit,
\begin{equation}
\begin{split}
\label{cyc_merit}
m_{\rm cyc}  & = \frac{\Phi}{\delta \Phi}=2\pi\frac{\tau_{\rm d}W_e}{P^2}(\textrm{S/N}) \sqrt{\sum_k k^2 a_k} \\ & \approx 2\pi(\textrm{S/N})\frac{\tau_{\rm d}}{\sqrt{P W_{\rm{e}}}},
\end{split}
\end{equation}

\noindent where $\Phi$ is the phase of the cyclic spectrum, $W_{\rm e}$ is the equivalent pulse width, S/N is the source signal-to-noise ratio, and $a_k \equiv A_k/A_0$, with a given $A_k$ being the $k^{\rm th}$ amplitude of the Fourier transform of the intensity pulse profile. In cyclic merit's original form, S/N in Equation \ref{cyc_merit} is given by

\begin{equation}
\label{s_n_eq}
    \textrm{S/N} = \frac{SG}{T}\sqrt{2\Delta t_{\rm d} B}\sqrt{\frac{P-W_{\rm e}}{W_{\rm e}}},
\end{equation}
where $S$ is source flux density, $G$ is instrument gain, $T$ is system temperature, and $B$ is observing bandwidth \citep{lorimer_kramer}. While cyclic merit in this form is useful for determining if one can recover the average phase slope of the cyclic spectrum, which itself is valuable for estimating $\tau_{\rm d}$, cyclic merit must only be considered over $\Delta \nu_{\rm d}$ when estimating full transfer function recovery \citep{dsj+20}. This, along with the tendency for the above form of cyclic merit to be overly optimistic with pulsars exhibiting large scattering delays but low S/N across individual scintles, motivated cyclic merit 2.0, $m_{\rm cyc,2.0}$ \citep{Turner_2025}. This metric can be used to estimate the likelihood of transfer function recovery from the cyclic spectrum, and has S/N given by 
\begin{equation}
\label{s_n_eq_2_0}
    \textrm{S/N} = \frac{SG}{T}\sqrt{2\Delta t_{\rm d} \Delta \nu_{\rm d}}\sqrt{\frac{P-W_{\rm e}}{W_{\rm e}}}.
\end{equation}
\par In addition to sufficient cyclic merit, a source must also be observed in its full deconvolution regime if one hopes to achieve complete cyclic phase retrieval. On the higher frequency end, this regime is bounded by the scintillation bandwidth being less than the inverse of the intrinsic pulse width, which will manifest as a visible scattering tail in the pulse profile \citep{dsj+20}. On the low frequency end, the cyclic spectrum can at most achieve frequency channelizations equivalent to the pulsar's spin frequency before becoming undersampled in cyclic frequency, able to capture signal components at most out to delays equal to half the pulse period \citep{wdv13}. That being said, significant phase retrieval can occur outside of this regime, and indeed increases the closer in frequency one observes to this regime, at least when approaching the upper bound from above. The dynamic wavefield power, which acts similarly to the pulsar's dynamic spectrum with additional harmonic information, can be fully recovered in this regime, as well as the secondary spectrum and portions of the secondary wavefield, which is the squared-modulus of the dynamic wavefield's Fourier transform \citep{Turner_2025}.
\par In this study, we examine 312 rapidly rotating pulsars to determine which sources are likely to achieve cyclic deconvolution using various telescopes and observing frequencies, as indicated by their cyclic merits and full deconvolution regime boundaries. In Section \ref{analysis}, we discuss how we selected our sources and calculated cyclic merits and deconvolution regime boundaries. In Section \ref{results}, we present our results and discuss which sources are likely to achieve cyclic deconvolution with different telescope-observing frequency combinations. Finally, in Section \ref{conclusions}, we summarize our conclusions, advocate for certain telescopes to consider developing cyclic spectroscopy backends, and discuss future research avenues to explore with the technique.

\section{Analysis}\label{analysis}
\subsection{Source Selection \& Acquisition of Observables}
\par Our source selection consisted of using \textsc{psrqpy} \citep{psrqpy} to search for all pulsars in the Australian National Telescope Facility (ATNF) pulsar catalog\footnote{\url{https://www.atnf.csiro.au/research/pulsar/psrcat/}} \citep{atnf_3} with pulse periods less than 40 ms (many millisecond pulsar population studies place an upper bound on spin period of around 30 ms, e.g., \cite{pulsar_max_period}, although we slightly relax this boundary to include accreting and transitional sources in our study) and pulse widths measured via either $W_{50}$ or $W_{10}$. We also included 10 additional pulsars from the MeerKAT Pulsar Timing Array (MPTA) \citep{meerpta} that did not have listed $W_{50}$ or $W_{10}$ values. However, we approximated $W_{50}$ and $W_{10}$ values for these pulsars using the profiles shown in \cite{meertiming}. This resulted in 312 pulsars, the vast majority of which are classified as fully recycled. In addition to pulse width sources listed in ATNF, we used $W_{50}$ and $W_{10}$ values from \cite{stairs}, \cite{Jacoby_2007}, \cite{2013_width}, \cite{2011_pol}, \cite{2020_width}, \cite{Kerr2020}, \cite{fast}, \cite{2024_deneva}, and \cite{2024_wang}. We ignore any potential frequency evolution in profile width, as this quantity is generally expected to stay fairly constant for millisecond pulsars \citep{kuzmin_width}.
\par We used weighted averages and standard deviations for scintillation bandwidths and timescales measured in the North American Nanohertz Observatory for Gravitational Waves' (NANOGrav) 9- \cite{Levin_Scat} and 12.5- \cite{turner_scat} year data releases, measured scintillation bandwidths from \cite{Bilous_2015}, \cite{epta_scint}, \cite{main_leap}, and \cite{Turner_1937}, and scattering delays listed in the ATNF catalog. For pulsars without measurements from these sources, we used scintillation bandwidths and timescales predicted by the NE2001 electron density model \citep{NE2001} by means of the \textsc{NE2001p} python package \citep{Ocker_2024}. When available, we used pulsar transverse velocities listed in ATNF to inform our NE2001 predictions. Otherwise, we assumed transverse velocities of 100 km s$^{-1}$, as this is NE2001's default assumption. To account for epoch-to-epoch variation in measurements that are not weighted averages or weighted standard deviations, we assumed a 50\% standard deviation on their scintillation bandwidths and/or timescales for our uncertainties. Additionally, we assumed thin screen scattering and a Kolmogorov medium ($C_1 = 0.957$) to switch between scintillation bandwidth and scattering delay \citep{Cordes_1998}, always starting from the observable that was initially measured. In keeping with our Kolmogorov turbulence assumption, we used an index of 4.4 for scaling values to relevant observing frequencies.
\par For pulsars in NANOGrav's 12.5-year data release, we used flux densities and spectral indices from \cite{alam2020nanograv}, and used the 16$^{\rm th}$ and 84$^{\rm th}$ percentile values as our uncertainties. For all other pulsars that did not have relevant flux densities listed in that paper or on ATNF, we used flux densities from the \textsc{pulsar\_spectra} database \citep{pulsar_spectra} and assumed simple power laws to acquire our spectral indices. These spectral index fits used data from \cite{lyne_1990}, \cite{Lorimer_1995b}, \cite{Manchester_1996}, \cite{toscano_1998}, \cite{stairs}, \cite{Malofeev_2000}, \cite{Kuzmin_2001}, \cite{Lewandowski_2004}, \cite{Freire_2007}, \cite{Stappers_2008}, \cite{Demorest_2013}, \cite{Manchester_2013}, \cite{Stovall_2014}, \cite{Dai_2015}, \cite{Kuniyoshi_2015}, \cite{Frail_2016}, \cite{Han_2016}, \cite{Kondratiev_2016}, \cite{Murphy_2017}, \cite{Swiggum_2017}, \cite{Gentile_2018}, \cite{Jankowski_2018}, \cite{Kaur_2019}, \cite{Sanidas_2019}, \cite{Zhang_2019}, \cite{Bondonneau_2020}, \cite{Crowter_2020}, \cite{2020_width}, \cite{alam2020nanograv}, \cite{Bondonneau_2021}, \cite{meertiming}, \cite{Bhat_2023}, and \cite{Gitika_2023}. If we were unable to perform a spectral index fit, we assumed a spectral index of $-1.7$.
\subsection{Determination of Cyclic Metrics}\label{cyc_deter}
\par To most thoroughly explore cyclic deconvolution feasibility, we calculated cyclic merits using Equation \ref{cyc_merit} with the S/N from Equation \ref{s_n_eq_2_0} for the selected population of 312 pulsars across 10 telescopes and 15 observing frequencies. These telescopes were chosen to maintain optimal sensitivity and sky coverage across observing frequency. We chose observing frequencies that would allow for continued coverage in different ranges while also considering whether a telescope had sufficient bandwidth above and below a given frequency to recover a sufficient portion of the transfer function, which is contingent on achieving an adequate number of scintles across the observing band. These telescope and observing frequency combinations can be seen in Table \ref{table_configs}, along with telescope gains and receiver temperatures, as well as declination ranges.
\par For the S/N determination in Equation \ref{s_n_eq}, we used linear additions of the receiver temperatures shown in Table \ref{table_configs} with the frequency and sky location-dependent contributions from the galactic background given by \textsc{PyGDSM} \citep{2008MNRAS.388..247D, 2015MNRAS.451.4311R, 2016ascl.soft03013P, 2017MNRAS.464.3486Z, 2017MNRAS.469.4537D}. For PSR B0531+21, we also included the frequency-dependent contribution from the Crab Nebula, as given by \cite{Macías-Pérez_2010}.
\par We have opted to use $W_{50}$ for our calculations of cyclic merit and S/N, partially because it is a better approximation of $W_{e}$ than $W_{10}$ and partially because it is a much more prevalent measurement in the literature, allowing us to calculate cyclic merits for more sources. This change required a slight rescaling of the cyclic merit 2.0 threshold from \cite{Turner_2025}, who used the effective pulse width, $W_{\rm eff}$, which they noted achieved very similar cyclic merits to $W_{10}$. For a self-consistent comparison, we opted for the same rescaling approach used by \cite{Turner_2025} for converting from $m_{\rm{cyc, 1.0}}$ to $m_{\rm{cyc, 2.0}}$, with the same pulsars, same observing frequencies, and same values on all observables, only switching out $W_{\rm eff}$ for $W_{50}$. This resulted in a slight change in threshold of $m_{\rm cyc}\gg 0.05$ to $m_{\rm cyc}\gg 0.07$. While it is somewhat subjective what cyclic merit would satisfy the $m_{\rm cyc} \gg 0.07$ criteria, we deemed an order of magnitude to be sufficient, and set our threshold for expected successful cyclic deconvolution at $m_{\rm cyc} \geq 0.7$. However, we also chose to set a more optimistic threshold of $m_{\rm cyc} \geq 0.3$ to explore whether such a relaxation of standards would significantly expand the potential number of sources. Additionally, it has been suggested for cyclic merit 1.0 that $m_{\rm cyc} \geq 10$ may not be necessary to satisfy the $m_{\rm cyc, 1.0} \gg 1$ criteria in some cases \citep{dsj+20}, and that logic should extend to cyclic merit 2.0. As such, we will be referring to our original $m_{\rm cyc, 2.0} \geq 0.7$ barrier as the conservative threshold and the relaxed barrier as the optimistic threshold.

\par We also determined full deconvolution regime boundaries for our pulsar population. To ensure we only considered cyclic deconvolution in instances where the transfer function was fully resolved, we set lower bounds in observing frequency for the full deconvolution regime based on whether a pulsar's measured scintillation bandwidth was greater than or equal to the pulsar spin frequency. For all pulsars that had $W_{10}$ measurements, we set upper bounds in observing frequency for the full deconvolution regime based on when a pulsar's scintillation bandwidth was less than or equal to its $W_{10}$.

\par It is important to note that, due to inter-epoch variations in flux density and scattering, depending on the observing frequency, there will be pulsars that are viable candidates for cyclic deconvolution in some epochs that may be too scatter-broadened, not scatter-broadened enough, or too dim in other epochs. As a disclaimer, some pulsars may have high cyclic merits for a given telescope-observing frequency combination but may have under-resolved scintles due to the limitations on cyclic spectroscopy's frequency channelization capabilities. However, given that such observations may still allow for partial recovery of the transfer function, we still calculate their cyclic merits for completeness. Additionally, our cyclic merits for LWA and NenuFAR may be biased high, as we assumed simple power law behavior for all of our spectral indices despite many pulsars experiencing spectral turnovers around 100 MHz \citep{Maron_2000, Murphy_2017, Bondonneau_2020}, which is higher than these telescopes' observing bands.

\par In addition to the cyclic merit analyses described above, we made projections for how many sources in the entire galaxy may be viable for cyclic deconvolution for a given instrument and observing frequency. This was accomplished by calculating cyclic merits for the ten simulated galactic millisecond pulsar populations created for \cite{Liu_2023} and averaging the results for each telescope-frequency combination used in the above analysis. These populations had a mean and standard deviation of 31,426 and 5,230 millisecond pulsars per population, respectively, and contained all information required to calculate cyclic merits except for scintillation timescale and transverse velocity. Consequently, for all sources, we assumed a transverse velocity of 100 km $^{-1}$ for the same reasons described above, and calculated scintillation timescales using the NE2001 electron density model.  

\subsection{Determination of Scintle Resolution}
\par While the primary focus of this paper is cyclic deconvolution, as mentioned in Section \ref{intro}, another significant benefit of cyclic spectroscopy is the ability to achieve frequency channelizations down to $1/P$ while maintaining one's desired pulse phase resolution. This allows for the recovery of scintles in dynamic spectra, which are essential for ISM studies, using instruments that would otherwise require baseband data to achieve. For this reason, we also used the quantities determined in Section \ref{cyc_deter} to predict which pulsars will have resolvable scintles for a given telescope-observing frequency combination. Since the scintillation bandwidth must be greater than $1/P$ for scintles to be resolved, it follows that all sources where $\Delta \nu_{\rm d}P>1$ should have resolvable scintles at a given frequency. We further disambiguated this population by using Equation \ref{s_n_eq_2_0} to determine which sources will have sufficient S/N for detection when scintles are integrated over. For simplicity, we set a S/N threshold of greater than 1. We then repeated this process using the simulated populations discussed in Section \ref{cyc_deter}.

\begin{deluxetable*}{CCCCCC}
\tablecolumns{6}

\tablecaption{Telescope Parameters \label{table_configs}}
\tablehead{ \colhead{Telescope} &
\colhead{Frequency} & \colhead{Receiver Temperature} & 
\colhead{Gain}&
\colhead{Minimum Declination} &
\colhead{Maximum Declination} 
\vspace{-.2cm}
 \\  \colhead{} & \colhead{\text{(MHz)}} & \colhead{\text{(K)}} & \colhead{(K/Jy)}& \colhead{($\degree$)}  &
\colhead{($\degree$)} }
\startdata
\textrm{GBT} & 350 & 25 & 2 & $-$45 & 90\\
\textrm{DSA} & 800 & 11 & 12.5 & $-40$ & 90\\
\textrm{MeerKAT} & 600 & 33 & 2.8 & $-90$ & 42\\ 
\textrm{FAST} & 550  & 33 & 13.67 & $-14$ & 65.6 \\ 
\textrm{MWA} & 150 & 50 & 1$^*$ & $-90$ & 25\\
\textrm{MWA} & 200 & 50 &  1$^*$ & $-90$ & 25\\
\textrm{LWA} & 74 & 300 & 1.67$^{*,d}$ & $-$29 & 90\\
\textrm{LOFAR} & 150 & 400 & 8.90$^{*,c}$ & 0 & 90\\
\textrm{LOFAR} & 200 & 400 & 5.01$^{*,c}$ & 0 & 90\\
\textrm{NenuFAR} & 48 & 470.6 & 8.60$^*$ & $-23$ & 90\\
\textrm{NenuFAR} & 85 & 3095.0 & 2.74$^*$ & $-23$ & 90\\
\textrm{CHIME} & 450 & 20 & 1.16$^*$ & $-11$ & 90 \\
\textrm{uGMRT} & 150 & 295 & 8.58$^\dagger$ & $-53$ & 90 \\
\textrm{uGMRT} & 200 & 120 & 8.58$^\dagger$ & $-53$ & 90 \\
\textrm{uGMRT} & 350 & 50 & 8.58$^\dagger$ & $-53$ & 90 
\enddata

\tablecomments{Telescopes, observing frequencies, sensitivity parameters, and sky coverage for all observing setups considered in this study. \\ $^*$Maximum possible gain. Actual gain for a given source was calculated depending on its angular distance away from zenith. \\$^d$Assumes the combined use of 1024 dipoles from the LWA1, LWA-SV, and LWA-OVRO sites using \cite{lwa_cite}. \\$^c$Assumes only LOFAR core stations were used. \\$^\dagger$Assumes that 26 of the 30 dishes will be available.}

\end{deluxetable*}

\section{Results \& Discussion}\label{results}

\par The general cyclic merit results for our survey of real sources are shown in Table \ref{table_gen_results}, while the cyclic merit results for our simulated survey shown in Table \ref{sim_gen_results}. The general scintle detection results from both real and simulated populations are shown in Table \ref{table_scint_resolve}. All calculated cyclic merits for real sources can be found for GBT, the DSA, MeerKAT, FAST, MWA, and LWA in Table \ref{table_cyc} and LOFAR, NenuFAR, CHIME, and uGMRT in Table \ref{table_cyc_2}, with markers indicating cyclic merits that pass certain thresholds.  We note that the results in all of these tables only considered the specific observing frequencies shown in Table \ref{table_configs}. However, all of these observing frequencies exist within observing bands that extend both above and below these frequencies by many tens of MHz. As a result, there are a number of pulsars for each telescope-observing frequency combination that may pass certain cyclic merit thresholds and are observable within the full deconvolution regime in a different part of the band than the observing frequencies we chose. We elaborate on these cases for real sources in the telescope subsections below.

\begin{figure}[!ht]
    \centering
    {\hspace*{-.5cm}\includegraphics[width=0.5\textwidth]{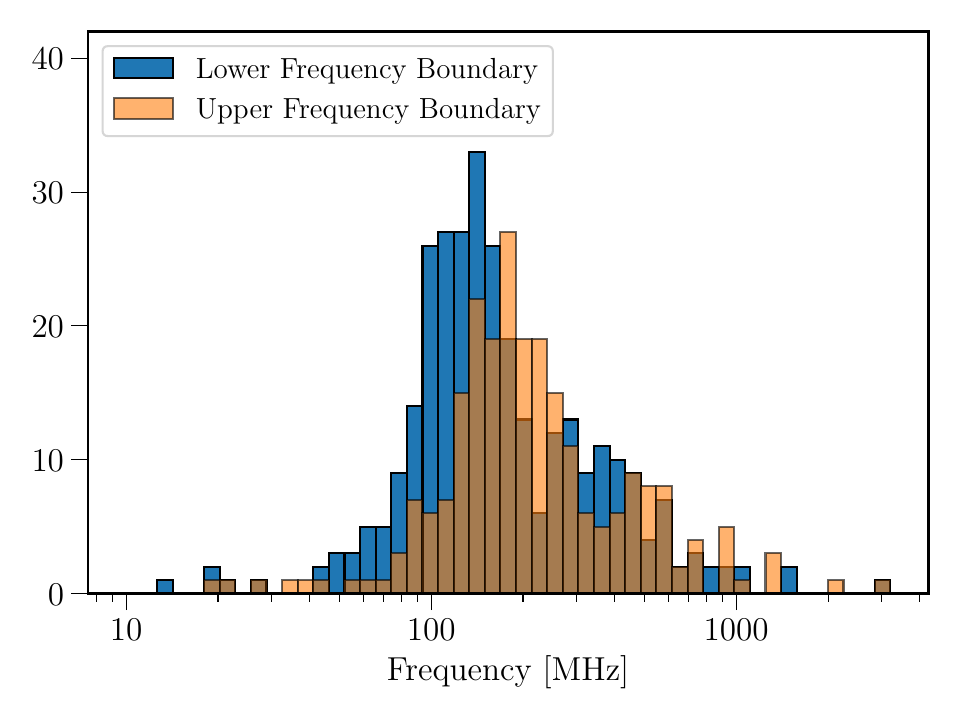}} %
    \caption{Histograms showing the lower (blue) and upper (orange) boundaries to the full deconvolution regime for all real pulsars in our survey. The lower frequency boundary has more instances due to the upper frequency boundary requiring $W_{10}$ meausrements, which were not available for all pulsars.}
    \label{decon_boundaries}%
\end{figure}

\par While a pulsar's cyclic merit will vary based on telescope and observing frequency, its upper and lower frequency boundaries for the full deconvolution regime should remain fairly constant, epoch-to-epoch variations aside. To determine which frequency ranges may be best-suited for cyclic deconvolution across the most pulsars, we examined the distributions for these upper and lower deconvolution regime boundaries, which can be seen in Figure \ref{decon_boundaries}. Based on these distributions, and the number of sources that simultaneously pass the conservative cyclic merit threshold while being observed in the full deconvolution regime, our findings suggest that the optimal frequency range for achieving cyclic deconvolution across the most pulsars is between  $\sim$80$-$300 MHz, notwithstanding our source count for pulsars passing certain cyclic merit thresholds being biased high below 100 MHz due to our spectral index assumptions. These results strongly indicate that instruments like the upgraded Giant Metrewave Radio Telescope (uGMRT), Low-Frequency Array (LOFAR), and Murchison Widefield Array (MWA) are the best current instruments for cyclic spectroscopy. Overall, based on number of unique sources that meet our criteria, uGMRT appears to be the best overall telescope for cyclic deconvolution, with LOFAR being the second best, especially considering the upcoming simultaneous low and high baseline array observing capabilities of the upgraded LOFAR2 \citep{lofar2_test}. That all being said, these instruments may be surpassed by telescopes like the DSA once a larger fraction of the galactic millisecond pulsar population is discovered.

\begin{deluxetable*}{CCCCCC}
\tablecolumns{6}

\tablefontsize{\footnotesize}

\tablecaption{General Survey Results for Real Sources \label{table_gen_results}}
\tablehead{ \colhead{Telescope} &
\colhead{Total Visible Sources} & \colhead{Outside Full Decon. Regime} & 
\colhead{In Full Decon. Regime}&
\colhead{Outside Full Decon. Regime} &
\colhead{In Full Decon. Regime} 
\vspace{-.3cm}
\\  \colhead{} & \colhead{\text{with Calculated $m_{\rm cyc}$}} & \colhead{\text{\& Passing Optimistic}} & \colhead{\text{\& Passing Optimistic}}& \colhead{\text{\& Passing Conservative}} & \colhead{\text{\& Passing Conservative}}  \vspace{-.3cm}
\\  \colhead{} & \colhead{Out of 312 Surveyed} & \colhead{\text{Threshold Only}} & \colhead{\text{Threshold Only}}& \colhead{\text{Threshold}} & \colhead{\text{Threshold}} }
\startdata
\text{GBT}$_{350}$ & 217 & 15 & 3 & 10 & 2\\
\text{DSA}$_{800}$ & 215 & 24 & 0 & 18 & 1\\
\text{MeerKAT}$_{600}$ & 252 & 12 & 0 & 6 & 1\\
\text{FAST}$_{550}$ & 144 & 19 & 0 & 16 & 2\\
\text{MWA}$_{150}$ & 222 & 8 & 8 & 17 & 7\\
\text{MWA}$_{200}$ & 222 & 11 & 7 & 10 & 6\\
\text{LWA}$_{74}$ & 189 & 18 & 1 & 35 & 2\\
\text{LOFAR}$_{150}$ & 121 & 12 & 9 & 13 & 13\\
\text{LOFAR}$_{200}$ & 121 & 2 & 7 & 7 & 2\\
\text{NenuFAR}$_{48}$ & 168 & 15 & 1 & 84 & 2\\
\text{NenuFAR}$_{85}$ & 168 & 16 & 1 & 19 & 3\\
\text{CHIME}$_{450}$ & 145 & 4 & 0 & 2 & 1 \\
\text{uGMRT}$_{150}$ & 229 & 22 & 15 & 50 & 29 \\
\text{uGMRT}$_{200}$ & 229 & 22 & 10 & 47 & 27 \\
\text{uGMRT}$_{350}$ & 229 & 30 & 2 & 39 & 9
\enddata

\tablecomments{Number of real sources passing cyclic deconvolution regime thresholds for each telescope and observing frequency. Note that the number of sources listed as being observable in the full deconvolution regime underestimates the total number of viable sources for a given telescope, as the criterion was only considered at the specific observing frequencies in each observing setup. However, these observing frequencies exist within observing bands that may span tens of MHz above and below the frequencies we considered. Consequently, there may be additional viable sources for a given telescope-observing frequency combination if that frequency was slightly adjusted. We elaborate on these instances for all of our observing configurations throughout Section \ref{results}.}

\end{deluxetable*}

\begin{deluxetable*}{CCCCCC}
\tablecolumns{6}

\tablefontsize{\footnotesize}

\tablecaption{General Survey Results for Simulated Galactic Population \label{sim_gen_results}}
\tablehead{ \colhead{Telescope} &
\colhead{Total Visible Sources} & \colhead{Outside Full Decon. Regime} & 
\colhead{In Full Decon. Regime}&
\colhead{Outside Full Decon. Regime} &
\colhead{In Full Decon. Regime} 
\vspace{-.3cm}
\\  \colhead{} & \colhead{\text{with Calculated $m_{\rm cyc}$}} & \colhead{\text{\& Passing Optimistic}} & \colhead{\text{\& Passing Optimistic}}& \colhead{\text{\& Passing Conservative}} & \colhead{\text{\& Passing Conservative}}  \vspace{-.3cm}
\\  \colhead{} & \colhead{\text{Out of 31426 $\pm$ 5230}} & \colhead{\text{Threshold Only}} & \colhead{\text{Threshold Only}}& \colhead{\text{Threshold}} & \colhead{\text{Threshold}} }
\startdata
\text{GBT}$_{350}$ & 24910 $\pm$ 4159 & 490$\pm$80 & 15$\pm$4 & 454$\pm$83 & 15 $\pm$ 4 \\
\text{DSA}$_{800}$ & 22952 $\pm$ 3843 & 812$\pm$137 & 67$\pm$11 & 837$\pm$140 & 60$\pm$10 \\
\text{MeerKAT}$_{600}$ & 30400$\pm$ 5072 & 444$\pm$82 & 15$\pm$3 & 353$\pm$65 & 11$\pm$2\\
\text{FAST}$_{550}$ & 8373$\pm$1406 & 360$\pm$62 & 30$\pm$6 & 410$\pm$75 & 34$\pm$5 \\
\text{MWA}$_{150}$ & 29421 $\pm$ 4917 & 679$\pm$159 & 4$\pm$1 & 810$\pm$156 & 6$\pm$3\\
\text{MWA}$_{200}$ & 29421 $\pm$ 4917 & 564$\pm$104 & 5$\pm$2 & 578 $\pm$105 & 7$\pm$3 \\
\text{LWA}$_{74}$ & 16367 $\pm$ 2718 & 611$\pm$100 & 3$\pm$1 & 948 $\pm$ 166 & 6$\pm$2 \\
\text{LOFAR}$_{150}$ & 4962 $\pm$ 813 & 278$\pm$46 & 9$\pm$4 & 444$\pm$69 & 14 $\pm$4 \\
\text{LOFAR}$_{200}$ & 4962 $\pm$ 813 & 161$\pm$30 & 7$\pm$2 & 200$\pm$34 & 7$\pm$2 \\
\text{NenuFAR}$_{48}$ & 12205$\pm$2026 & 876$\pm$173 & 2$\pm$1 & 2200$\pm$349 & 5$\pm$3\\
\text{NenuFAR}$_{85}$ & 12205$\pm$2026 & 426$\pm$67 & 3$\pm$1 & 647$\pm$110 & 5$\pm$2\\
\text{CHIME}$_{450}$ & 7527 $\pm$ 1258 & 61$\pm$10 & 5$\pm$2 & 41 $\pm$ 9 & 4 $\pm$ 1 \\
\text{uGMRT}$_{150}$ & 27516$\pm$ 4561 & 1734$\pm$286 & 16$\pm$7 & 2935$\pm$483 & 28$\pm$7\\
\text{uGMRT}$_{200}$ & 27516$\pm$ 4561 & 1604$\pm$267 & 24$\pm$7 & 2477$\pm$415 & 35$\pm$7 \\
\text{uGMRT}$_{350}$ & 27516$\pm$ 4561 & 1254$\pm$198 & 40$\pm$11 & 1576$\pm$275 & 43$\pm$8
\enddata

\tablecomments{Mean and standard deviation of sources passing cyclic deconvolution regime thresholds from ten simulated galactic populations for each telescope and observing frequency. Note that the number of sources listed as being observable in the full deconvolution regime underestimates the total number of viable sources for a given telescope, as the criterion was only considered at the specific observing frequencies in each observing setup. However, these observing frequencies exist within observing bands that may span tens of MHz above and below the frequencies we considered. Consequently, there may be additional viable sources for a given telescope-observing frequency combination if that frequency was slightly adjusted.}

\end{deluxetable*}

\begin{deluxetable}{CCC}
\tablecolumns{3}

\tablecaption{Detectable \& Resolvable Scintles \label{table_scint_resolve}}
\tablehead{ \colhead{Telescope} &
\colhead{Total from} & \colhead{Total from} 
\vspace{-.2cm}
 \\ \colhead{} & \colhead{Real Population} & \colhead{Simulated Galactic Population} \vspace{-.2cm}
 \\\colhead{} & \colhead{of 312 Sources} & \colhead{of 31426 $\pm$ 5230 Sources} }
\startdata
\text{GBT}$_{350}$ & 87 & 81$\pm$19\\
\text{DSA}$_{800}$ & 178 & 442$\pm$85 \\
\text{MeerKAT}$_{600}$ & 151 & 127$\pm$28\\
\text{FAST}$_{550}$ & 112 & 200$\pm$38 \\
\text{MWA}$_{150}$ & 13 & 10$\pm$3 \\\
\text{MWA}$_{200}$ & 26 & 17$\pm$6  \\
\text{LWA}$_{74}$ & 5 & 6$\pm$3  \\
\text{LOFAR}$_{150}$ & 21 & 21$\pm$5 \\
\text{LOFAR}$_{200}$ & 21 & 21$\pm$6 \\
\text{NenuFAR}$_{48}$ & 5 & 4$\pm$2\\
\text{NenuFAR}$_{85}$ & 5 & 6$\pm$3 \\
\text{CHIME}$_{450}$ & 70 & 60$\pm$16 \\
\text{uGMRT}$_{150}$ & 47 & 36$\pm$12 \\
\text{uGMRT}$_{200}$ & 72 & 62$\pm$15 \\
\text{uGMRT}$_{350}$ & 127 & 137$\pm$28
\enddata

\tablecomments{Total number of sources from our real and simulated simulated populations that have resolvable and detectable scintles.}

\end{deluxetable}

\subsection{GBT 350 MHz}

\par
\begin{figure}[!ht]
    \centering
    \captionsetup[subfigure]{labelformat=empty}
    {\hspace*{-.4cm}\includegraphics[width=0.5\textwidth]{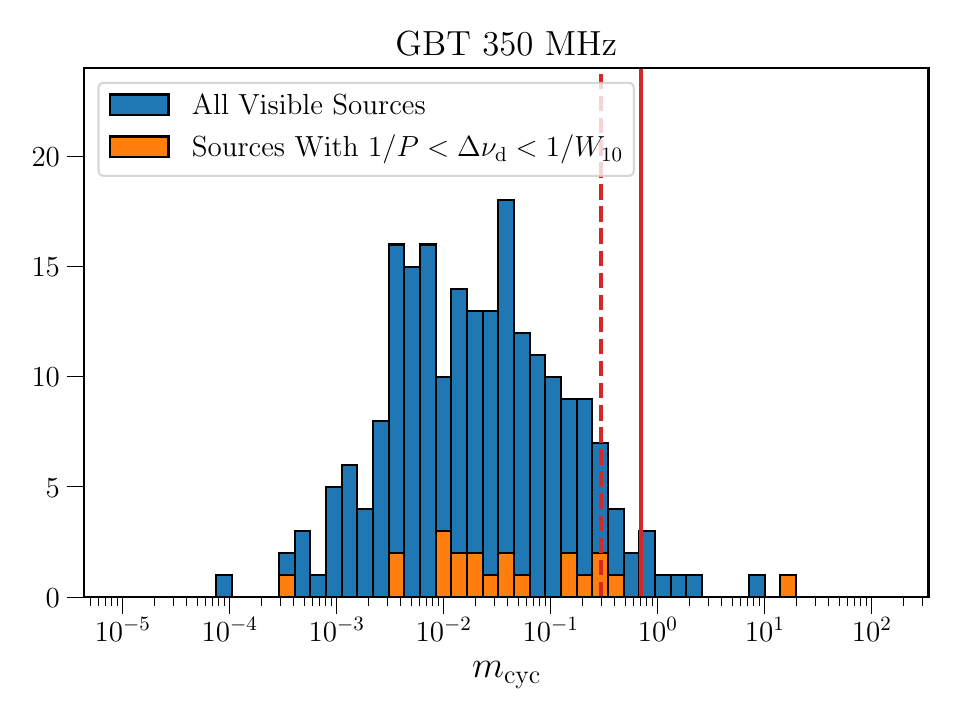} }%
    \caption{Cyclic merits for all real sources analyzed for GBT at 350 MHz (blue), and the subset of sources in the full deconvolution regime (orange). The two outliers on the right are PSRs J2205+6012 (blue) and  B1937+21 (orange). The dashed and solid red lines indicate the optimistic and conservative cyclic merit thresholds, respectively. Note that some sources passing certain cyclic merit thresholds may do so within error, and thus may not appear to meet those criteria as visualized by histogram. Similarly, some sources may pass certain cyclic merit thresholds but are just outside of the full deconvolution regime, and would be strong candidates for cyclic deconvolution if observed slightly lower or higher than the specific observing frequency we chose for our analysis. }%
    \label{GBT_hist}%
\end{figure}

\begin{figure}[!ht]
    \centering
    \captionsetup[subfigure]{labelformat=empty}
    {\hspace*{-.4cm}\includegraphics[width=0.5\textwidth]{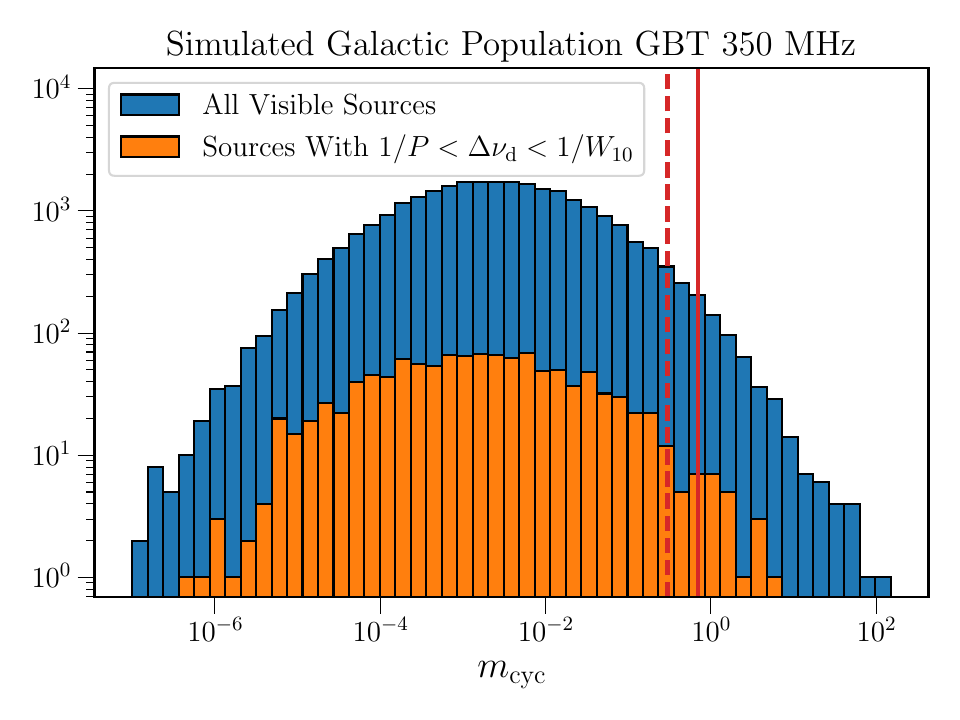} }%
    \caption{Cyclic merits for all sources from a realization of our simulations analyzed for GBT at 350 MHz (blue), and the subset of sources in the full deconvolution regime (orange). Plot description same as that in Figure \ref{GBT_hist}.}%
    \label{GBT_sim_hist}%
\end{figure}

\begin{figure}[!ht]
    \centering
    \captionsetup[subfigure]{labelformat=empty}
    {\hspace*{-.4cm}\includegraphics[width=0.5\textwidth]{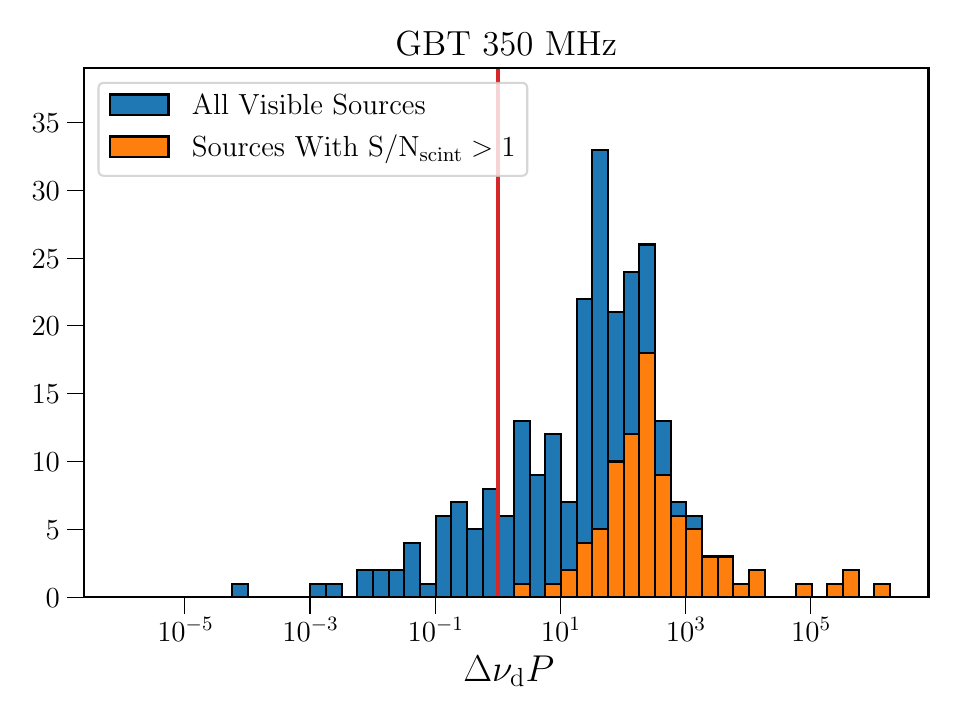} }%
    \caption{Expected scintle resolvability for all sources analyzed for GBT at 350 MHz (blue), and the subset of sources with S/N$>$1 (orange). Scintles should be resolvable when $\Delta \nu_{\rm d}P>1$, as indicated the by the solid red vertical line. Note that some sources passing the scintle resolution threshold may do so within error, and thus may not appear to meet the above criterion visualized by histogram. Similarly, some sources may pass our S/N threshold but are just outside of scintle resolvability, although resolution may be achieved if observed slightly higher in frequency.}%
    \label{GBT_scint_hist}%
\end{figure}

\begin{figure}[!ht]
    \centering
    \captionsetup[subfigure]{labelformat=empty}
    {\hspace*{-.4cm}\includegraphics[width=0.5\textwidth]{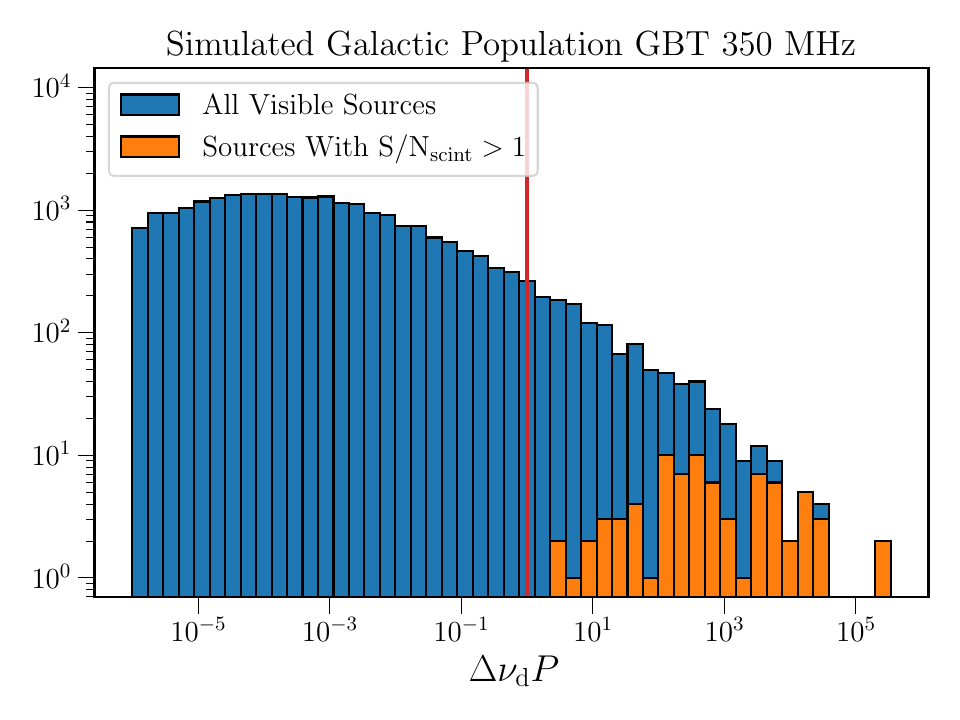} }%
    \caption{Expected scintle resolvability from a realization of our simulations analyzed for GBT at 350 MHz (blue), and the subset of sources with S/N$>$1 (orange). Plot description same as that in Figure \ref{GBT_scint_hist}.}%
    \label{sim_GBT_scint_hist}%
\end{figure}

 Arguably the most relevant instrument for discussions related to cyclic deconvolution, the GBT will soon have the world's first cyclic spectroscopy backend, allowing for near real-time processing of observations with this technique. Based on our results, when this backend becomes available, currently expected for the second half of 2026, we argue that observing campaigns dedicated to cyclic deconvolution and phase retrieval should prioritize PSRs J0751+1807, J1643$-$1224, J1811$-$2405, and B1937+21. However, we note that, based on our calculated frequency boundaries for their full deconvolution regimes, PSR J1643$-$1224 will likely achieve cyclic deconvolution only in the top half of the band and PSR J1811$-$2405 in the bottom half of the band. Additionally, it is entirely possible that PSR J2215+5135 may also be able to achieve cyclic decovolution with the GBT at 350 MHz, although, given that we were unable to find $W_{10}$ values in the literature, we were unable to determine the upper bound to its full deconvolution regime. Interestingly, PSR B0531+21, the Crab Pulsar, both passes the optimistic cyclic merit threshold and is in the full deconvolution regime at 350 MHz, suggesting that cyclic deconvolution may be possible for some canonical pulsars. While there is justifiably some concern regarding whether its significant giant pulse fraction \citep{Doskoch_2024} may  inhibit phase retrieval, PSR B1937+21, the only pulsar in the literature currently demonstrated to achieve cyclic deconvolution \citep{wdv13}, also exhibits giant pulses \citep{1937_gp_1, 1937_gp_2}, indicating this may not be a significant issue. The full histogram for all real sources analyzed using the GBT at 350 MHz is shown in Figure \ref{GBT_hist}. We should also expect a proportionally significant increase in viable sources for cyclic deconvolution with the availability of a larger class of radio telescopes, as our simulations predict 15$\pm$4 pulsars that pass our conservative threshold and are observable within the full deconvolution regime. An example realization from these simulations is shown in Figure \ref{GBT_sim_hist}. Additionally, we find that the GBT should be excellent for recovering scintillation in pulsars at 350 MHz, with 87 real sources having resolvable scintles that pass our S/N threshold. However, our simulations indicate that most, if not all, sources that could have detectable scintles at 350 MHz for the GBT have likely already been discovered. The scintle resolution prospects for all real sources with the GBT at 350 MHz are shown in Figure \ref{GBT_scint_hist}, while an example realization for our simulated population is shown in Figure \ref{sim_GBT_scint_hist}.

\subsection{DSA 800 MHz}

\par
\begin{figure}[!ht]
    \centering
    \captionsetup[subfigure]{labelformat=empty}
    {\hspace*{-.8cm}\includegraphics[width=0.5\textwidth]{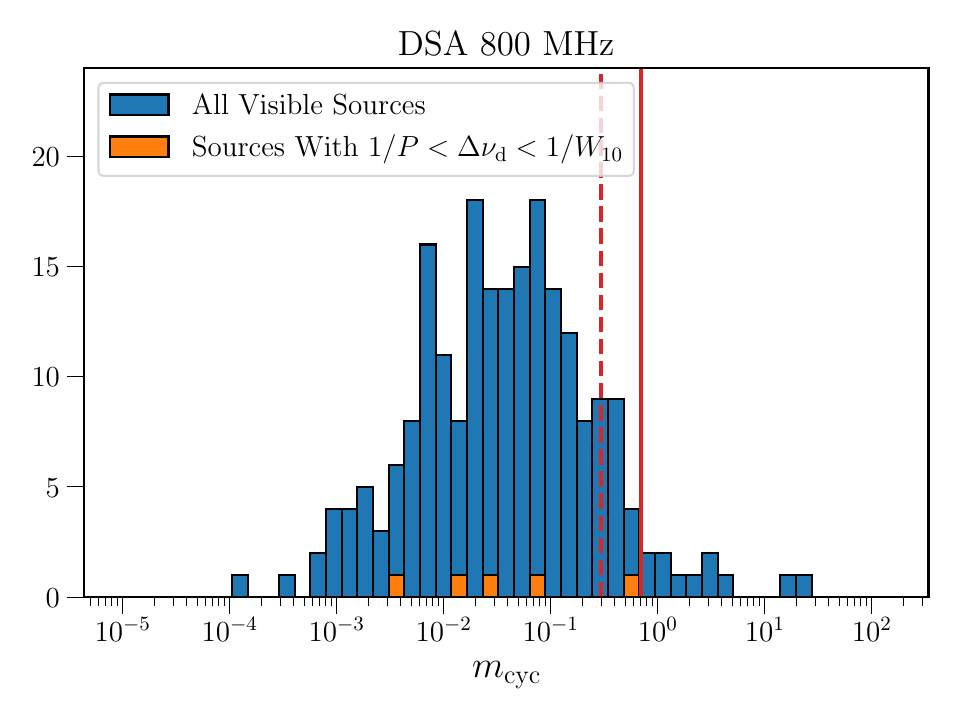} }%
    \caption{Cyclic merits for all real sources analyzed for the DSA at 800 MHz (blue), and the subset of sources in the full deconvolution regime (orange). Plot description same as that in Figure \ref{GBT_hist}. The two outliers on the right are PSRs J2205+6012 (lower cyclic merit) and B1937+21 (higher cyclic merit).}%
    \label{DSA_hist}%
\end{figure}

\begin{figure}[!ht]
    \centering
    \captionsetup[subfigure]{labelformat=empty}
    {\hspace*{-.4cm}\includegraphics[width=0.5\textwidth]{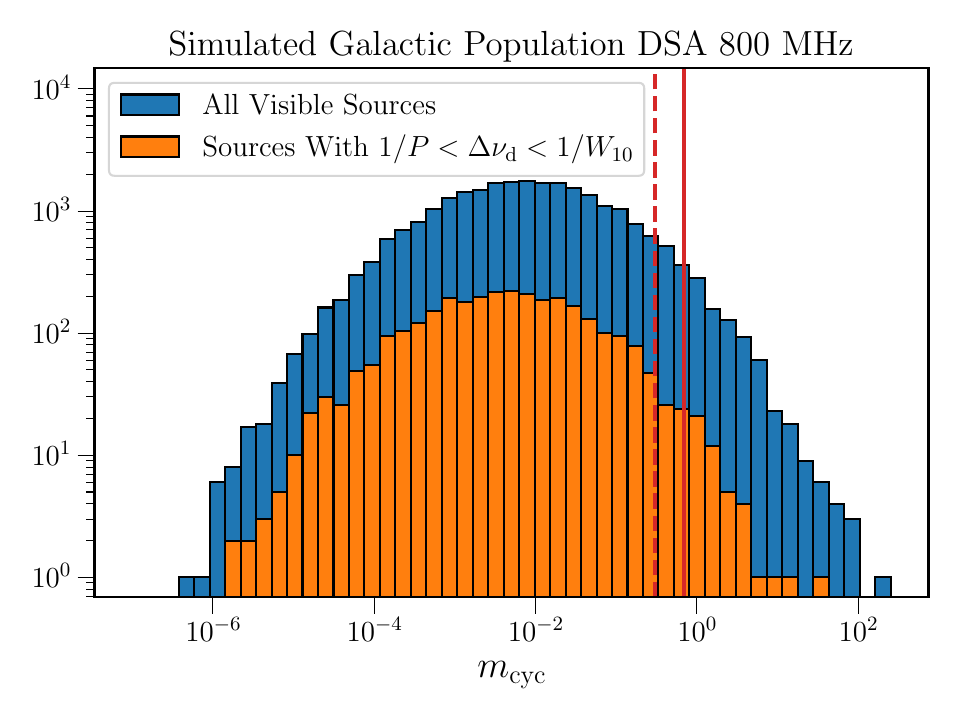} }%
    \caption{Cyclic merits for all sources from a realization of our simulations analyzed for the DSA at 800 MHz (blue), and the subset of sources in the full deconvolution regime (orange). Plot description same as that in Figure \ref{GBT_hist}.}%
    \label{DSA_sim_hist}%
\end{figure}

\begin{figure}[!ht]
    \centering
    \captionsetup[subfigure]{labelformat=empty}
    {\hspace*{-.4cm}\includegraphics[width=0.5\textwidth]{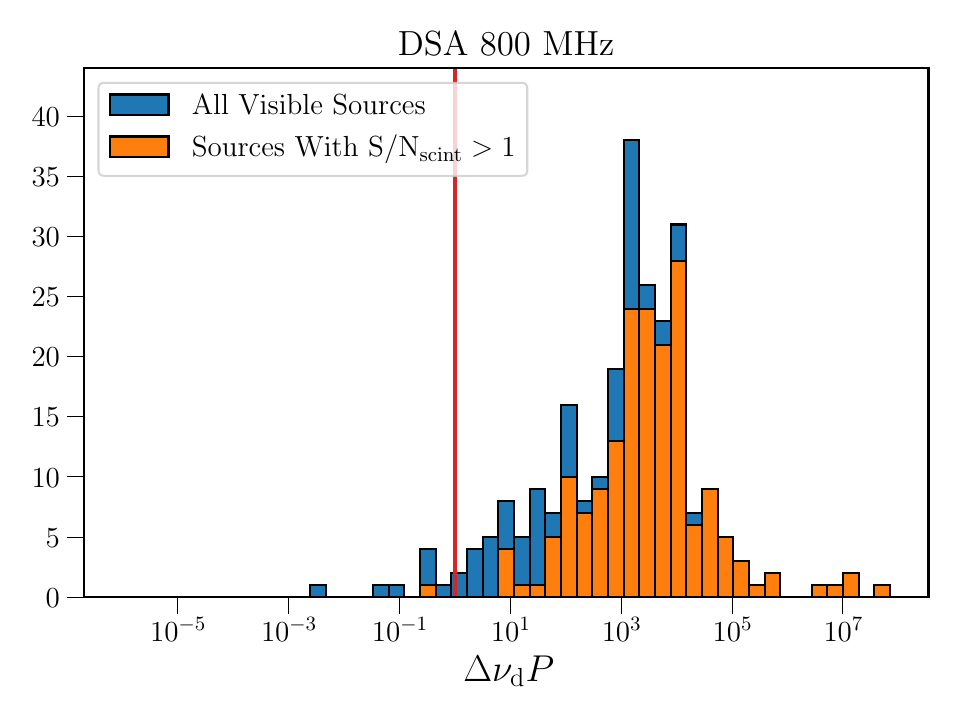} }%
    \caption{Expected scintle resolvability for all sources analyzed for the DSA at 800 MHz (blue), and the subset of sources with S/N$>$1 (orange). Plot description same as that in Figure \ref{GBT_scint_hist}.}%
    \label{DSA_scint_hist}%
\end{figure}

\begin{figure}[!ht]
    \centering
    \captionsetup[subfigure]{labelformat=empty}
    {\hspace*{-.4cm}\includegraphics[width=0.5\textwidth]{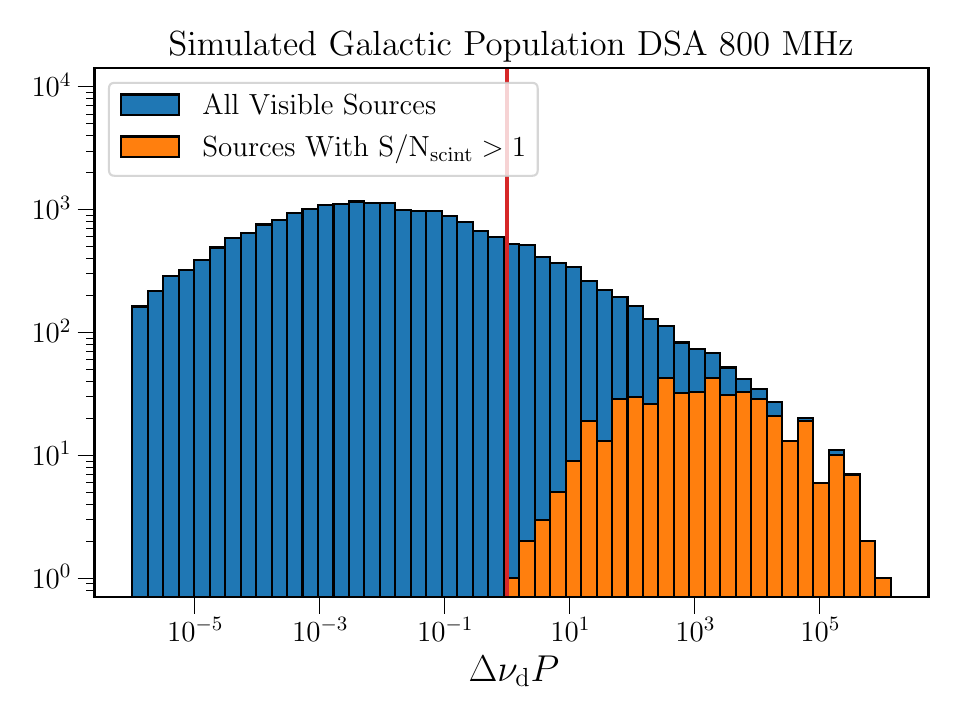} }%
    \caption{Expected scintle resolvability from a realization of our simulations analyzed for the DSA at 800 MHz (blue), and the subset of sources with S/N$>$1 (orange). Plot description same as that in Figure \ref{GBT_scint_hist}.}%
    \label{sim_DSA_scint_hist}%
\end{figure}

 Expected to be a primary instrument for NANOGrav beginning in the late 2020s, the incredible sensitivity and significant sky coverage of the Schmidt Sciences DSA would make it an outstanding instrument for cyclic spectroscopy. That capability is currently limited by its frequency coverage, which, expected to span from 0.7$-$2 GHz, is too high for the vast majority of sources. Even so, we would expect cyclic deconvolution to be achievable with PSR J1802$-$2124, which is already timed by NANOGrav \citep{Agazie_2023_timing}, and PSR J2205+6012 if observed 150-200 MHz higher than the frequency we chose for this study. The full histogram for all real sources analyzed using the DSA at 800 MHz is shown in Figure \ref{DSA_hist}. That all being said, we should expect a significant increase in viable sources for cyclic deconvolution once the DSA comes online, as our simulations predict that the vast majority of pulsars for this telescope that pass our conservative threshold and are observable within the full deconvolution regime likely have not yet been discovered. Indeed, if the predictions of our simulations prove correct, they would make the DSA the best telescope for cyclic spectroscopy out of all instruments considered, despite not operating in the ideal frequency range for this technique. An example realization from these simulations is shown in Figure \ref{DSA_sim_hist}. Additionally, we find that the DSA should be superb for recovering scintillation in pulsars at 800 MHz, with 178 sources having resolvable scintles that pass our S/N threshold, by far the most out of any setup we have considered, and the highest percentage relative to sources that were visible (83\%). Furthermore, our simulations predict that over half of all sources with detectable scintles at 800 MHz for the DSA have likely not yet been discovered. The scintle resolution prospects for all real sources with the DSA at 800 MHz are shown in Figure \ref{DSA_scint_hist}, while an example realization for our simulated population is shown in Figure \ref{sim_DSA_scint_hist}.

\subsection{MeerKAT 600 MHz}

\begin{figure}[!ht]
    \centering
    \captionsetup[subfigure]{labelformat=empty}
    {\hspace{-.8cm}\includegraphics[width=0.5\textwidth]{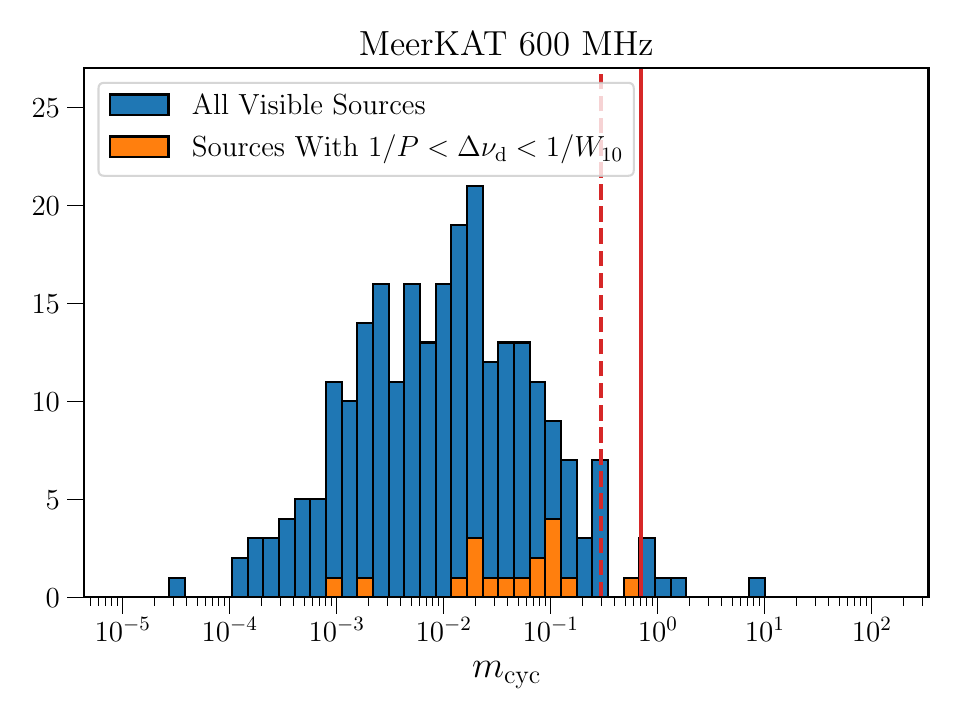} }%
    \caption{Cyclic merits for all real sources analyzed for MeerKAT at 600 MHz (blue), and the subset of sources in the full deconvolution regime (orange). Plot description same as that in Figure \ref{GBT_hist}. The outlier on the right is PSR B1937+21.}%
    \label{Meer_hist}%
\end{figure}

\begin{figure}[!ht]
    \centering
    \captionsetup[subfigure]{labelformat=empty}
    {\hspace*{-.4cm}\includegraphics[width=0.5\textwidth]{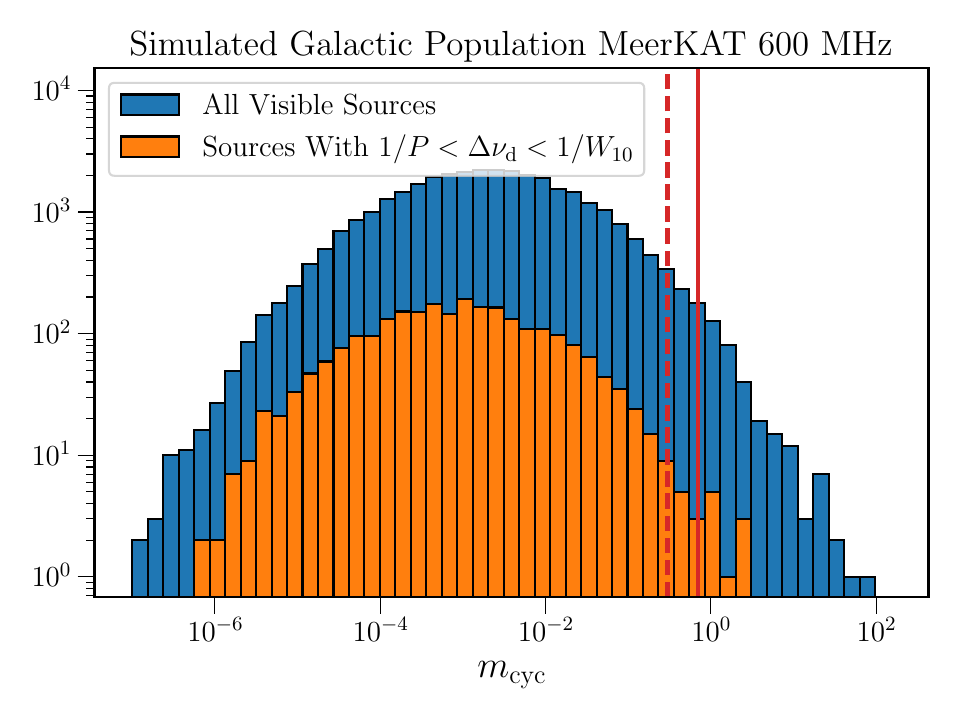} }%
    \caption{Cyclic merits for all sources from a realization of our simulations analyzed for MeerKAT at 600 MHz (blue), and the subset of sources in the full deconvolution regime (orange). Plot description same as that in Figure \ref{GBT_hist}.}%
    \label{Meer_sim_hist}%
\end{figure}

\begin{figure}[!ht]
    \centering
    \captionsetup[subfigure]{labelformat=empty}
    {\hspace*{-.4cm}\includegraphics[width=0.5\textwidth]{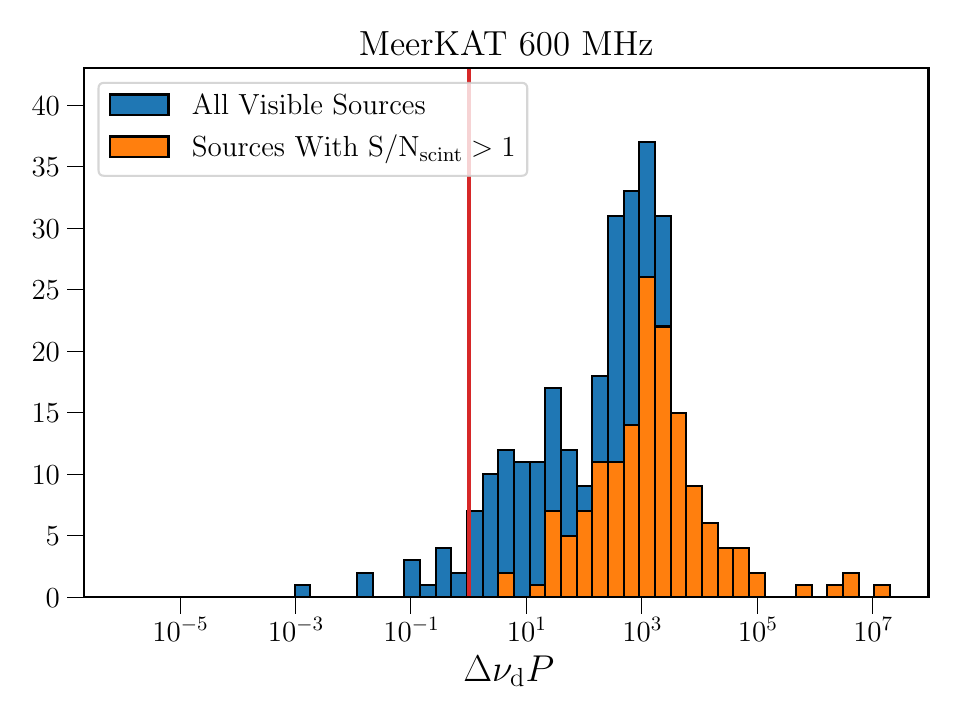} }%
    \caption{Expected scintle resolvability for all sources analyzed for MeerKAT at 600 MHz (blue), and the subset of sources with S/N$>$1 (orange). Plot description same as that in Figure \ref{GBT_scint_hist}.}%
    \label{Meer_scint_hist}%
\end{figure}

\begin{figure}[!ht]
    \centering
    \captionsetup[subfigure]{labelformat=empty}
    {\hspace*{-.4cm}\includegraphics[width=0.5\textwidth]{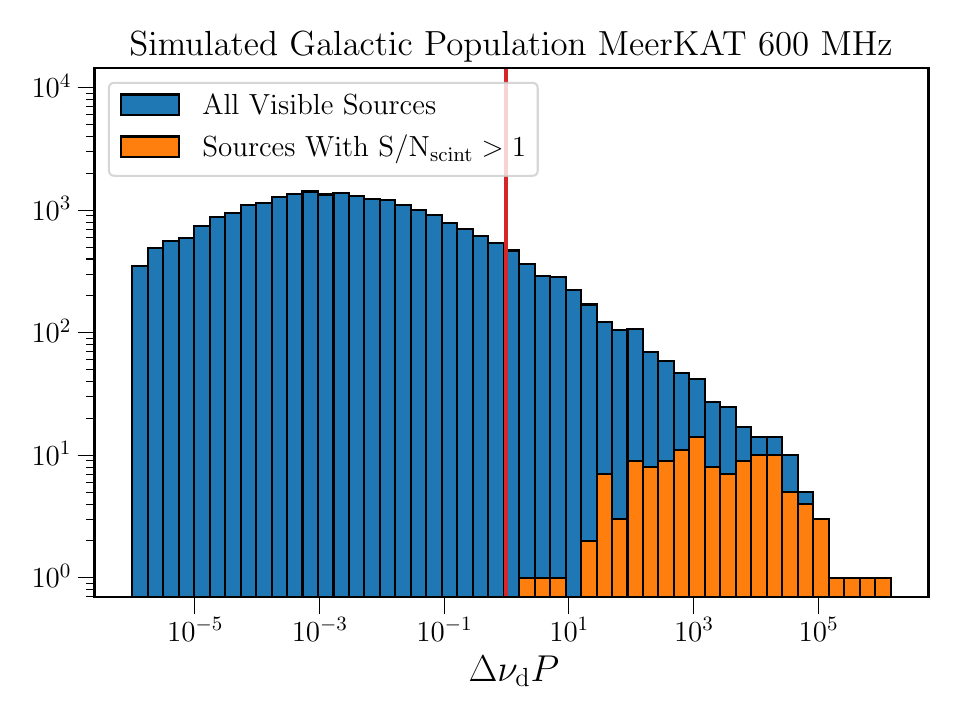} }%
    \caption{Expected scintle resolvability from a realization of our simulations analyzed for MeerKAT at 600 MHz (blue), and the subset of sources with S/N$>$1 (orange). Plot description same as that in Figure \ref{GBT_scint_hist}.}%
    \label{sim_Meer_scint_hist}%
\end{figure}

Another telescope with fantastic sensitivity, like the DSA, MeerKAT's limiting constraint for cyclic spectroscopy is its frequency coverage, likely relegating most cyclic spectroscopy-based studies done with the Square Kilometer Array (SKA) precursors to SKA-Low pathfinders rather than SKA-Mid pathfinders. However, once fully operational, SKA-Mid should be able to observe as low as 350 MHz, making it a much more compelling instrument for cyclic spectroscopy. As it stands, MeerKAT should be able to achieve cyclic deconvolution for PSR J1643$-$1224 in the UHF band (580-1015 MHz). Additionally, PSRs J0955$-$6150, J1017$-$7156, and B1937+21, while having full deconvolution regime upper bounds a few tens of MHz lower than the center frequency we chose for our analysis, would also be very likely to achieve cyclic deconvolution if observed near the bottom of that band. The full histogram for all real sources analyzed using MeerKAT at 600 MHz is shown in Figure \ref{Meer_hist}. As with the DSA, our simulations predict that the vast majority of viable cyclic deconvolution sources for MeerKAT have likely not been discovered yet. An example realization from these simulations is shown in Figure \ref{Meer_sim_hist}. Additionally, we find that MeerKAT should be fantastic for recovering scintillation in pulsars at 600 MHz, with 151 sources having resolvable scintles that pass our S/N threshold. Despite this, our simulations indicate that most, if not all, sources that could have detectable scintles at 600 MHz for MeerKAT have likely already been discovered. The scintle resolution prospects for all real sources with MeerKAT at 600 MHz are shown in Figure \ref{Meer_scint_hist}, while an example realization for our simulated population is shown in Figure \ref{sim_Meer_scint_hist}.

\subsection{FAST 550 MHz}

\par
\begin{figure}[!ht]
    \centering
    \captionsetup[subfigure]{labelformat=empty}
    {\hspace*{-.5cm}\includegraphics[width=0.5\textwidth]{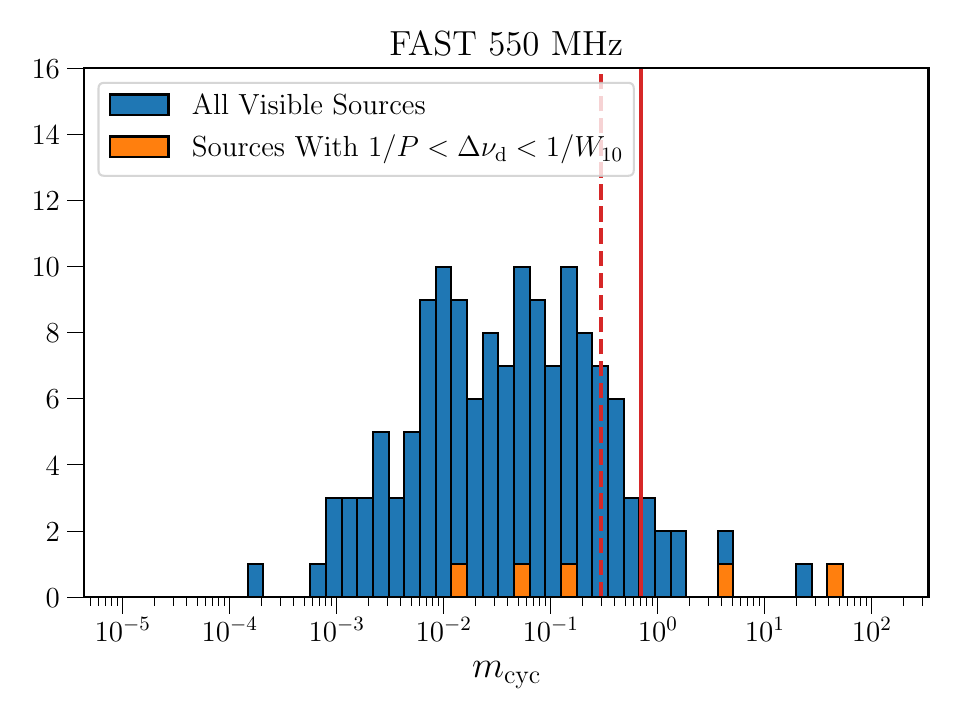} }%
    \caption{Cyclic merits for all real sources analyzed for FAST at 550 MHz (blue), and the subset of sources in the full deconvolution regime (orange). Plot description same as that in Figure \ref{GBT_hist}. The two outliers on the right are PSRs J2205+6012 (blue) and  B1937+21 (orange).}%
    \label{FAST_hist}%
\end{figure}

\begin{figure}[!ht]
    \centering
    \captionsetup[subfigure]{labelformat=empty}
    {\hspace*{-.4cm}\includegraphics[width=0.5\textwidth]{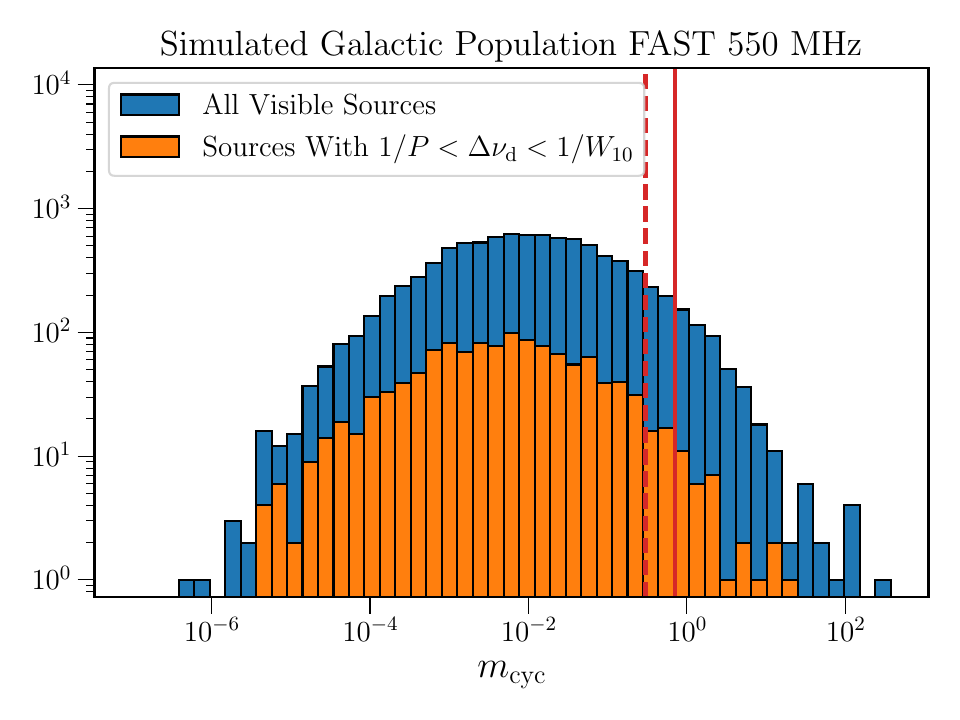} }%
    \caption{Cyclic merits for all sources from a realization of our simulations analyzed for FAST at 550 MHz (blue), and the subset of sources in the full deconvolution regime (orange). Plot description same as that in Figure \ref{GBT_hist}.}%
    \label{FAST_sim_hist}%
\end{figure}

\begin{figure}[!ht]
    \centering
    \captionsetup[subfigure]{labelformat=empty}
    {\hspace*{-.4cm}\includegraphics[width=0.5\textwidth]{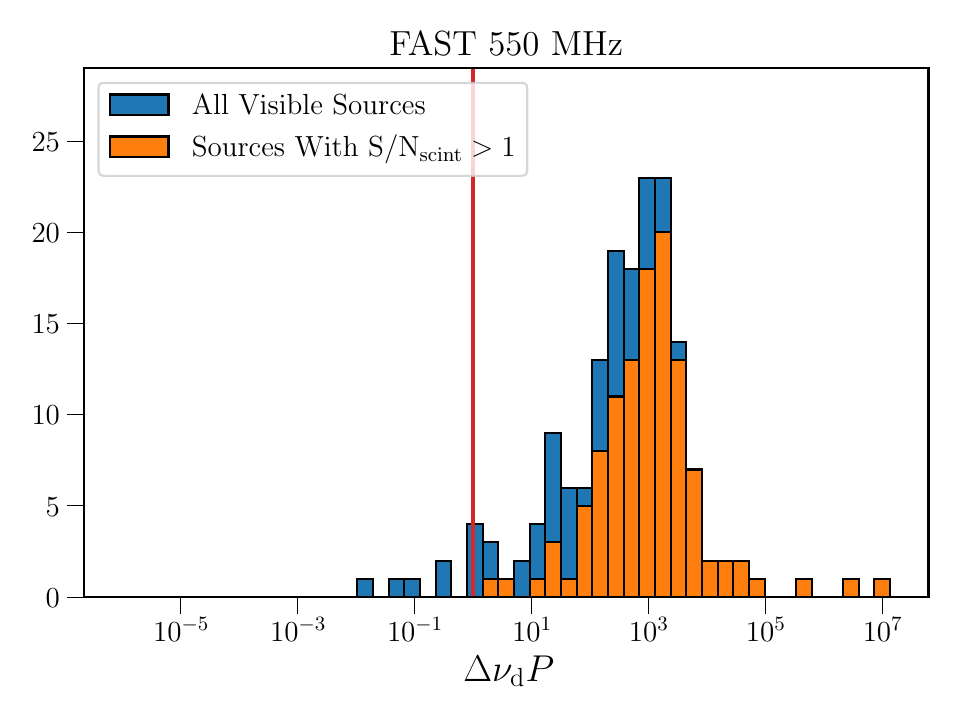} }%
    \caption{Expected scintle resolvability for all sources analyzed for FAST at 550 MHz (blue), and the subset of sources with S/N$>$1 (orange). Plot description same as that in Figure \ref{GBT_scint_hist}.}%
    \label{FAST_scint_hist}%
\end{figure}

\begin{figure}[!ht]
    \centering
    \captionsetup[subfigure]{labelformat=empty}
    {\hspace*{-.4cm}\includegraphics[width=0.5\textwidth]{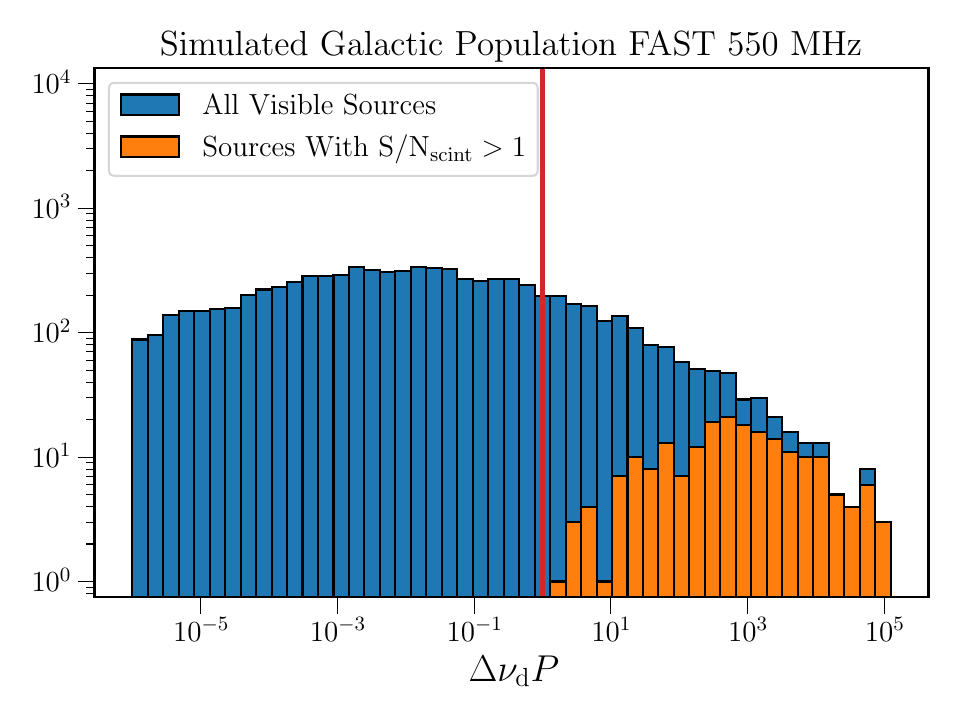} }%
    \caption{Expected scintle resolvability from a realization of our simulations analyzed for FAST at 550 MHz (blue), and the subset of sources with S/N$>$1 (orange). Plot description same as that in Figure \ref{GBT_scint_hist}.}%
    \label{sim_FAST_scint_hist}%
\end{figure}

 With appropriate frequency coverage, FAST could potentially be the best telescope in the northern hemisphere for cyclic spectroscopy. However, while lower frequency receivers for FAST exist, including one spanning 0.27$-$1.6 GHz with a considerable gain of around 10 K/Jy and system temperatures between 60 and 70 K \citep{fast_first}, they may not be currently available outside of commissioning and early science observations. Recently, another ultra-wideband receiver spanning 0.5$-$3 GHz was tested that appears extremely promising for science use cases \citep{Liu_2022}. Given its low-frequency coverage, and significantly greater sensitivity than other existing telescopes at 550 MHz, we opted to analyze cyclic spectroscopy potential with this receiver instead. We find that at this observing frequency FAST would be excellent for cyclic deconvolution of PSRs J1643$-$1224 and B1937+21, as well as J2205+6012 if it were observed at around 800 MHz rather than the center frequency used for this analysis. Additionally, PSRs J1737-0811 and J2215+5135 may also be able to achieve cyclic deconvolution with this receiver, although we were not able to estimate upper bounds for their full deconvolution regimes. The full histogram for all real sources analyzed using FAST at 550 MHz is shown in Figure \ref{FAST_hist}. As with the previous two instruments, our simulations predict that the vast majority of viable cyclic deconvolution sources for FAST have likely not been discovered yet, or were only recently recently discovered, as in \citep{fast_survey_1,fast_survey_2}, and so were not included in this study. An example realization from these simulations is shown in Figure \ref{FAST_sim_hist}. Additionally, despite its limited sky coverage, FAST is incredibly efficient for recovering scintillation for the pulsars it can observe, with 112 (78\%) sources having resolvable scintles that pass our S/N threshold. Furthermore, our simulations predict that about half of all sources with detectable scintles at 550 MHz for FAST have likely not yet been discovered. The scintle resolution prospects for all real sources with FAST at 550 MHz are shown in Figure \ref{FAST_scint_hist}, while an example realization for our simulated population is shown in Figure \ref{sim_FAST_scint_hist}.

\subsection{MWA 150 and 200 MHz}

\begin{figure}[!ht]
    \centering
    \captionsetup[subfigure]{labelformat=empty}
    {\hspace*{-.5cm}\includegraphics[width=0.5\textwidth]{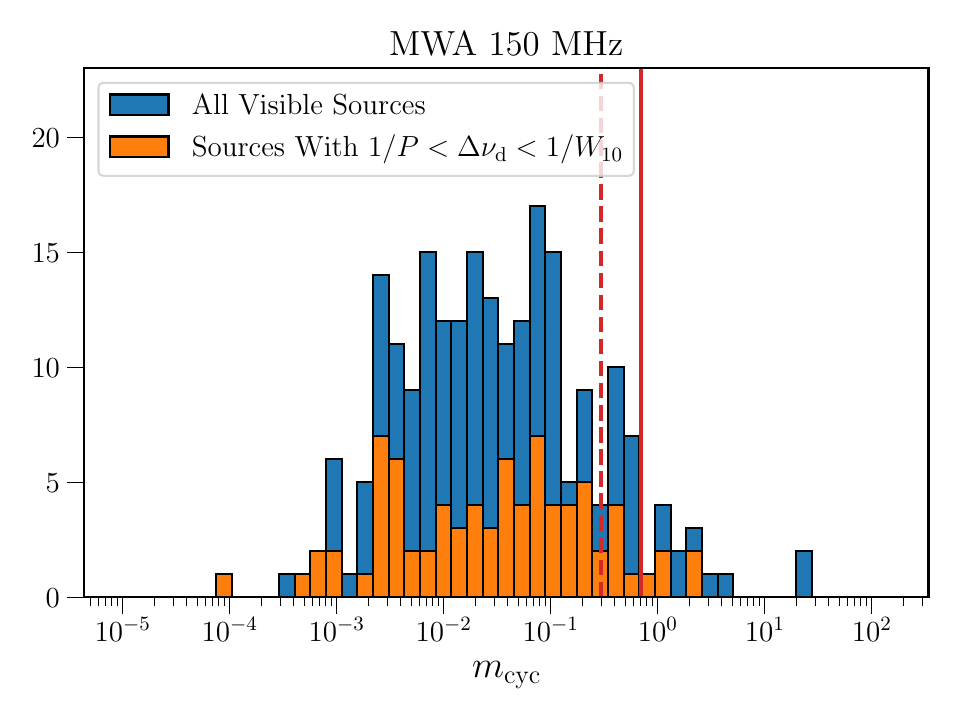} }%
    \caption{Cyclic merits for all real sources analyzed for MWA at 150 MHz (blue), and the subset of sources in the full deconvolution regime (orange). Plot description same as that in Figure \ref{GBT_hist}. The two outliers on the right are PSRs J0955$-$6150 and B1937+21.}%
    \label{MWA_150_hist}%
\end{figure}

\begin{figure}[!ht]
    \centering
    \captionsetup[subfigure]{labelformat=empty}
    {\hspace*{-.5cm}\includegraphics[width=0.5\textwidth]{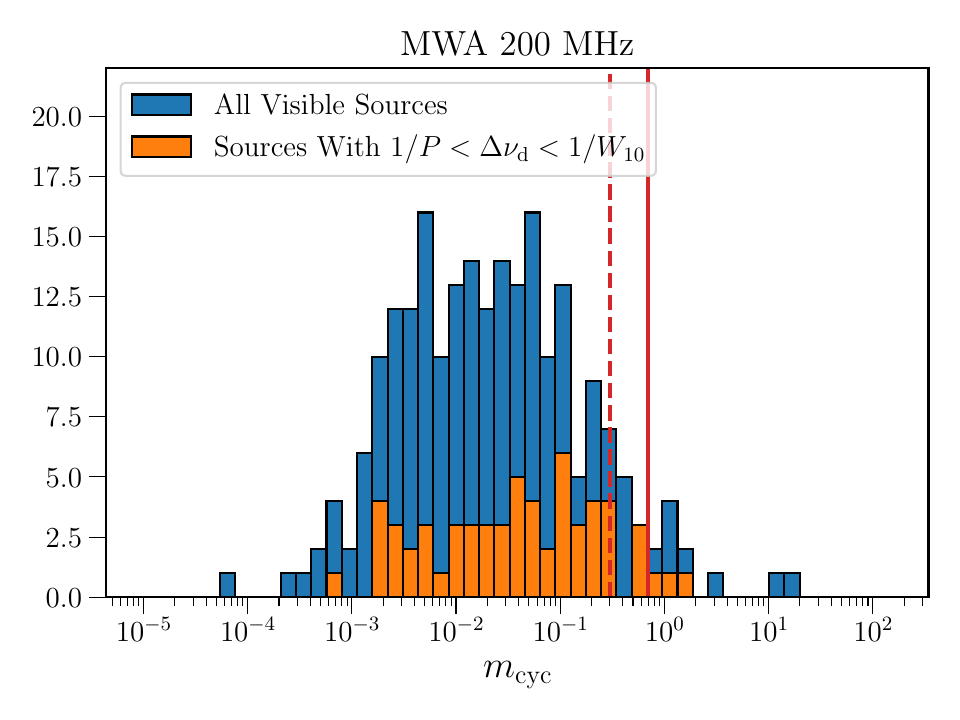}} %
    \caption{Cyclic merits for all real sources analyzed for MWA at 200 MHz (blue), and the subset of sources in the full deconvolution regime (orange). Plot description same as that in Figure \ref{GBT_hist}. The two outliers on the right are PSRs J0955$-$6150 and B1937+21.}%
    \label{MWA_200_hist}%
\end{figure}

\begin{figure}[!ht]
    \centering
    \captionsetup[subfigure]{labelformat=empty}
    {\hspace*{-.4cm}\includegraphics[width=0.5\textwidth]{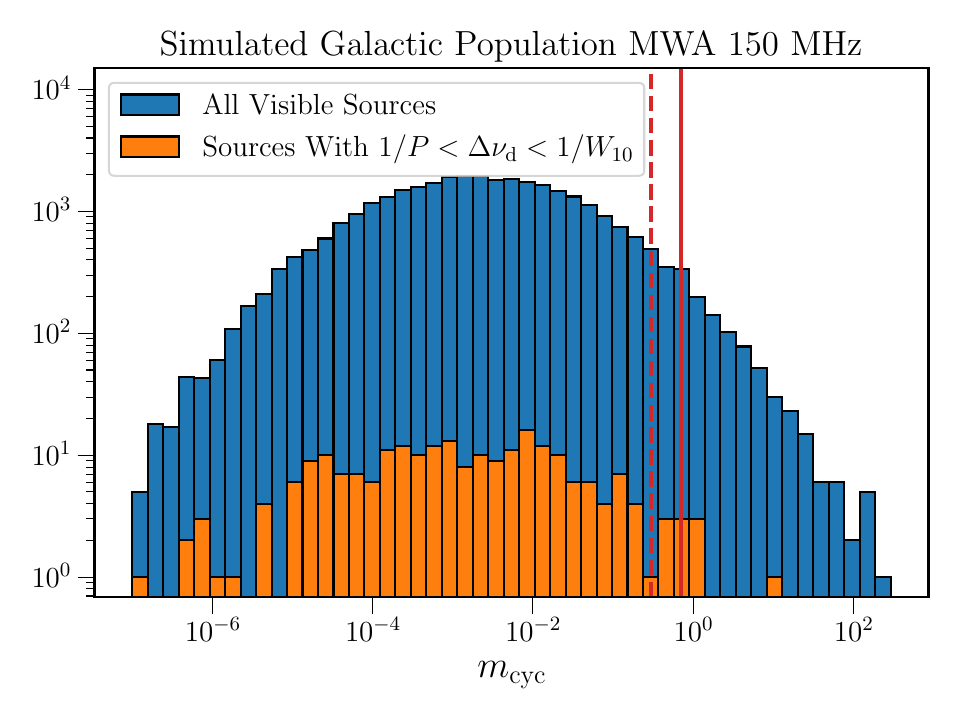} }%
    \caption{Cyclic merits for all sources from a realization of our simulations analyzed for MWA at 150 MHz (blue), and the subset of sources in the full deconvolution regime (orange). Plot description same as that in Figure \ref{GBT_hist}.}%
    \label{MWA_150_sim_hist}%
\end{figure}

\begin{figure}[!ht]
    \centering
    \captionsetup[subfigure]{labelformat=empty}
    {\hspace*{-.4cm}\includegraphics[width=0.5\textwidth]{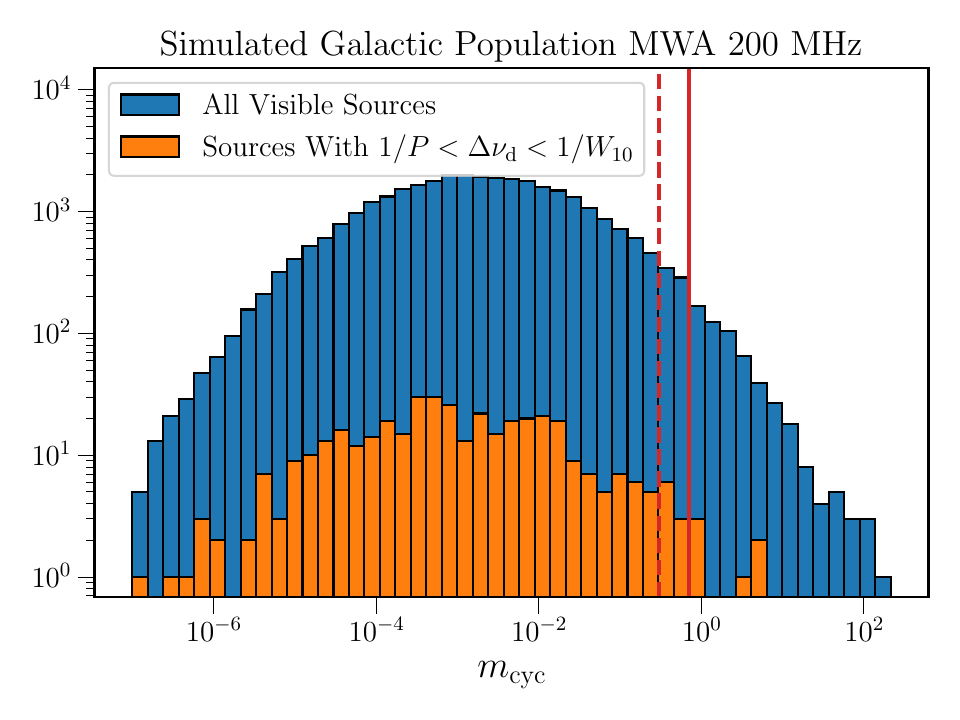} }%
    \caption{Cyclic merits for all sources from a realization of our simulations analyzed for MWA at 200 MHz (blue), and the subset of sources in the full deconvolution regime (orange). Plot description same as that in Figure \ref{GBT_hist}.}%
    \label{MWA_200_sim_hist}%
\end{figure}

\begin{figure}[!ht]
    \centering
    \captionsetup[subfigure]{labelformat=empty}
    {\hspace*{-.4cm}\includegraphics[width=0.5\textwidth]{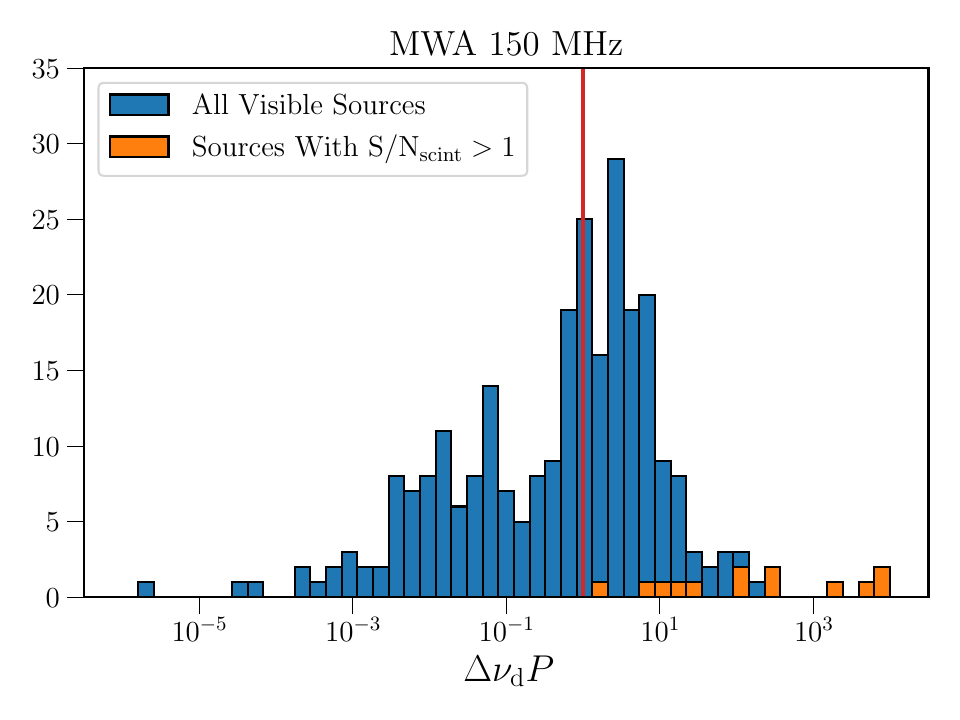} }%
    \caption{Expected scintle resolvability for all sources analyzed for MWA at 150 MHz (blue), and the subset of sources with S/N$>$1 (orange). Plot description same as that in Figure \ref{GBT_scint_hist}.}%
    \label{MWA_150_scint_hist}%
\end{figure}

\begin{figure}[!ht]
    \centering
    \captionsetup[subfigure]{labelformat=empty}
    {\hspace*{-.4cm}\includegraphics[width=0.5\textwidth]{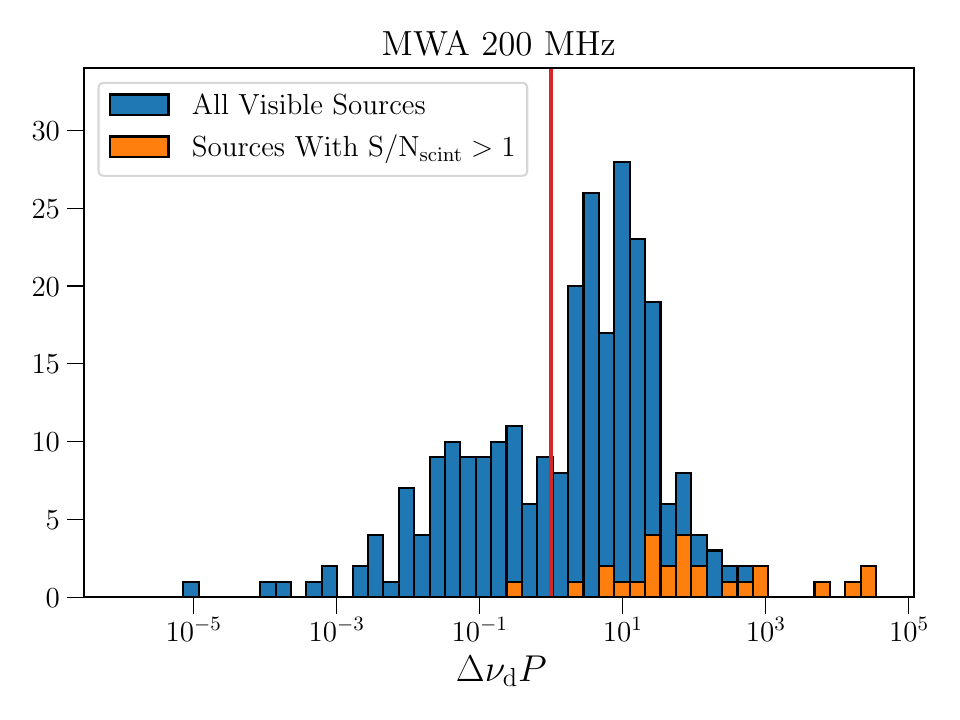} }%
    \caption{Expected scintle resolvability for all sources analyzed for MWA at 200 MHz (blue), and the subset of sources with S/N$>$1 (orange). Plot description same as that in Figure \ref{GBT_scint_hist}.}%
    \label{MWA_200_scint_hist}%
\end{figure}

\begin{figure}[!ht]
    \centering
    \captionsetup[subfigure]{labelformat=empty}
    {\hspace*{-.4cm}\includegraphics[width=0.5\textwidth]{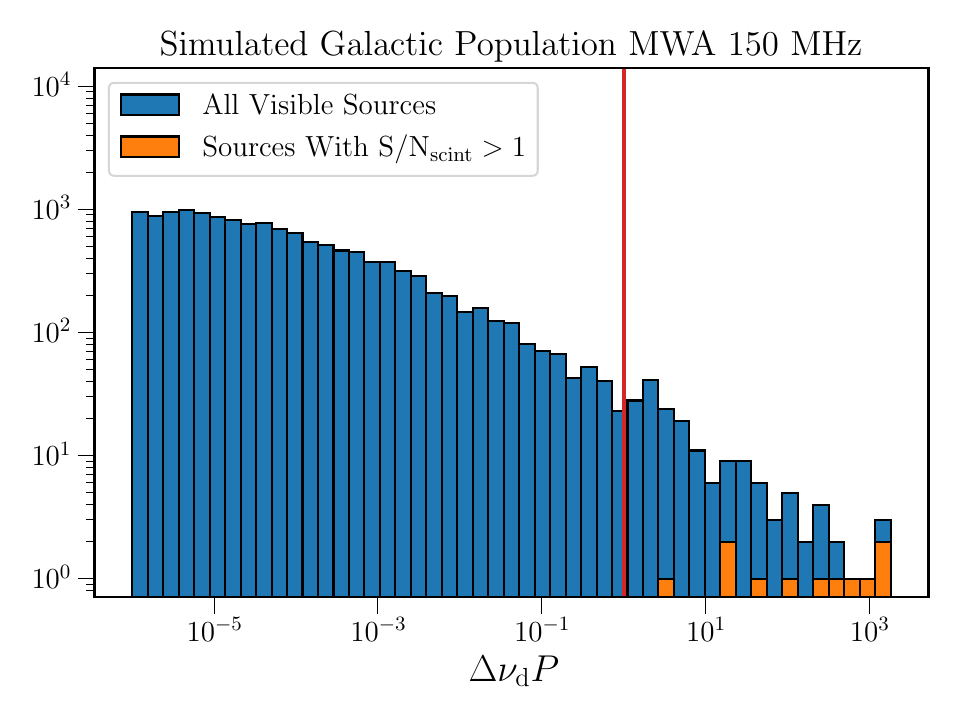} }%
    \caption{Expected scintle resolvability from a realization of our simulations analyzed for MWA at 150 MHz (blue), and the subset of sources with S/N$>$1 (orange). Plot description same as that in Figure \ref{GBT_scint_hist}.}%
    \label{sim_MWA_150_scint_hist}%
\end{figure}

\begin{figure}[!ht]
    \centering
    \captionsetup[subfigure]{labelformat=empty}
    {\hspace*{-.4cm}\includegraphics[width=0.5\textwidth]{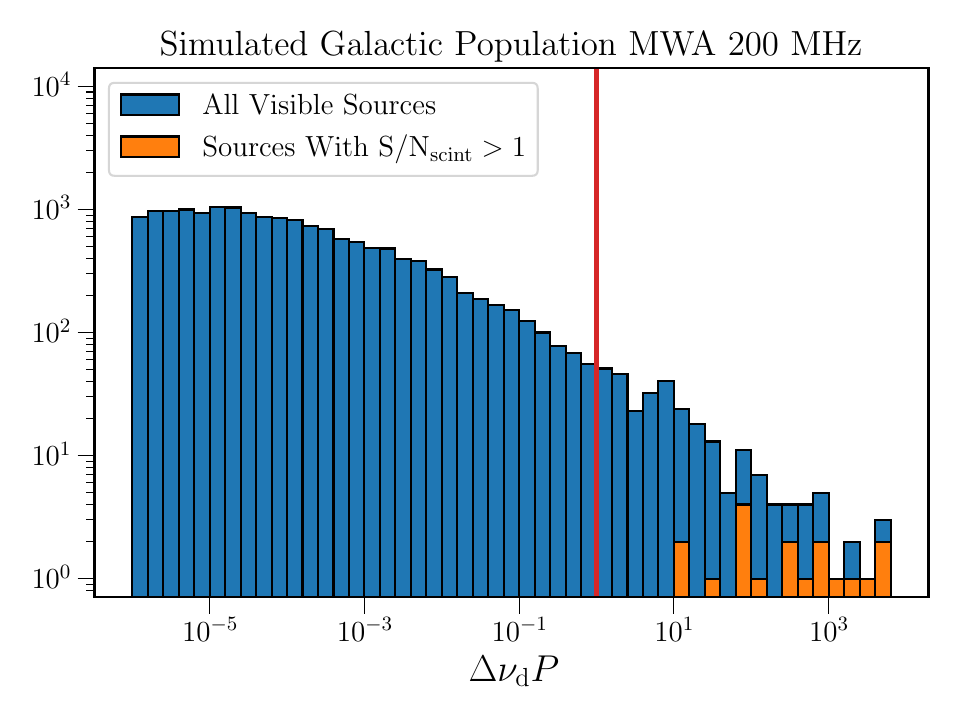} }%
    \caption{Expected scintle resolvability from a realization of our simulations analyzed for MWA at 200 MHz (blue), and the subset of sources with S/N$>$1 (orange). Plot description same as that in Figure \ref{GBT_scint_hist}.}%
    \label{sim_MWA_200_scint_hist}%
\end{figure}

Based on total sources that pass the conservative cyclic merit threshold while simultaneously being observed in the full deconvolution regime, MWA is likely the best current instrument for cyclic spectroscopy in the southern hemisphere, and shows strong promise for the cyclic spectroscopy capabilities of a fully functional SKA-Low. In total, we find 9 unique sources that meet the above criteria when considering both our 150 and 200 MHz measurements. Additionally, as with GBT, the Crab Pulsar has a very high cyclic merit with MWA at 200 MHz (although it would likely need to be observed 50 MHz higher), suggesting cyclic spectroscopy may be possible for some canonical pulsars. The full histograms for all real sources analyzed using these telescope-frequency combinations are shown in Figures \ref{MWA_150_hist} and \ref{MWA_200_hist}. Our simulations also indicate that most, if not all, sources that pass the conservative threshold and are observable within the full deconvolution regime are already known for MWA at both frequencies considered. Example realizations from these simulations at 150 and 200 MHz are shown in Figure \ref{MWA_150_sim_hist} and \ref{MWA_200_sim_hist}, respectively. Additionally, we find that there are around two times as many sources at 150 MHz and four times as many 200 MHz that have resolvable scintles and pass our S/N threshold as there are sources at these respective frequencies that are in their full deconvolution regimes and pass our conservative cyclic merit threshold. However, our simulations indicate that most, if not all, sources that could have detectable scintles at 150 and 200 MHz for MWA have likely already been discovered. The scintle resolution prospects for all real sources with MWA at 150 and 200 MHz are shown in Figures \ref{MWA_150_scint_hist} and \ref{MWA_200_scint_hist}, respectively, while example realizations for our simulated population are shown in Figures \ref{sim_MWA_150_scint_hist} and \ref{sim_MWA_200_scint_hist}.


\subsection{LWA 74 MHz}
\begin{figure}[!ht]
    \centering
    \captionsetup[subfigure]{labelformat=empty}
    {\hspace*{-.5cm}\includegraphics[width=0.5\textwidth]{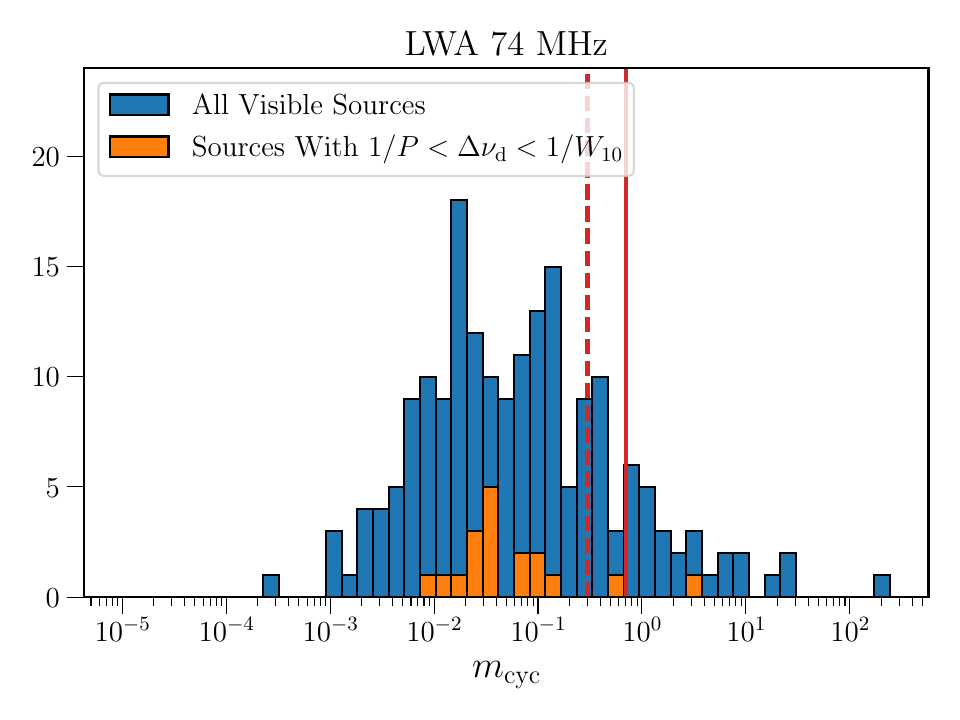}} %
    \caption{Cyclic merits for all real sources analyzed for LWA at 74 MHz (blue), and the subset of sources in the full deconvolution regime (orange). Plot description same as that in Figure \ref{GBT_hist}. The outlier on the right is PSR B1937+21.}%
    \label{LWA_74_hist}%
\end{figure}

\begin{figure}[!ht]
    \centering
    \captionsetup[subfigure]{labelformat=empty}
    {\hspace*{-.4cm}\includegraphics[width=0.5\textwidth]{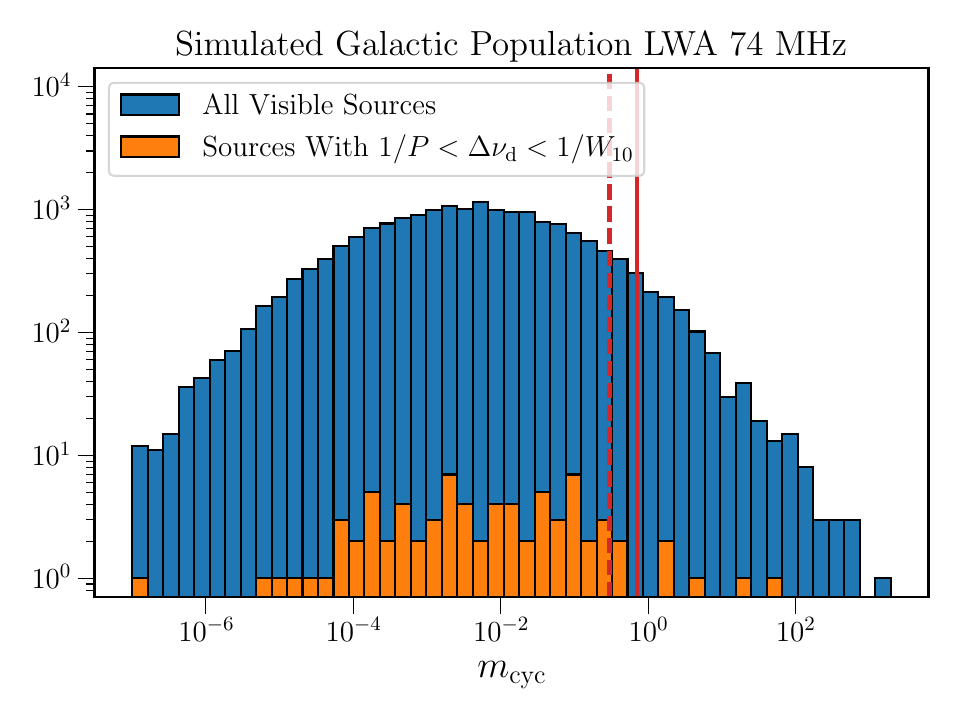} }%
    \caption{Cyclic merits for all sources from a realization of our simulations analyzed for LWA at 74 MHz (blue), and the subset of sources in the full deconvolution regime (orange). Plot description same as that in Figure \ref{GBT_hist}.}%
    \label{LWA_sim_hist}%
\end{figure}

\begin{figure}[!ht]
    \centering
    \captionsetup[subfigure]{labelformat=empty}
    {\hspace*{-.4cm}\includegraphics[width=0.5\textwidth]{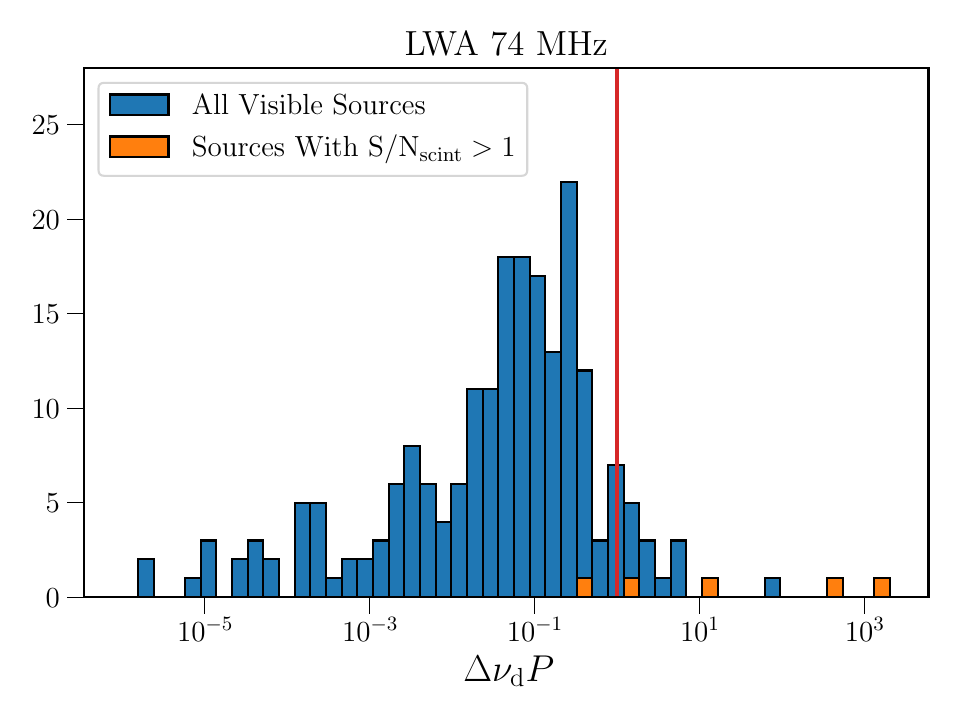} }%
    \caption{Expected scintle resolvability for all sources analyzed for LWA at 74 MHz (blue), and the subset of sources with S/N$>$1 (orange). Plot description same as that in Figure \ref{GBT_scint_hist}.}%
    \label{LWA_scint_hist}%
\end{figure}

\begin{figure}[!ht]
    \centering
    \captionsetup[subfigure]{labelformat=empty}
    {\hspace*{-.4cm}\includegraphics[width=0.5\textwidth]{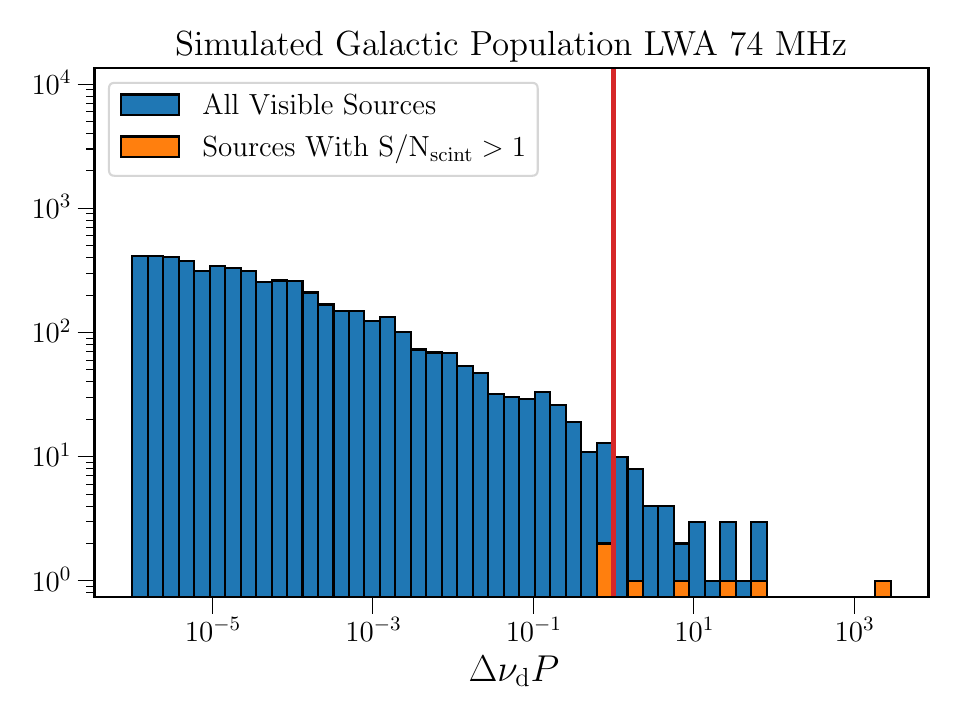} }%
    \caption{Expected scintle resolvability from a realization of our simulations analyzed for LWA at 74 MHz (blue), and the subset of sources with S/N$>$1 (orange). Plot description same as that in Figure \ref{GBT_scint_hist}.}%
    \label{sim_LWA_scint_hist}%
\end{figure}

Arguably the premier instrument below 100 MHz in the Western Hemisphere, LWA could be excellent for cyclic spectroscopy if not for the number of sources observable within full deconvolution regimes falling off rapidly after around 80 MHz. This falloff is likely due to a combination of pulses being scattered out, less overall S/N as scintles decrease in size faster than pulsars get brighter, and the galactic background temperature beginning to dominate. On the note of pulsars getting brighter, as mentioned earlier, it is likely that we are overestimating cyclic merits for many pulsars in this frequency band, as we assume simple power laws that imply pulsars continue to get brighter at lower frequencies, when in reality many pulsars experience spectral turnovers around 100 MHz \citep{Kumar_2025}. However, we still calculate that two pulsars, PSRs J1022+1001 and J2145$-$0750 pass the conservative cyclic merit threshold and are observable within their full deconvolution regimes. Additionally, we also find that PSR J1630+3734 has a very high cyclic merit with LWA at 74 MHz, and may be a strong candidate for cyclic deconvolution if observed around 7 MHz higher. In the next few years, LWA may become even better at cyclic deconvolution for the handful of sources that exist within their full deconvolution regimes below 80 MHz, as there are plans to rapidly expand the number of available dipoles through the commissioning of ``Swarm" stations, significantly increasing the overall collecting area of the instrument if using all stations for simultaneous observations is practically tractable \citep{swarm_1, swarm_2}. The full histograms for all real sources analyzed using LWA at 74 MHz combination are shown in Figures \ref{LWA_74_hist}. Our simulations also indicate that while most sources that pass the conservative threshold and are observable within the full deconvolution regime are already known for LWA at 74 MHz, a handful more may exist that have not been discovered. An example realization from these simulations is shown in Figure \ref{LWA_sim_hist}.  Additionally, we find that, without the upgrade of further Swarm stations, LWA will likely not be very useful for scintillation studies, with only five sources passing our S/N threshold. Our simulations also indicate that most, if not all, sources that could have detectable scintles at 74 MHz for LWA have likely already been discovered. The scintle resolution prospects for all real sources with LWA at 74 MHz are shown in Figure \ref{LWA_scint_hist}, while an example realization for our simulated population is shown in Figure \ref{sim_LWA_scint_hist}.

\subsection{LOFAR 150 and 200 MHz}

\begin{figure}[!ht]
    \centering
    \captionsetup[subfigure]{labelformat=empty}
    {\hspace*{-.5cm}\includegraphics[width=0.5\textwidth]{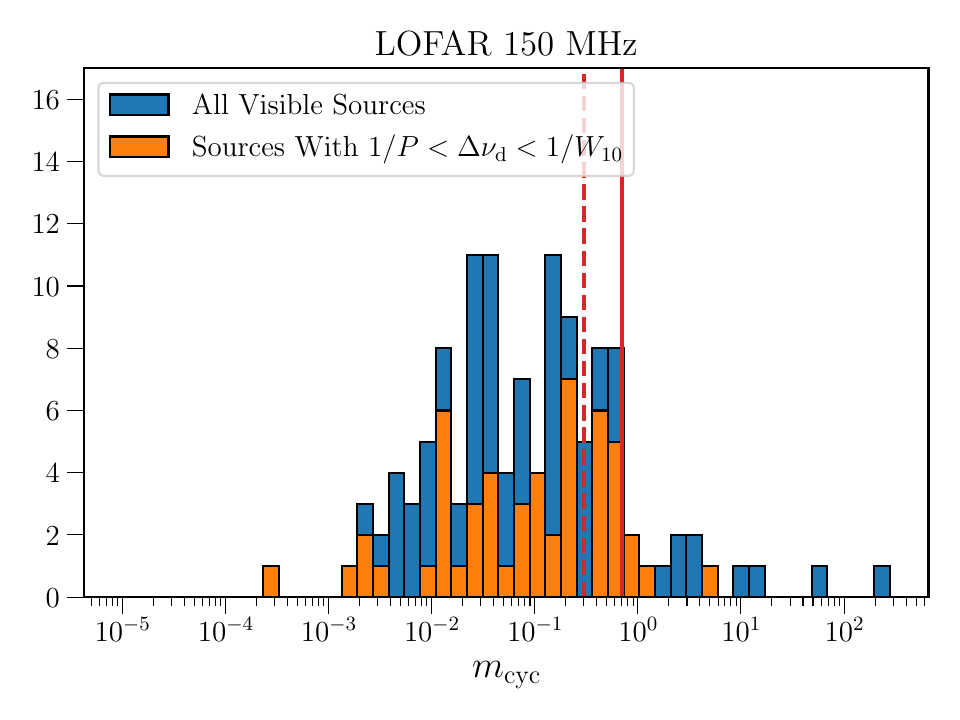} }%
    \caption{Cyclic merits for all real sources analyzed for LOFAR at 150 MHz (blue), and the subset of sources in the full deconvolution regime (orange). Plot description same as that in Figure \ref{GBT_hist}. The two outliers on the right are PSRs J2205+6012 (lower cyclic merit) and B1937+21 (higher cyclic merit).}%
    \label{LOFAR_150_hist}%
\end{figure}

\begin{figure}[!ht]
    \centering
    \captionsetup[subfigure]{labelformat=empty}
    {\hspace*{-.5cm}\includegraphics[width=0.5\textwidth]{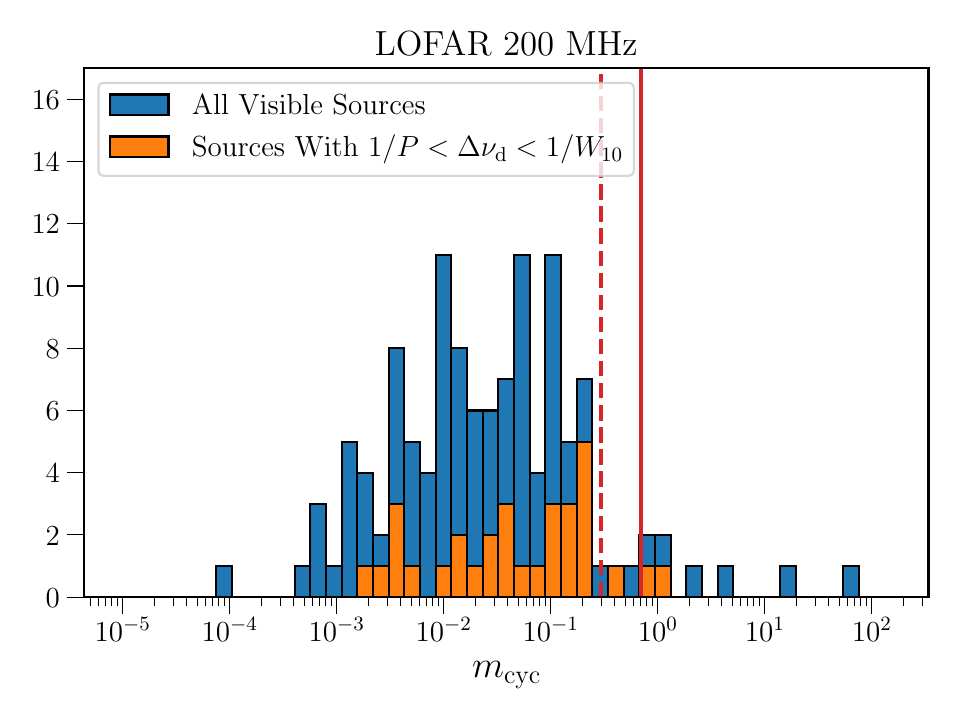}} %
    \caption{Cyclic merits for all real sources analyzed for LOFAR at 200 MHz (blue), and the subset of sources in the full deconvolution regime (orange). Plot description same as that in Figure \ref{GBT_hist}. The two outliers on the right are PSRs J2205+6012 (lower cyclic merit) and B1937+21 (higher cyclic merit).}%
    \label{LOFAR_200_hist}%
\end{figure}

\begin{figure}[!ht]
    \centering
    \captionsetup[subfigure]{labelformat=empty}
    {\hspace*{-.4cm}\includegraphics[width=0.5\textwidth]{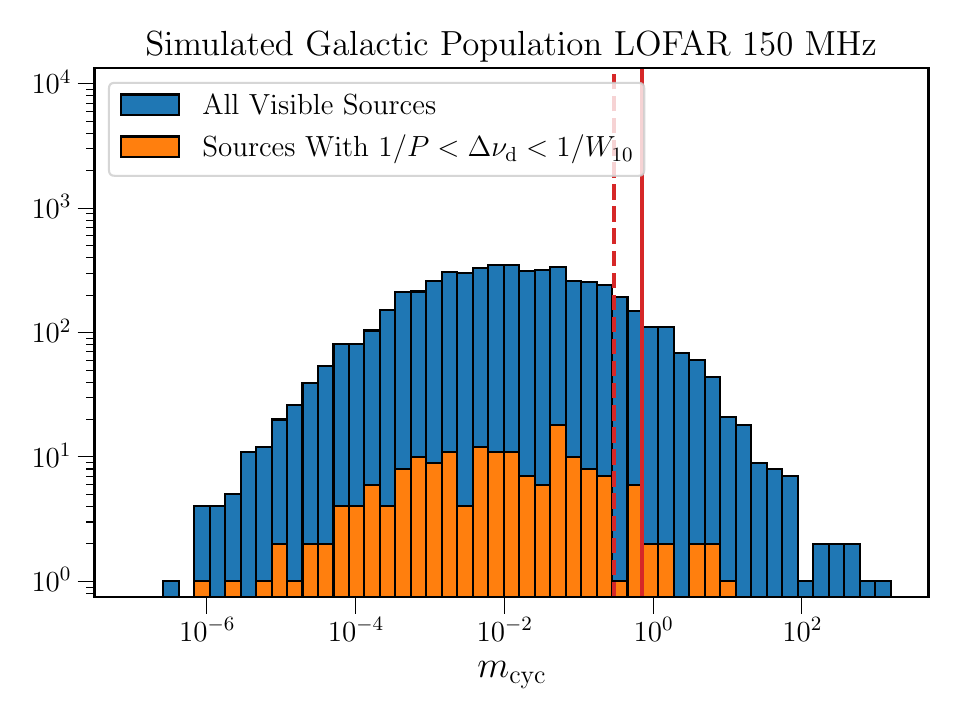} }%
    \caption{Cyclic merits for all sources from a realization of our simulations analyzed for LOFAR at 150 MHz (blue), and the subset of sources in the full deconvolution regime (orange). Plot description same as that in Figure \ref{GBT_hist}.}%
    \label{LOFAR_150_sim_hist}%
\end{figure}

\begin{figure}[!ht]
    \centering
    \captionsetup[subfigure]{labelformat=empty}
    {\hspace*{-.4cm}\includegraphics[width=0.5\textwidth]{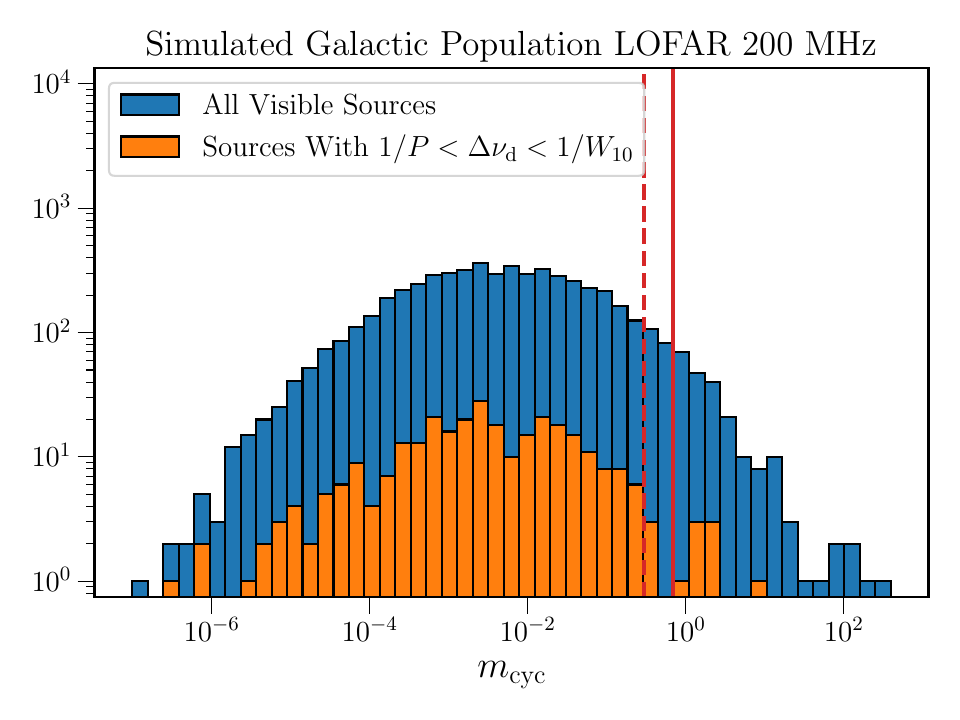} }%
    \caption{Cyclic merits for all sources from a realization of our simulations analyzed for LOFAR at 200 MHz (blue), and the subset of sources in the full deconvolution regime (orange). Plot description same as that in Figure \ref{GBT_hist}.}%
    \label{LOFAR_200_sim_hist}%
\end{figure}

\begin{figure}[!ht]
    \centering
    \captionsetup[subfigure]{labelformat=empty}
    {\hspace*{-.4cm}\includegraphics[width=0.5\textwidth]{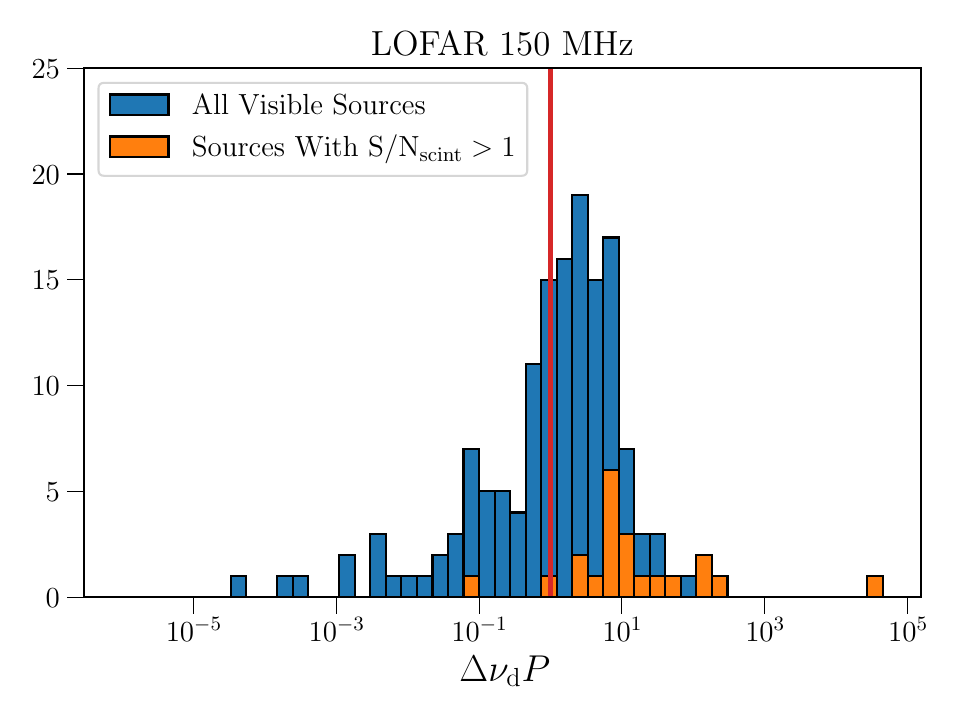} }%
    \caption{Expected scintle resolvability for all sources analyzed for LOFAR at 150 MHz (blue), and the subset of sources with S/N$>$1 (orange). Plot description same as that in Figure \ref{GBT_scint_hist}.}%
    \label{LOFAR_150_scint_hist}%
\end{figure}

\begin{figure}[!ht]
    \centering
    \captionsetup[subfigure]{labelformat=empty}
    {\hspace*{-.4cm}\includegraphics[width=0.5\textwidth]{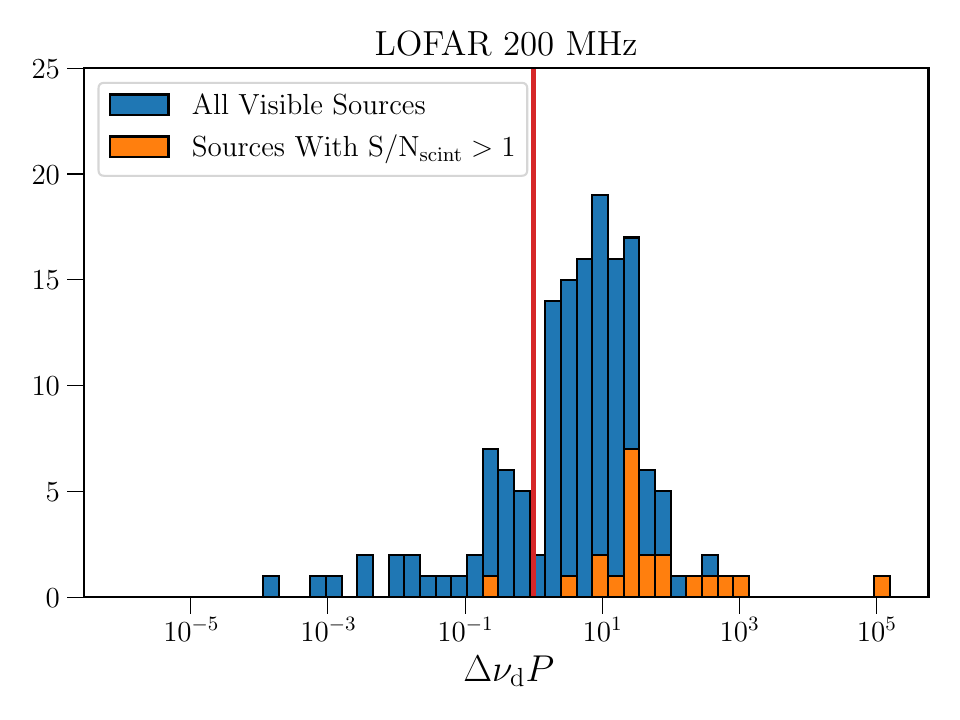} }%
    \caption{Expected scintle resolvability for all sources analyzed for LOFAR at 150 MHz (blue), and the subset of sources with S/N$>$1 (orange). Plot description same as that in Figure \ref{GBT_scint_hist}.}%
    \label{LOFAR_200_scint_hist}%
\end{figure}

\begin{figure}[!ht]
    \centering
    \captionsetup[subfigure]{labelformat=empty}
    {\hspace*{-.4cm}\includegraphics[width=0.5\textwidth]{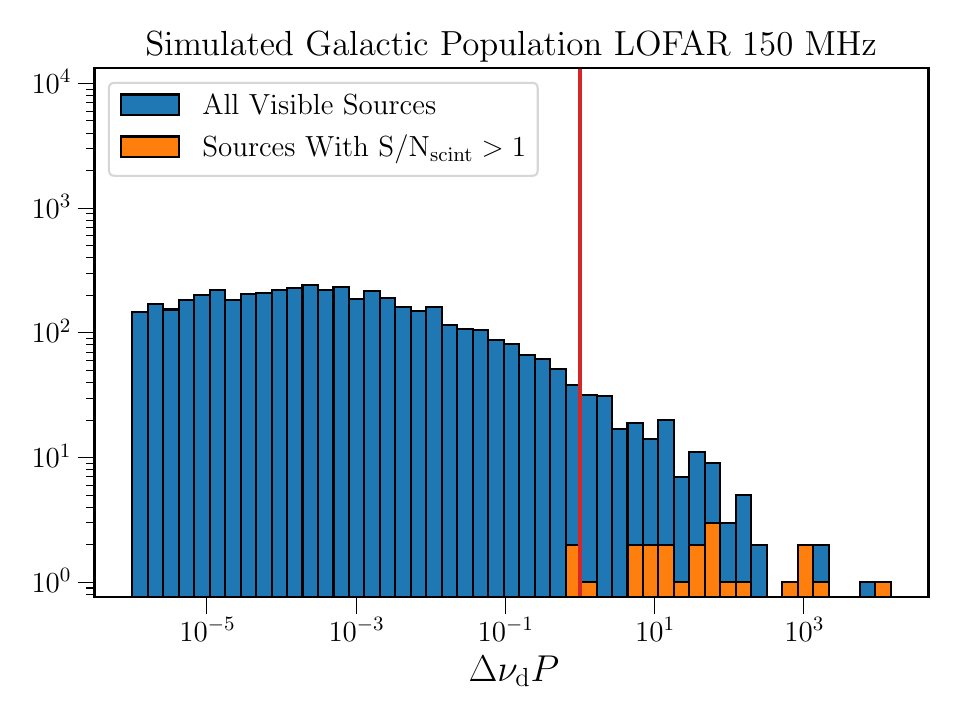} }%
    \caption{Expected scintle resolvability from a realization of our simulations analyzed for LOFAR at 150 MHz (blue), and the subset of sources with S/N$>$1 (orange). Plot description same as that in Figure \ref{GBT_scint_hist}.}%
    \label{sim_LOFAR_150_scint_hist}%
\end{figure}

\begin{figure}[!ht]
    \centering
    \captionsetup[subfigure]{labelformat=empty}
    {\hspace*{-.4cm}\includegraphics[width=0.5\textwidth]{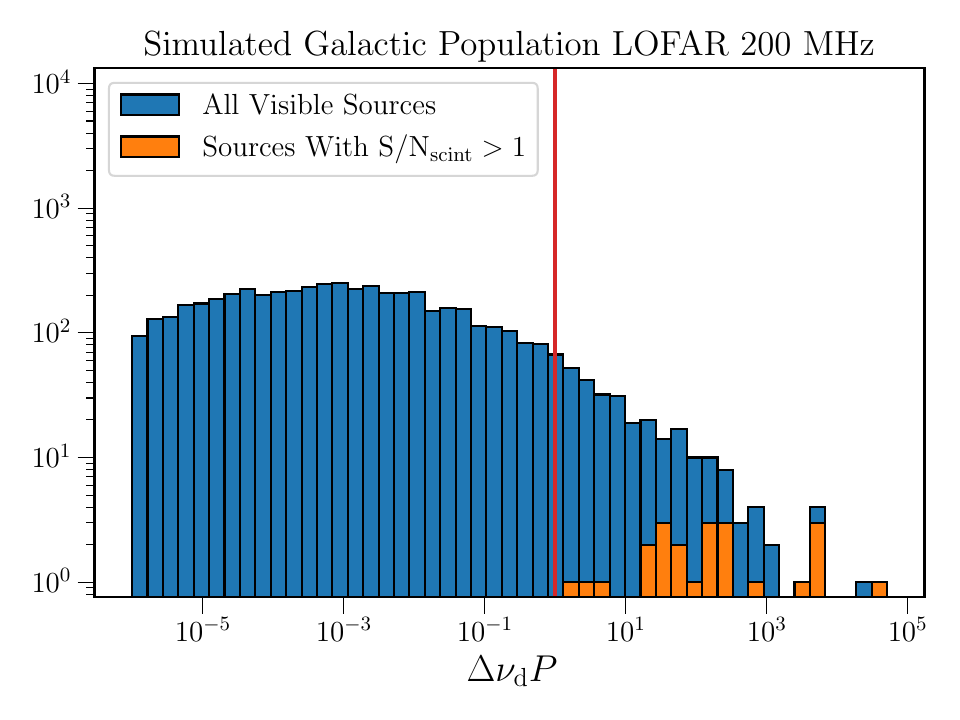} }%
    \caption{Expected scintle resolvability from a realization of our simulations analyzed for LOFAR at 200 MHz (blue), and the subset of sources with S/N$>$1 (orange). Plot description same as that in Figure \ref{GBT_scint_hist}.}%
    \label{sim_LOFAR_200_scint_hist}%
\end{figure}

Based on total sources that pass the conservative cyclic merit threshold while simultaneously being observed in the full deconvolution regime, among the nine telescopes we considered, LOFAR is likely the second best current instrument for cyclic spectroscopy, with 15 unique sources meeting the above criteria when accounting for both the 150 and 200 MHz bands, and 13 unique sources just at 150 MHz. Our survey likely undercounts the true number of viable sources LOFAR can observe, as we only considered the sensitivity of the LOFAR core, meaning there could potentially be twice as many viable sources if we also accounted for the remote and international stations. In addition to all sources meeting our criteria, PSR J2215+5135, while not having a determined upper bound for its full deconvolution regime, has a lower bound of around 190$-$200 MHz while exhibiting a large cyclic merit, indicating it would also very likely achieve cyclic deconvolution slightly above 200 MHz with LOFAR. Similarly, PSRs J1012+5307, J1022+1001, J1630+3734 have very large cyclic merits at 150 MHz, and would be in the full deconvolution regime if observed around 10$-$20 MHz lower. Additionally, PSR J0621+1002, while already meeting our criteria at 200 MHz, would very likely achieve cyclic deconvolution if observed around 160$-$170 MHz. Finally, as with GBT and MWA, the Crab Pulsar has a very high cyclic merit with LOFAR at 200 MHz (although would likely need to be observed 50 MHz higher), suggesting cyclic spectroscopy may be possible for some canonical pulsars. The full histograms for all real sources analyzed using these telescope-frequency combinations are shown in Figures \ref{LOFAR_150_hist} and \ref{LOFAR_200_hist}. Our simulations also indicate that most, if not all, sources that pass the conservative threshold and are observable within the full deconvolution regime are already known for LOFAR at 150 MHz, although we predict a handful more may exist at 200 MHz that have likely not been discovered. Example realizations from these simulations at 150 and 200 MHz are shown in Figure \ref{LOFAR_150_sim_hist} and \ref{LOFAR_200_sim_hist}, respectively. Additionally, we find that there are 21 sources both at 150 and 200 MHz that have resolvable scintles and pass our S/N threshold. Despite that, our simulations indicate that most, if not all, sources that could have detectable scintles at 150 and 200 MHz for LOFAR have likely already been discovered. The scintle resolution prospects for all real sources with LOFAR at 150 and 200 MHz are shown in Figures \ref{LOFAR_150_scint_hist} and \ref{LOFAR_200_scint_hist}, respectively, while example realizations for our simulated population are shown in Figures \ref{sim_LOFAR_150_scint_hist} and \ref{sim_LOFAR_200_scint_hist}.

\break
\subsection{NenuFAR 48 and 85 MHz}

\begin{figure}[!ht]
    \centering
    \captionsetup[subfigure]{labelformat=empty}
    {\hspace*{-.5cm}\includegraphics[width=0.5\textwidth]{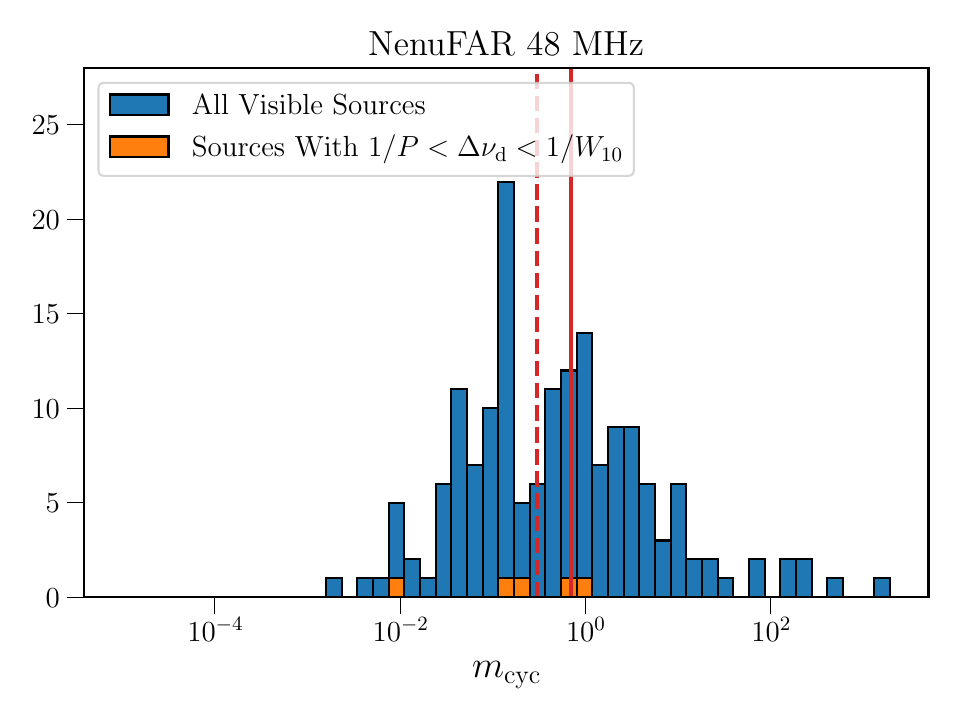}} %
    \caption{Cyclic merits for all real sources analyzed for NenuFAR at 48 MHz (blue), and the subset of sources in the full deconvolution regime (orange). Plot description same as that in Figure \ref{GBT_hist}.}%
    \label{NenuFAR_48_hist}%
\end{figure}

\begin{figure}
    \centering
    \captionsetup[subfigure]{labelformat=empty}
    {\hspace*{-.5cm}\includegraphics[width=0.5\textwidth]{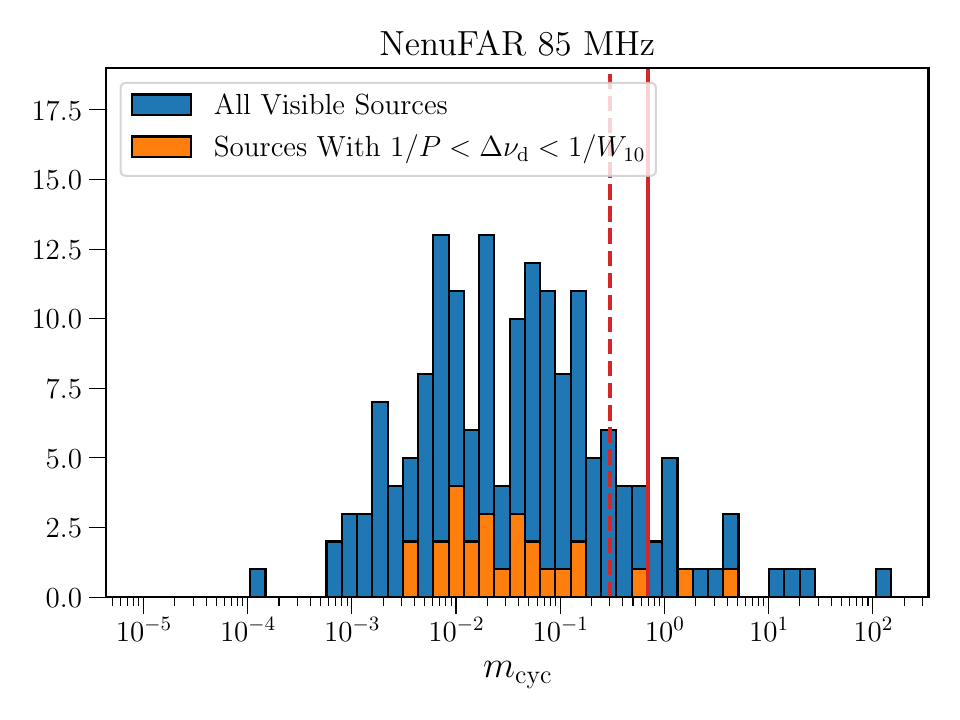}} %
    \caption{Cyclic merits for all real sources analyzed for NenuFAR at 85 MHz (blue), and the subset of sources in the full deconvolution regime (orange). Plot description same as that in Figure \ref{GBT_hist}. The outlier on the right is PSR B1937+21.}%
    \label{NenuFAR_85_hist}%
\end{figure}

\begin{figure}[!ht]
    \centering
    \captionsetup[subfigure]{labelformat=empty}
    {\hspace*{-.4cm}\includegraphics[width=0.5\textwidth]{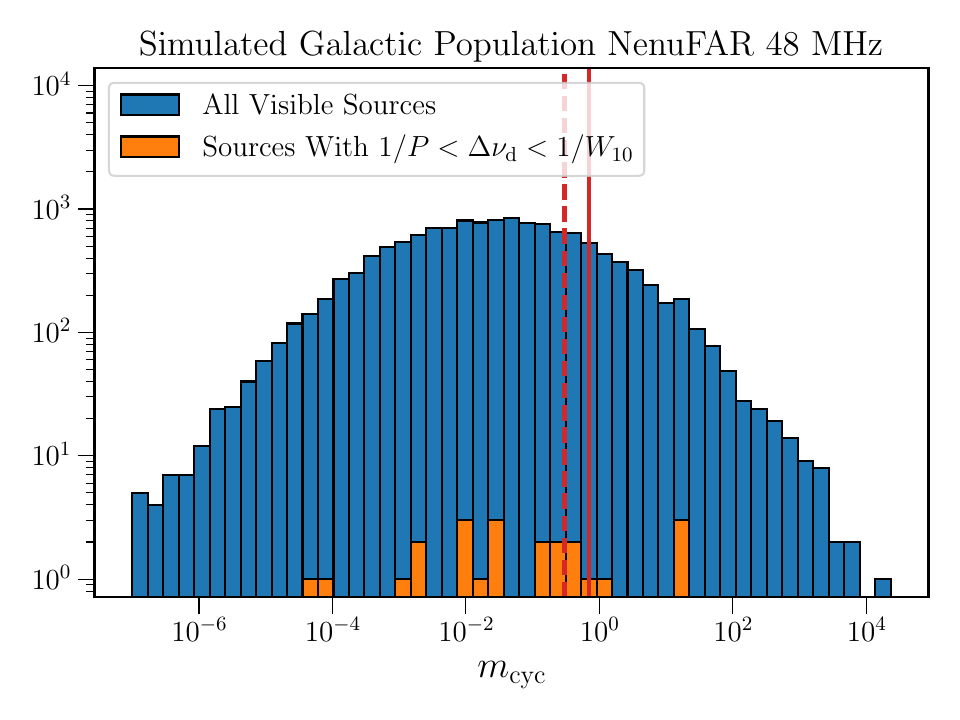} }%
    \caption{Cyclic merits for all sources from a realization of our simulations analyzed for NenuFAR at 48 MHz (blue), and the subset of sources in the full deconvolution regime (orange). Plot description same as that in Figure \ref{GBT_hist}.}%
    \label{NenuFAR_48_sim_hist}%
\end{figure}

\begin{figure}[!ht]
    \centering
    \captionsetup[subfigure]{labelformat=empty}
    {\hspace*{-.4cm}\includegraphics[width=0.5\textwidth]{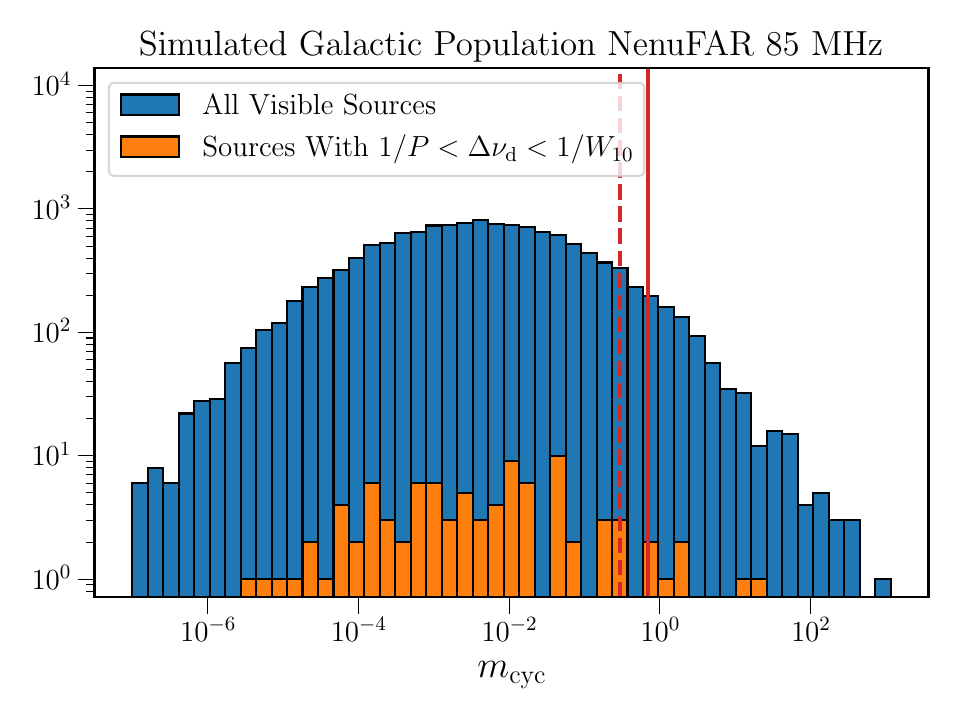} }%
    \caption{Cyclic merits for all sources from a realization of our simulations analyzed for NenuFAR at 85 MHz (blue), and the subset of sources in the full deconvolution regime (orange). Plot description same as that in Figure \ref{GBT_hist}.}%
    \label{NenuFAR_85_sim_hist}%
\end{figure}

\begin{figure}[!ht]
    \centering
    \captionsetup[subfigure]{labelformat=empty}
    {\hspace*{-.4cm}\includegraphics[width=0.5\textwidth]{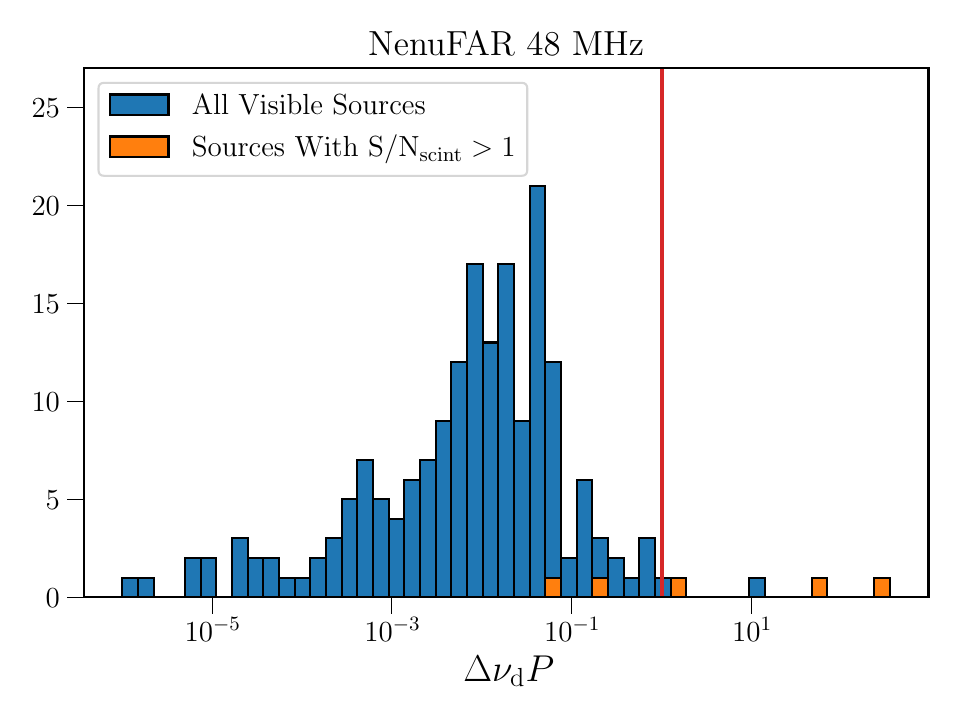} }%
    \caption{Expected scintle resolvability for all sources analyzed for NenuFAR at 48 MHz (blue), and the subset of sources with S/N$>$1 (orange). Plot description same as that in Figure \ref{GBT_scint_hist}.}%
    \label{NenuFAR_48_scint_hist}%
\end{figure}

\begin{figure}[!ht]
    \centering
    \captionsetup[subfigure]{labelformat=empty}
    {\hspace*{-.4cm}\includegraphics[width=0.5\textwidth]{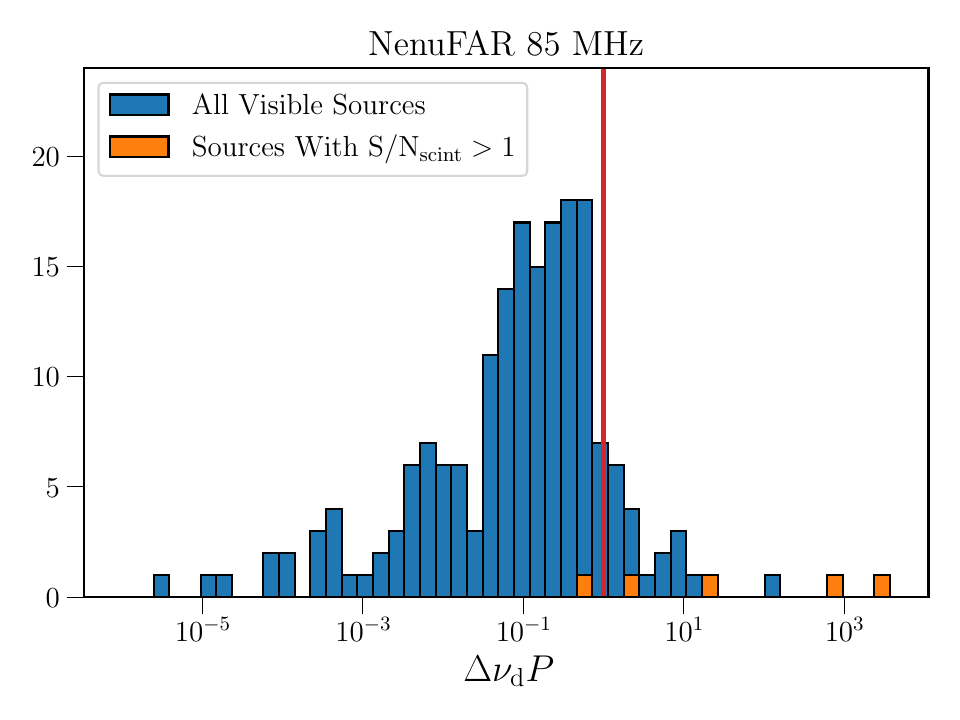} }%
    \caption{Expected scintle resolvability for all sources analyzed for NenuFAR at 85 MHz (blue), and the subset of sources with S/N$>$1 (orange). Plot description same as that in Figure \ref{GBT_scint_hist}.}%
    \label{NenuFAR_85_scint_hist}%
\end{figure}

\begin{figure}[!ht]
    \centering
    \captionsetup[subfigure]{labelformat=empty}
    {\hspace*{-.4cm}\includegraphics[width=0.5\textwidth]{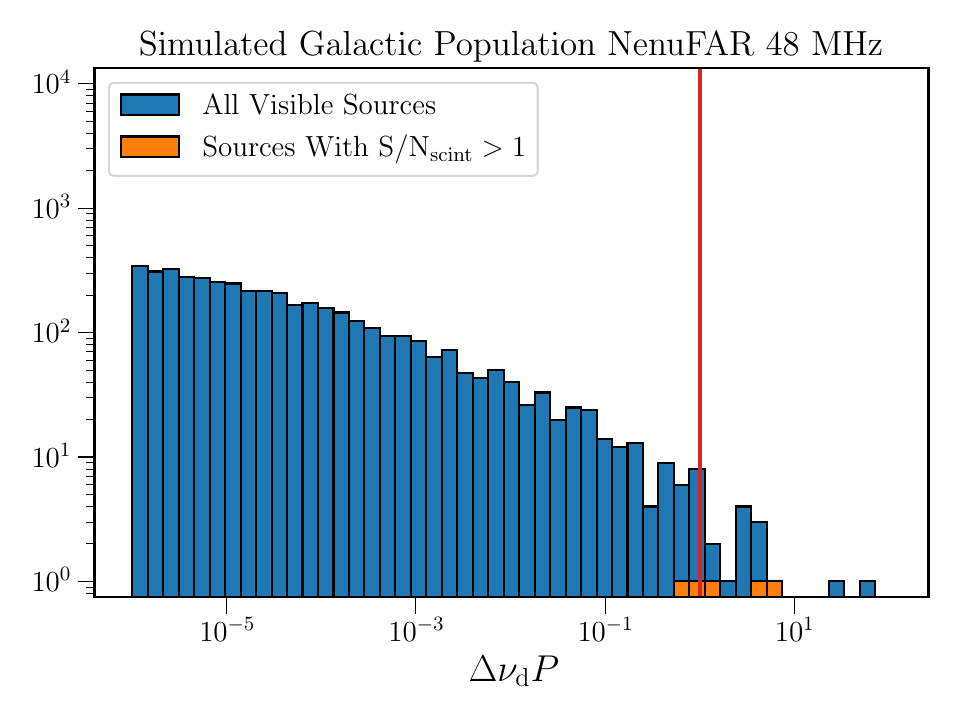} }%
    \caption{Expected scintle resolvability from a realization of our simulations analyzed for NenuFAR at 48 MHz (blue), and the subset of sources with S/N$>$1 (orange). Plot description same as that in Figure \ref{GBT_scint_hist}.}%
    \label{sim_NenuFAR_48_scint_hist}%
\end{figure}

\begin{figure}[!ht]
    \centering
    \captionsetup[subfigure]{labelformat=empty}
    {\hspace*{-.4cm}\includegraphics[width=0.5\textwidth]{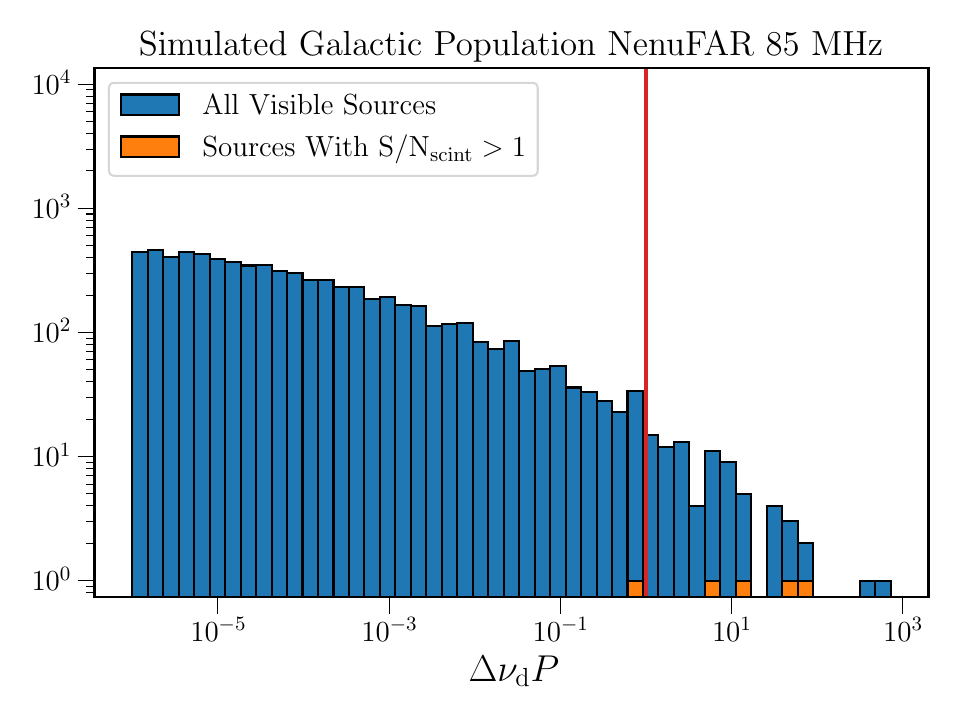} }%
    \caption{Expected scintle resolvability from a realization of our simulations analyzed for NenuFAR at 85 MHz (blue), and the subset of sources with S/N$>$1 (orange). Plot description same as that in Figure \ref{GBT_scint_hist}.}%
    \label{sim_NenuFAR_85_scint_hist}%
\end{figure}

While an incredibly sensitive instrument, as we saw with LWA, NenuFAR's frequency range may be where we start to see diminishing returns for observing at lower frequencies. In contrast to the $\sim$10 sources seen for MWA and LOFAR, we only get 5 total unique sources between the two bands that are simultaneously in the full deconvolution regime and pass the conservative cyclic merit threshold. Additionally, as with LWA, the frequency ranges at which NenuFAR operates may also introduce a bias in our results, where some pulsars may be viewed more favorably than they should due to our assumption of a simple power law governing flux density rather than one with a spectral turnover near 100 MHz like we see in many pulsars. The full histograms for all real sources analyzed using these telescope-frequency combinations are shown in Figures \ref{NenuFAR_48_hist} and \ref{NenuFAR_85_hist}. Our simulations also indicate that most, if not all, sources that pass the conservative threshold and are observable within the full deconvolution regime are already known for NenuFAR at both frequencies considered. Example realizations from these simulations at 48 and 85 MHz are shown in Figure \ref{NenuFAR_48_sim_hist} and \ref{NenuFAR_85_sim_hist}, respectively. 
Unfortunately, we find that NenuFAR will likely not be very useful for scintillation studies, with only five sources at both 48 MHz and 85 MHz passing our S/N threshold. Our simulations also predict that most, if not all, sources that could have detectable scintles at 48 and 85 MHz for NenuFAR have likely already been discovered. The scintle resolution prospects for all real sources with NenuFAR at 48 and 85 MHz are shown in Figures \ref{NenuFAR_48_scint_hist} and \ref{NenuFAR_85_scint_hist}, respectively, while example realizations for our simulated population are shown in Figures \ref{sim_NenuFAR_48_scint_hist} and \ref{sim_NenuFAR_85_scint_hist}.

\subsection{CHIME 450 MHz}

\begin{figure}[!ht]
    \centering
    \captionsetup[subfigure]{labelformat=empty}
    {\hspace*{-.5cm}\includegraphics[width=0.5\textwidth]{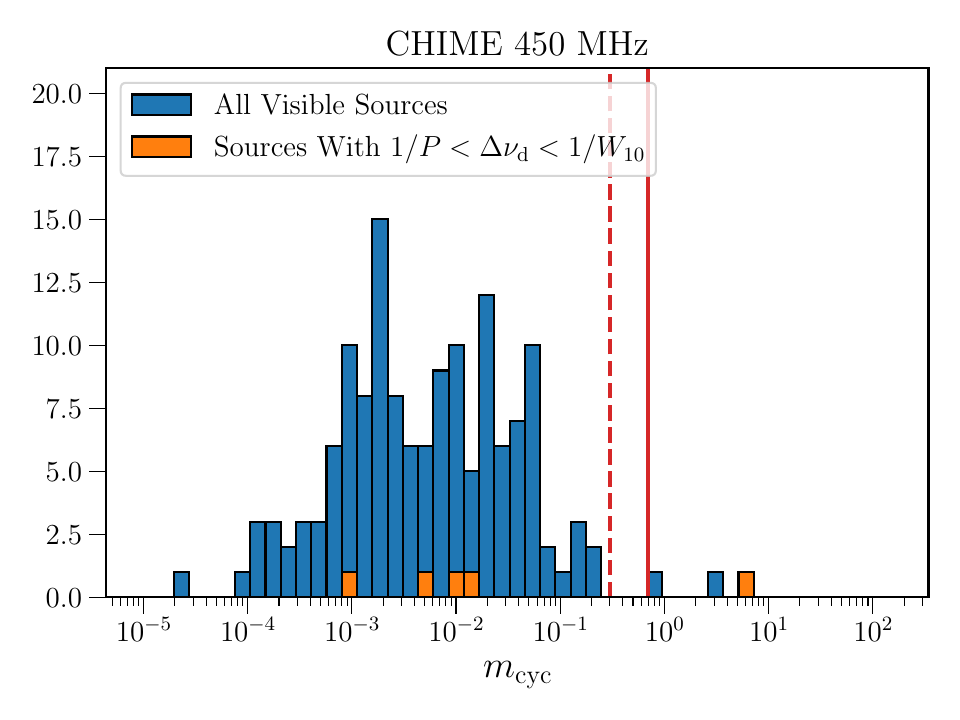} }%
    \caption{Cyclic merits for all real sources analyzed for CHIME at 450 MHz (blue), and the subset of sources in the full deconvolution regime (orange). Plot description same as that in Figure \ref{GBT_hist}. The three outliers on the right (from left to right) are PSRs J1630+3734, J2205+6012, and B1937+21.}%
    \label{CHIME_hist}%
\end{figure}

\begin{figure}[!ht]
    \centering
    \captionsetup[subfigure]{labelformat=empty}
    {\hspace*{-.4cm}\includegraphics[width=0.5\textwidth]{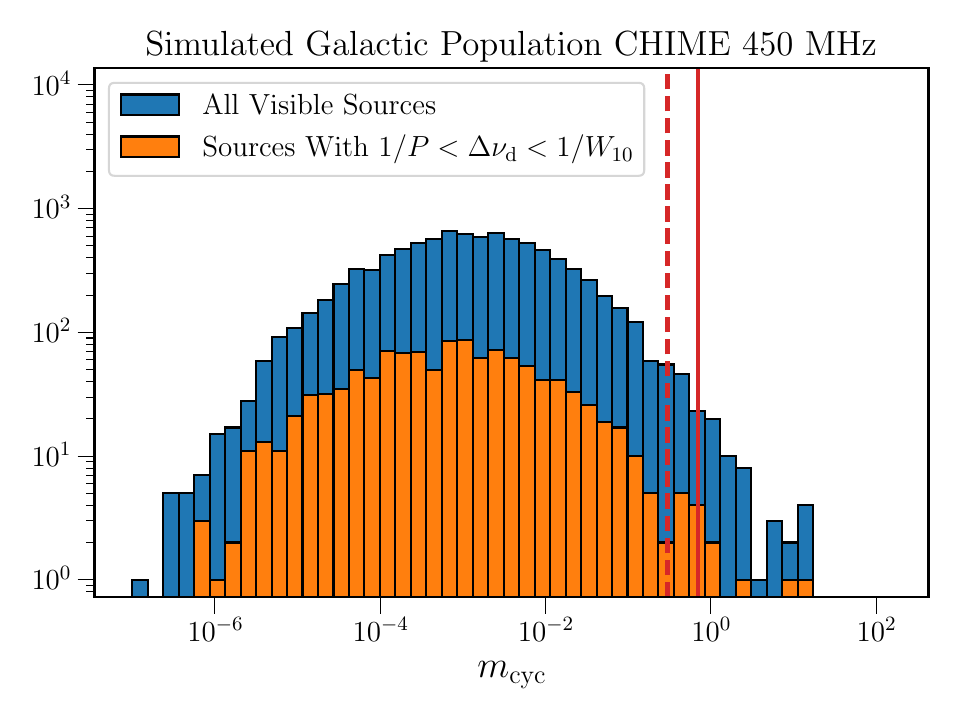} }%
    \caption{Cyclic merits for all sources from a realization of our simulations analyzed for CHIME at 450 MHz (blue), and the subset of sources in the full deconvolution regime (orange). Plot description same as that in Figure \ref{GBT_hist}.}%
    \label{CHIME_sim_hist}%
\end{figure}

\begin{figure}[!ht]
    \centering
    \captionsetup[subfigure]{labelformat=empty}
    {\hspace*{-.4cm}\includegraphics[width=0.5\textwidth]{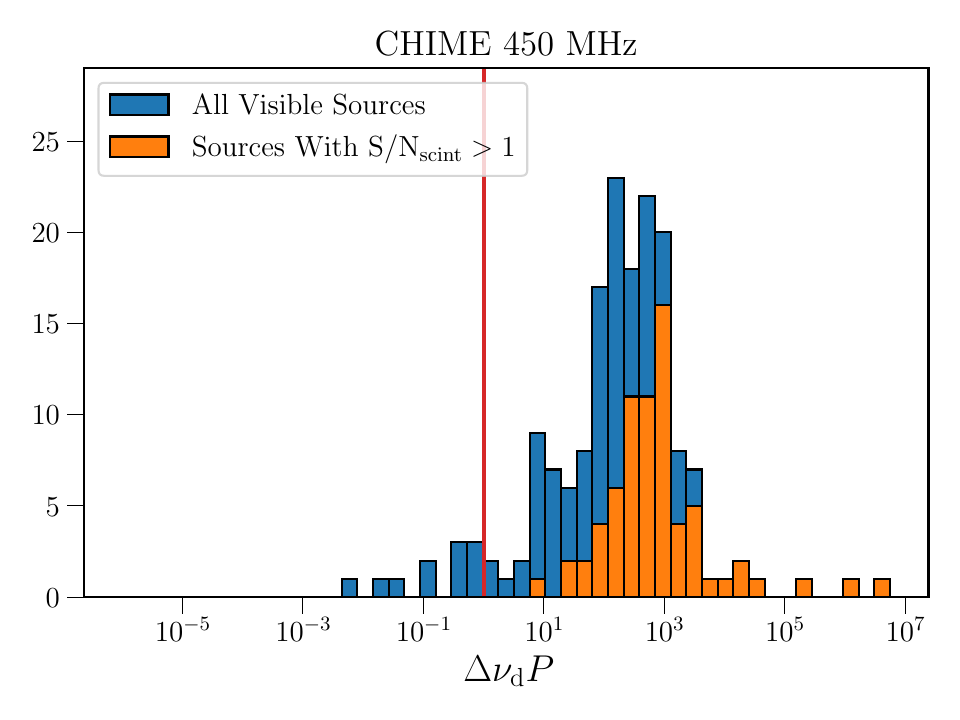} }%
    \caption{Expected scintle resolvability for all sources analyzed for CHIME at 450 MHz (blue), and the subset of sources with S/N$>$1 (orange). Plot description same as that in Figure \ref{GBT_scint_hist}.}%
    \label{CHIME_scint_hist}%
\end{figure}

\begin{figure}[!ht]
    \centering
    \captionsetup[subfigure]{labelformat=empty}
    {\hspace*{-.4cm}\includegraphics[width=0.5\textwidth]{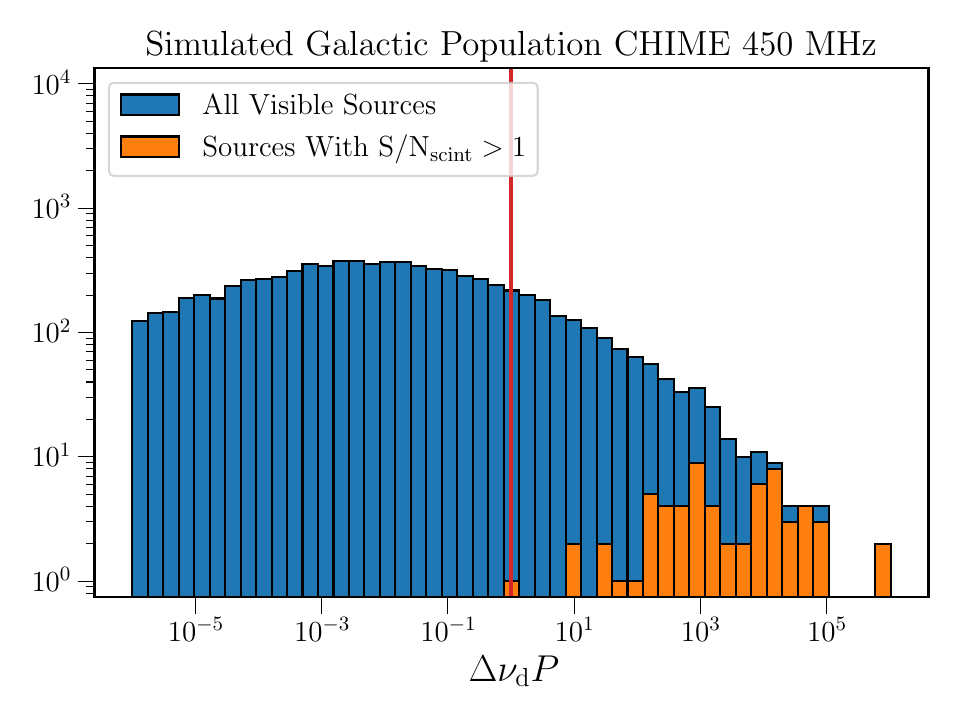} }%
    \caption{Expected scintle resolvability from a realization of our simulations analyzed for CHIME at 450 MHz (blue), and the subset of sources with S/N$>$1 (orange). Plot description same as that in Figure \ref{GBT_scint_hist}.}%
    \label{sim_CHIME_scint_hist}%
\end{figure}

CHIME observes in a somewhat sparse frequency range for cyclic deconvolution, as a small fraction of pulsars in our survey begin to exhibit scatter broadening toward the bottom of its band. Unfortunately, CHIME is not sensitive enough to take advantage of most of these sources. Additionally, due to its limited observing time of 10 minutes per source per day, CHIME would limited almost exclusively to analyzing recovered pulse broadening functions rather than full dynamic wavefields, as most pulsars would not be observable for a sufficient number of scintillation timescales to see well-resolved features in secondary spectra or secondary wavefields. Further minimizing CHIME's impact on cyclic spectroscopy, only one pulsar, PSR B1937+21, both passes our conservative cyclic merit threshold and is observed in the full deconvolution regime at 450 MHz. However, the unique observing program of CHIME does allow for an interesting opportunity: Since cyclic spectroscopy should be possible for PSR B1937+21 with CHIME, one could in principle run a long-term cyclically-deconvolved scattering delay analysis for this pulsar, with the daily cadence allowing for an unprecedented level of temporal detail. Additionally, one could make simultaneous measurements of scintillation bandwidth to constrain $C_1$, allowing for high temporal monitoring of this line of sight's scattering geometry and turbulence. The full histogram for all real sources analyzed using CHIME at 450 MHz is shown in Figure \ref{CHIME_hist}. Our simulations also predict that most, if not all, sources that pass the conservative threshold and are observable within the full deconvolution regime are already known for CHIME. An example realization from these simulations is shown in Figure \ref{CHIME_sim_hist}. Additionally, we find that CHIME would be a useful instrument for scintillation studies, with just under half (70) of its sources likely to have resolvable scintles that pass our S/N threshold. That being said, our simulations indicate that most, if not all, sources that could have detectable scintles at 450 MHz for CHIME have likely already been discovered. The scintle resolution prospects for all real sources with CHIME at 450 MHz are shown in Figure \ref{CHIME_scint_hist}, while an example realization for our simulated population is shown in Figure \ref{sim_CHIME_scint_hist}.

\subsection{uGMRT 150, 200, \& 350 MHz}
With its combination of high sensitivity, significant sky visibility, and low frequency coverage, uGMRT is by far the best instrument in our study for cyclic spectroscopy. With 49 sources that pass our conservative cyclic merit threshold while being observable in the full deconvolution regime, uGMRT has over three times as many viable sources for cyclic spectroscopy as LOFAR, which comes in second. In addition to all sources meeting the above criteria, it is incredibly likely that PSR J1811$-$2405 will be able to achieve cyclic deconvolution with uGMRT, as, while its full deconvolution regime is surrounded by, but not included within, the frequencies used for this study, measurements at all observing frequencies considered pass our conservative cyclic merit threshold. Furthermore, PSR B1855+09 has a very high cyclic merit at 150 MHz and would likely achieve full cyclic deconvolution if we were to observe around 10 MHz lower. Finally, PSR J1843$-$1113 would likely achieve full cyclic deconvolution at 350 MHz, assuming its full deconvoluiton regime spans more than 100 MHz above its lower bound of 268$\pm$30 MHz.
\par
\begin{figure}
    \centering
    \captionsetup[subfigure]{labelformat=empty}
    {\hspace*{-.5cm}\includegraphics[width=0.5\textwidth]{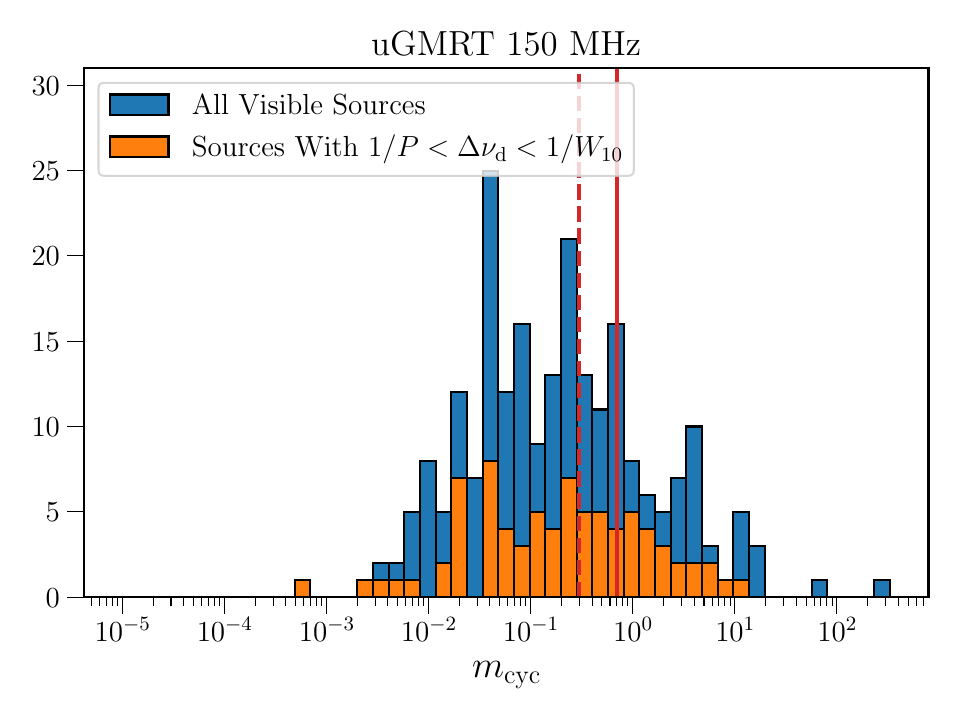}} %
    \caption{Cyclic merits for all real sources analyzed for uGMRT at 150 MHz (blue), and the subset of sources in the full deconvolution regime (orange). Plot description same as that in Figure \ref{GBT_hist}. The two outliers on the right are PSRs J2205+6012 (lower cyclic merit) and B1937+21 (higher cyclic merit).}%
    \label{uGMRT_150_hist}%
\end{figure}

\begin{figure}
    \centering
    \captionsetup[subfigure]{labelformat=empty}
    {\hspace*{-.5cm}\includegraphics[width=0.5\textwidth]{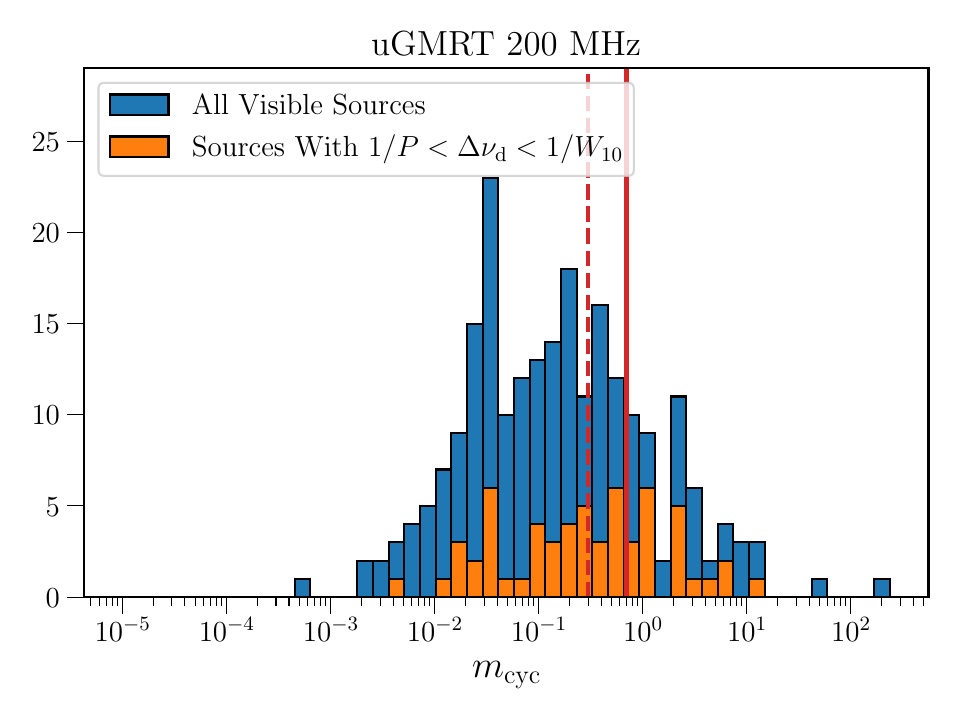}} %
    \caption{Cyclic merits for all real sources analyzed for uGMRT at 200 MHz (blue), and the subset of sources in the full deconvolution regime (orange). Plot description same as that in Figure \ref{GBT_hist}. The two outliers on the right are PSRs J2205+6012 (lower cyclic merit) and B1937+21 (higher cyclic merit).}%
    \label{uGMRT_200_hist}%
\end{figure}

\begin{figure}
    \centering
    \captionsetup[subfigure]{labelformat=empty}
    {\hspace*{-.5cm}\includegraphics[width=0.5\textwidth]{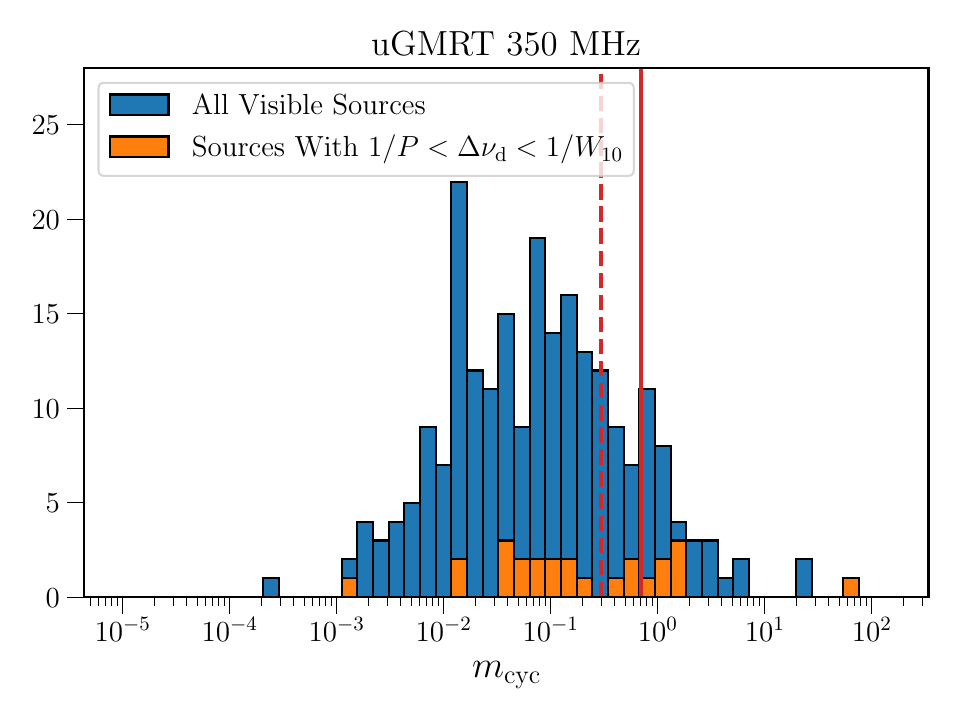}} %
    \caption{Cyclic merits for all real sources analyzed for uGMRT at 350 MHz (blue), and the subset of sources in the full deconvolution regime (orange). Plot description same as that in Figure \ref{GBT_hist}. The three outliers on the right are PSRs J0437$-$4715 (one of the two blues), J2205+6012 (one of the two blues), and B1937+21 (orange).}%
    \label{uGMRT_350_hist}%
\end{figure}

\begin{figure}[!ht]
    \centering
    \captionsetup[subfigure]{labelformat=empty}
    {\hspace*{-.4cm}\includegraphics[width=0.5\textwidth]{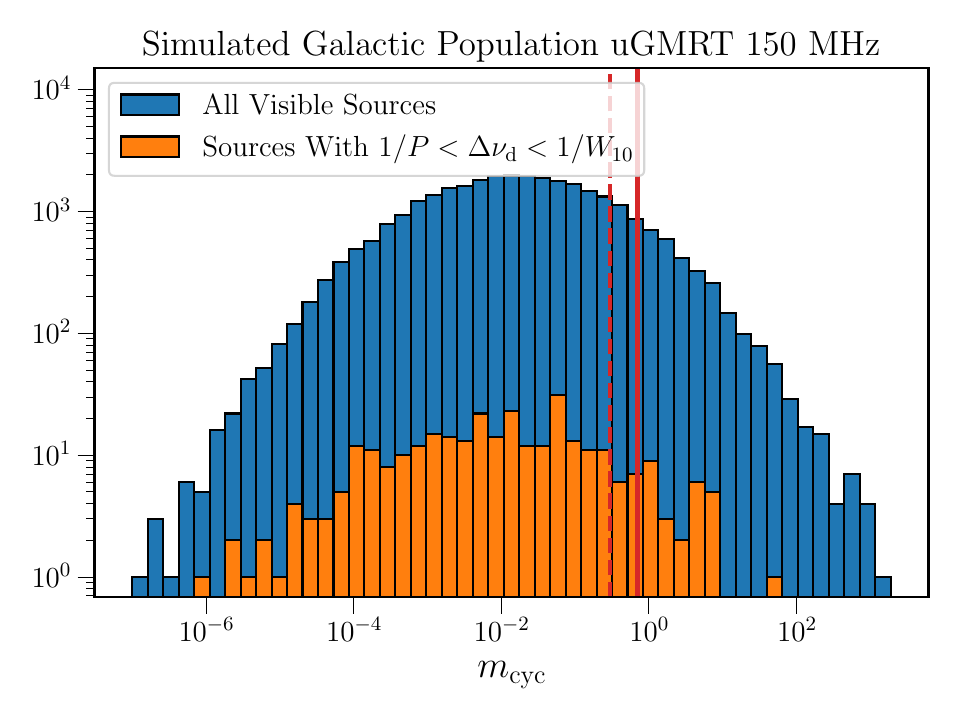} }%
    \caption{Cyclic merits for all sources from a realization of our simulations analyzed for uGMRT at 150 MHz (blue), and the subset of sources in the full deconvolution regime (orange). Plot description same as that in Figure \ref{GBT_hist}.}%
    \label{uGMRT_150_sim_hist}%
\end{figure}

\begin{figure}[!ht]
    \centering
    \captionsetup[subfigure]{labelformat=empty}
    {\hspace*{-.4cm}\includegraphics[width=0.5\textwidth]{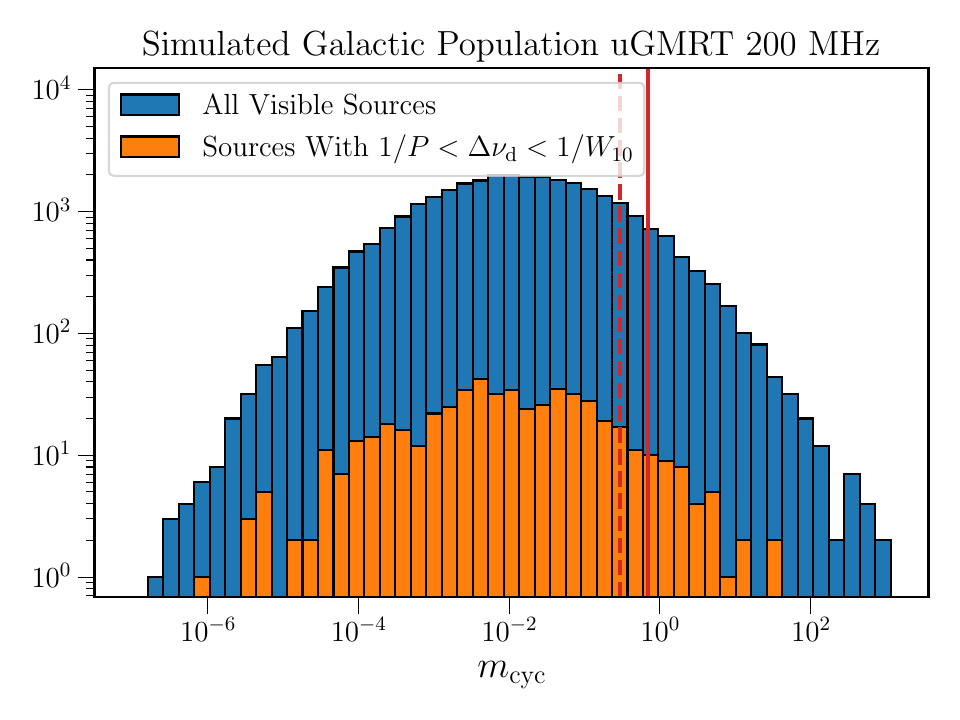} }%
    \caption{Cyclic merits for all sources from a realization of our simulations analyzed for uGMRT at 200 MHz (blue), and the subset of sources in the full deconvolution regime (orange). Plot description same as that in Figure \ref{GBT_hist}.}%
    \label{uGMRT_200_sim_hist}%
\end{figure}

\begin{figure}[!ht]
    \centering
    \captionsetup[subfigure]{labelformat=empty}
    {\hspace*{-.4cm}\includegraphics[width=0.5\textwidth]{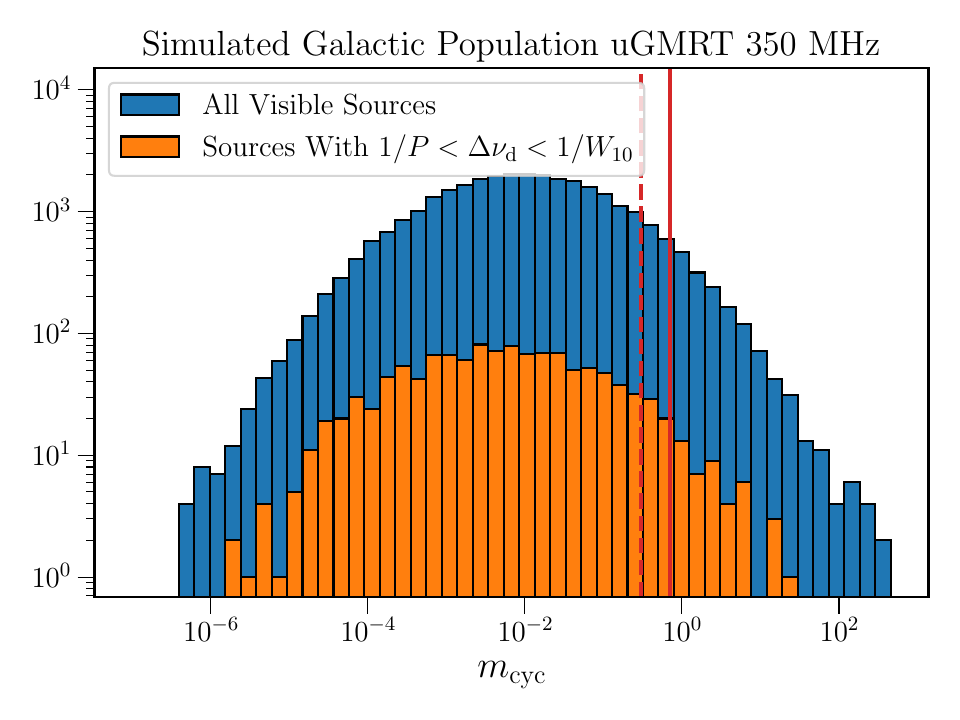} }%
    \caption{Cyclic merits for all sources from a realization of our simulations analyzed for uGMRT at 350 MHz (blue), and the subset of sources in the full deconvolution regime (orange). Plot description same as that in Figure \ref{GBT_hist}.}%
    \label{uGMRT_350_sim_hist}%
\end{figure}

\begin{figure}[!ht]
    \centering
    \captionsetup[subfigure]{labelformat=empty}
    {\hspace*{-.4cm}\includegraphics[width=0.5\textwidth]{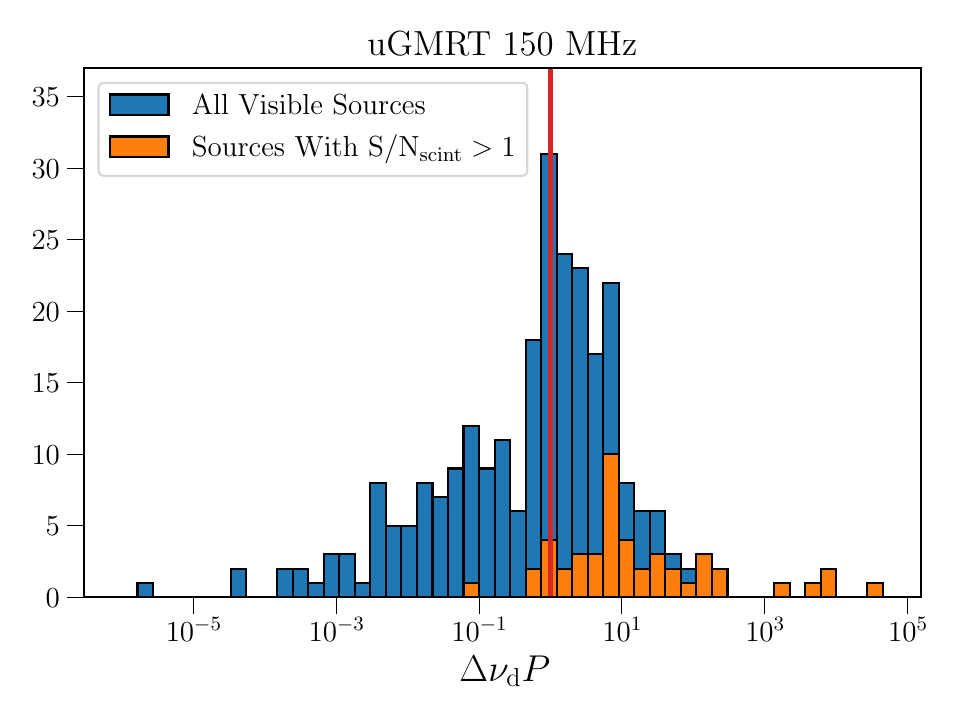} }%
    \caption{Expected scintle resolvability for all sources analyzed for uGMRT at 150 MHz (blue), and the subset of sources with S/N$>$1 (orange). Plot description same as that in Figure \ref{GBT_scint_hist}.}%
    \label{uGMRT_150_scint_hist}%
\end{figure}

\begin{figure}[!ht]
    \centering
    \captionsetup[subfigure]{labelformat=empty}
    {\hspace*{-.4cm}\includegraphics[width=0.5\textwidth]{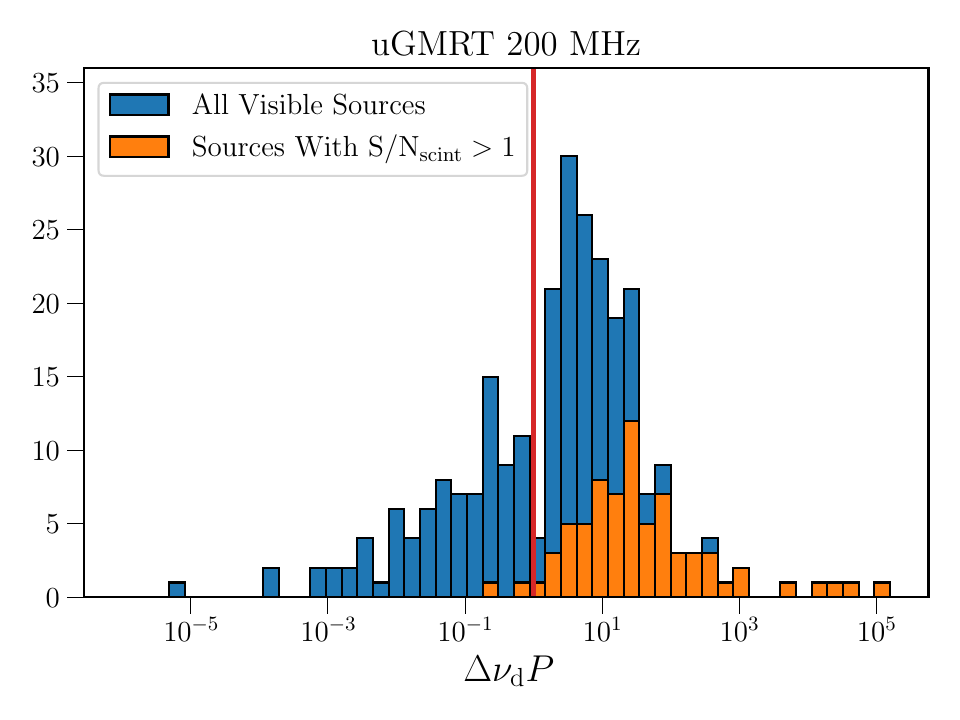} }%
    \caption{Expected scintle resolvability for all sources analyzed for uGMRT at 200 MHz (blue), and the subset of sources with S/N$>$1 (orange). Plot description same as that in Figure \ref{GBT_scint_hist}.}%
    \label{uGMRT_200_scint_hist}%
\end{figure}

\begin{figure}[!ht]
    \centering
    \captionsetup[subfigure]{labelformat=empty}
    {\hspace*{-.4cm}\includegraphics[width=0.5\textwidth]{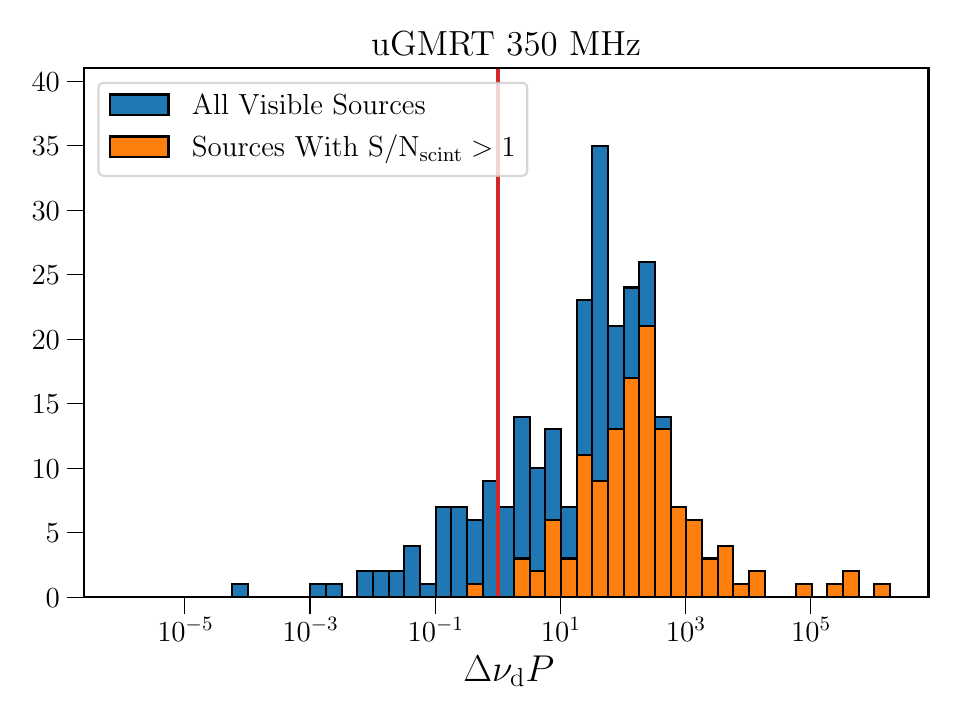} }%
    \caption{Expected scintle resolvability for all sources analyzed for uGMRT at 350 MHz (blue), and the subset of sources with S/N$>$1 (orange). Plot description same as that in Figure \ref{GBT_scint_hist}.}%
    \label{uGMRT_350_scint_hist}%
\end{figure}

\begin{figure}[!ht]
    \centering
    \captionsetup[subfigure]{labelformat=empty}
    {\hspace*{-.4cm}\includegraphics[width=0.5\textwidth]{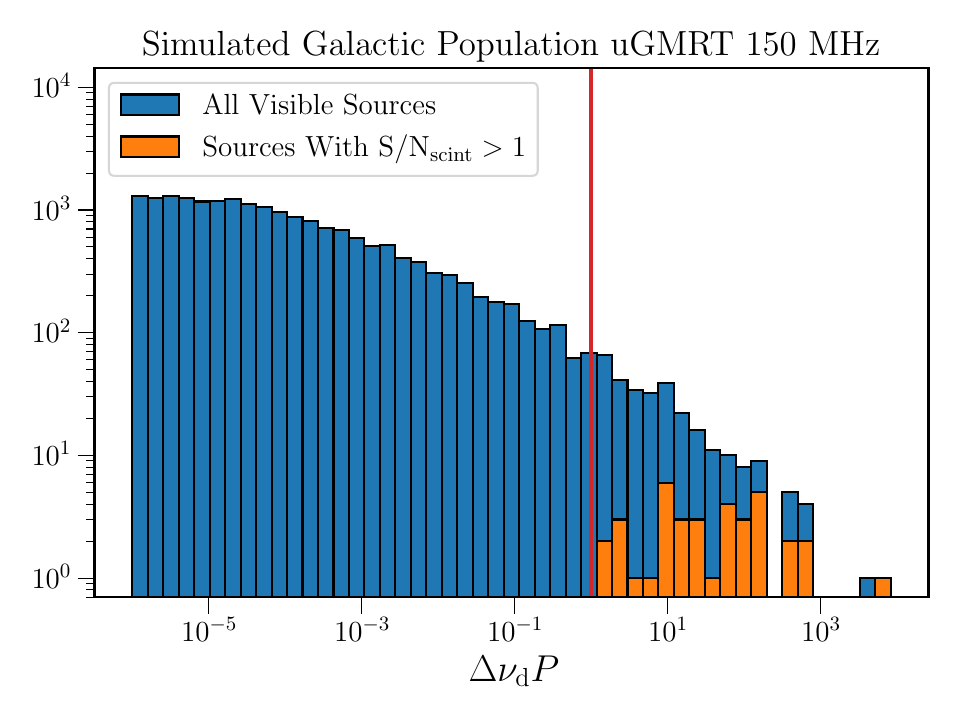} }%
    \caption{Expected scintle resolvability from a realization of our simulations analyzed for uGMRT at 150 MHz (blue), and the subset of sources with S/N$>$1 (orange). Plot description same as that in Figure \ref{GBT_scint_hist}.}%
    \label{sim_uGMRT_150_scint_hist}%
\end{figure}

\begin{figure}[!ht]
    \centering
    \captionsetup[subfigure]{labelformat=empty}
    {\hspace*{-.4cm}\includegraphics[width=0.5\textwidth]{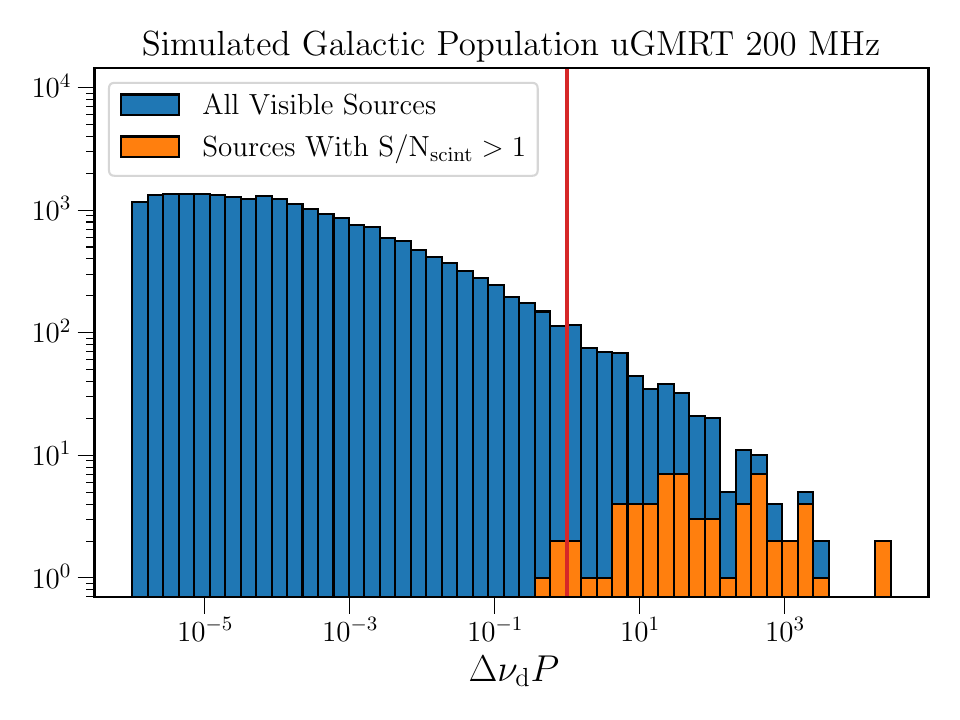} }%
    \caption{Expected scintle resolvability from a realization of our simulations analyzed for uGMRT at 200 MHz (blue), and the subset of sources with S/N$>$1 (orange). Plot description same as that in Figure \ref{GBT_scint_hist}.}%
    \label{sim_uGMRT_200_scint_hist}%
\end{figure}

\begin{figure}[!ht]
    \centering
    \captionsetup[subfigure]{labelformat=empty}
    {\hspace*{-.4cm}\includegraphics[width=0.5\textwidth]{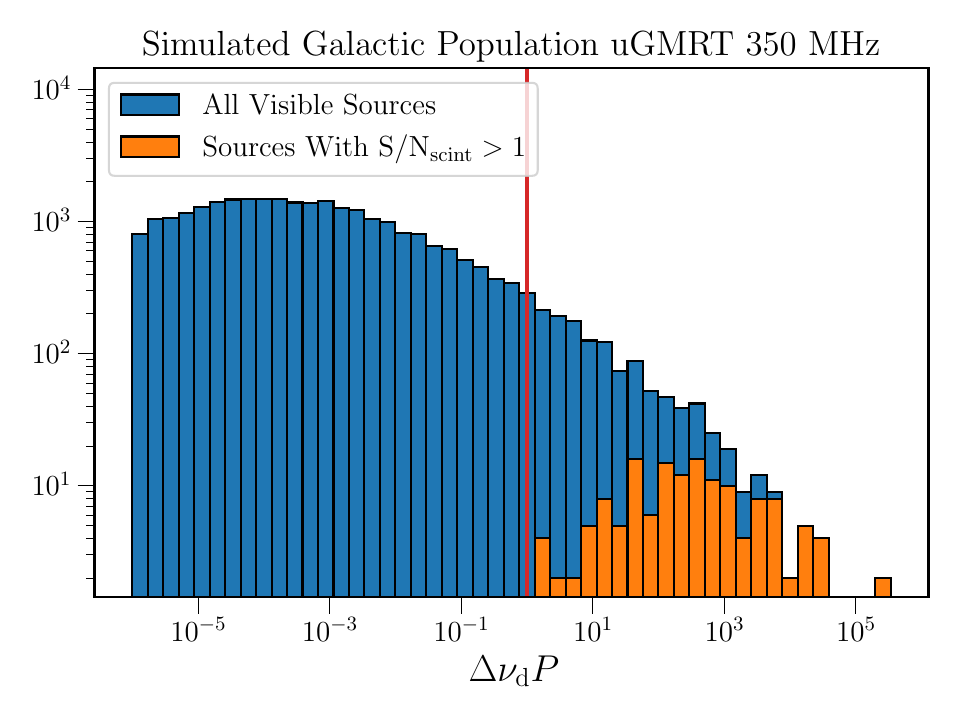} }%
    \caption{Expected scintle resolvability from a realization of our simulations analyzed for uGMRT at 350 MHz (blue), and the subset of sources with S/N$>$1 (orange). Plot description same as that in Figure \ref{GBT_scint_hist}.}%
    \label{sim_uGMRT_350_scint_hist}%
\end{figure}

 Given the significant cyclic merits for some viable sources, it may be possible to take advantage of uGMRT's subarraying capabilities to perform simultaneous scans across multiple observing bands, allowing for cyclic deconvolution across a larger range of frequencies than would be possible with a single uGMRT receiver. Strong potential candidates for this observing configuration include PSR J0613$-$0200 at 150 and 200 MHz, PSR J1012+5307 at 150 and 200 MHz, PSR J1306$-$4035 at 150 and 200 MHz, PSR J1744$-$1134 at 150 and 200 MHz, PSR J1909$-$3744 at 150 and 200 MHz, and PSR J1911$-$1114 at 150 and 200 MHz. The full histograms for all real sources analyzed using these telescope-frequency combinations are shown in Figures \ref{uGMRT_150_hist}, \ref{uGMRT_200_hist} and \ref{uGMRT_350_hist}. Our simulations also indicate that most, if not all, sources that pass the conservative threshold and are observable within the full deconvolution regime are already known for uGMRT at 150, and 200 MHz, although we predict significantly more may exist at 350 MHz that have likely not yet been discovered. Example realizations from these simulations at 150, 200, and 350 MHz are shown in Figure \ref{uGMRT_150_sim_hist}, \ref{uGMRT_200_sim_hist}, and \ref{uGMRT_350_sim_hist},  respectively. Additionally, we find that uGMRT should be excellent for scintillation studies, with 47 sources expected to have resolvable scintles while passing our S/N threshold at 150 MHz, 72 sources at 200 MHz, and 127 sources at 350 MHz. However, our simulations indicate that most, if not all, sources that could have detectable scintles at 150, 200, and 350 MHz for uGMRT have likely already been discovered. The scintle resolution prospects for all real sources with uGMRT at 150, 200, and 350 MHz are shown in Figures \ref{uGMRT_150_scint_hist}, \ref{uGMRT_200_scint_hist} and \ref{uGMRT_350_scint_hist}, respectively, while example realizations for our simulated population are shown in Figures \ref{sim_uGMRT_150_scint_hist}, \ref{sim_uGMRT_200_scint_hist} and \ref{sim_uGMRT_350_scint_hist}.

\newpage

\section{Conclusions \& Future Work}\label{conclusions}
\par We have calculated cyclic merits across 312 pulsars using 15 different telescope-frequency combinations to determine which sources, instruments, and observing frequencies are best suited for cyclic deconvolution and phase retrieval. We find that observing frequencies between $\sim$ 80$-$300 MHz are optimal both for observing the most pulsars in their full deconvolution regimes and for achieving cyclic deconvolution for the most pulsars with currently available instruments. Based on these findings, telescopes like LOFAR, MWA, uGMRT, and eventually the DSA once a greater fraction of the galactic millisecond pulsar population is discovered, appear to currently be the most capable for cyclic spectroscopy, and we strongly believe these telescopes should consider developing near real-time cyclic spectroscopy systems, like the one currently being implemented for the GBT, to increase the accessibility of this technique. As such, we feel it would be extremely beneficial to the pulsar community for these instruments to consider developing cyclic spectroscopy backends of their own during the early stages of commissioning.
\par While cyclic deconvolution will likely primarily benefit millisecond pulsars, our findings of high cyclic merit for the Crab Pulsar suggest some canonical pulsars may also be able to take advantage of this technique. Future work will determine cyclic merits for a subset of faster spinning canonical pulsars to ascertain whether cyclic deconvolution attempts with real data are worth pursuing in this population.
\section{Acknowledgments} 
\par The National Radio Astronomy Observatory and Green Bank Observatory are facilities of the U.S. National Science Foundation operated under cooperative agreement by Associated Universities, Inc. Thanks to Fengqiu Adam Dong for suggesting the inclusion of CHIME in this analysis. Thanks to Ue-Li Pen for suggesting we include uGMRT in this analysis. Thanks to the referee for suggesting the inclusion of LWA in this analysis. Thanks to Tyler Cohen for the simulated galactic pulsar populations. Thanks to Timothy Dolch and Ryan Lynch for providing paper feedback.

\par \textit{Software}: \textsc{PyGDSM} \citep{2016ascl.soft03013P}, \textsc{psrqpy} \citep{psrqpy}, \textsc{pulsar\_spectra} \citep{pulsar_spectra}, \textsc{NE2001p} \citep{Ocker_2024}, \textsc{astropy} \citep{astropy}, \textsc{numpy} \citep{numpy}, and \textsc{matplotlib} \citep{matplotlib}.

\newpage

\startlongtable
\begin{deluxetable*}{CCCCCCCCCCC}
\tablefontsize{\notsotiny}
\tabletypesize{\notsotiny}

\tablecolumns{10
}

\tablecaption{Cyclic Merits for GBT, DSA, MeerKAT, FAST, MWA, \& LWA \label{table_cyc}}
\vspace{-1cm}
\tablehead{ \colhead{$\textrm{Pulsar}$} &
\colhead{\text{Max Freq}} & \colhead{\text{Min Freq}} & 
\colhead{\text{GBT}$_{350}$}&
\colhead{\text{DSA}$_{800}$} &
\colhead{\text{MeerKAT}$_{600}$} & 
\colhead{\text{FAST}$_{550}$} & 
\colhead{\text{MWA}$_{150}$} &
\colhead{\text{MWA}$_{200}$} & 
\colhead{\text{LWA}$_{74}$} 
\vspace{-.2cm}
 \\  \colhead{} & \colhead{\text{(MHz)}} & \colhead{\text{(MHz)}} & \colhead{}& \colhead{}  &
\colhead{} & \colhead{} & \colhead{} & \colhead{} & & \colhead{} }
\setlength{\extrarowheight}{0.2cm}
\startdata 
\textrm{B}0021\text{$--$}72\textrm{C} & 148(17) & 103(11) & \textrm{---} & \textrm{---} & 0.0011(6) & \textrm{---} & 0.08(5) & 0.03(2) & \textrm{---}\\
\textrm{B}0021\text{$--$}72\textrm{D} & 183(21) & 105(12) & \textrm{---} & \textrm{---} & 0.003(2) & \textrm{---} & 0.02(1) & 0.01(1)  & \textrm{---}\\
\textrm{B}0021\text{$--$}72\textrm{E} & \textrm{---} & 115(13) & \textrm{---} & \textrm{---} & \textrm{---} & \textrm{---} & \textrm{---} & \textrm{---}  & \textrm{---}\\
\textrm{B}0021\text{$--$}72\textrm{F} & \textrm{---} & 123(14) & \textrm{---} & \textrm{---} & \textrm{---} & \textrm{---} & \textrm{---} & \textrm{---}  & \textrm{---}\\
\textrm{B}0021\text{$--$}72\textrm{G} & \textrm{---} & 112(13) & \textrm{---} & \textrm{---} & \textrm{---} & \textrm{---} & \textrm{---} & \textrm{---}  & \textrm{---}\\
\textrm{B}0021\text{$--$}72\textrm{H} & \textrm{---} & 117(13) & \textrm{---} & \textrm{---} & \textrm{---} & \textrm{---} & \textrm{---} & \textrm{---}  & \textrm{---}\\
\textrm{B}0021\text{$--$}72\textrm{J} & 198(23) & 130(15) & \textrm{---} & \textrm{---} & 0.19(12) & \textrm{---} & 2.3(1.4)$^!$ & 1.2(7)$^!$   & \textrm{---}\\
\textrm{B}0021\text{$--$}72\textrm{L} & \textrm{---} & 110(12) & \textrm{---} & \textrm{---} & \textrm{---} & \textrm{---} & \textrm{---} & \textrm{---}  & \textrm{---}\\
\enddata
\tablecomments{Cyclic merits and upper and lower observing frequency limits on the full deconvolution regime for GBT, DSA, MeerKAT, FAST, MWA, and LWA. Note that some values that may pass either the optimistic or conservative cyclic merit threshold may not have additional markers in cases where we could not find reported $W_{10}$ values, and so their full deconvolution regime maximum frequencies could not be estimated. Estimated cyclic merits for NANOGrav pulsars may be listed with asymmetric errors due to their flux density uncertainties originating from 16$^{\rm th}$ and 84$^{\rm th}$ percentile measurements of flux density.
\\ $^\$$Cyclic that pass the only the optimistic threshold for cyclic deconvolution within error but are not within the full deconvolution regime. \\ $^{*}$Cyclic merits that pass only the optimistic threshold for cyclic deconvolution within error and are within the full deconvolution regime. \\ $^{\&}$Cyclic merits that pass the conservative cyclic merit threshold within error but are not within the full deconvolution regime. \\ $^!$Cyclic merits that pass the conservative threshold for cyclic deconvolution within error and are within the full deconvolution regime.}
\end{deluxetable*}

\startlongtable
\begin{deluxetable*}{CCCCCCCCCCC}
\tablefontsize{\notsotiny}
\tabletypesize{\notsotiny}

\tablecolumns{11} 

\tablecaption{Cyclic Merits for LOFAR, NenuFAR, CHIME, \& uGMRT \label{table_cyc_2}}
\vspace{-1cm}
\tablehead{ \colhead{$\textrm{Pulsar}$} &
\colhead{\text{Max Freq}} & \colhead{\text{Min Freq}} & 
\colhead{\text{LOFAR}$_{150}$}&
\colhead{\text{LOFAR}$_{200}$} &
\colhead{\text{NenuFAR}$_{48}$} & 
\colhead{\text{NenuFAR}$_{85}$} & 
\colhead{\text{CHIME}$_{450}$} & 
\colhead{\text{uGMRT}$_{150}$} & 
\colhead{\text{uGMRT}$_{200}$} &
\colhead{\text{uGMRT}$_{350}$}
\vspace{-.2cm}
 \\  \colhead{} & \colhead{\text{(MHz)}} & \colhead{\text{(MHz)}} & \colhead{}& \colhead{}  &
\colhead{} & \colhead{} & \colhead{} & \colhead{} & \colhead{}& \colhead{} }
\setlength{\extrarowheight}{0.2cm}
\startdata 
\textrm{B}0021\text{$--$}72\textrm{C} & 148(17) & 103(11)  & \textrm{---} & \textrm{---} & \textrm{---} & \textrm{---} & \textrm{---} & \textrm{---} & \textrm{---}  & \textrm{---}\\
\textrm{B}0021\text{$--$}72\textrm{D} & 183(21) & 105(12) & \textrm{---} & \textrm{---} & \textrm{---} & \textrm{---} & \textrm{---} & \textrm{---} & \textrm{---}  & \textrm{---}\\
\textrm{B}0021\text{$--$}72\textrm{E} & \textrm{---} & 115(13)  & \textrm{---} & \textrm{---} & \textrm{---} & \textrm{---} & \textrm{---} & \textrm{---} & \textrm{---}  & \textrm{---}\\
\textrm{B}0021\text{$--$}72\textrm{F} & \textrm{---} & 123(14)& \textrm{---} & \textrm{---} & \textrm{---} & \textrm{---} & \textrm{---}  & \textrm{---} & \textrm{---}  & \textrm{---}\\
\textrm{B}0021\text{$--$}72\textrm{G} & \textrm{---} & 112(13)  & \textrm{---} & \textrm{---} & \textrm{---} & \textrm{---} & \textrm{---} & \textrm{---} & \textrm{---}  & \textrm{---}\\
\textrm{B}0021\text{$--$}72\textrm{H} & \textrm{---} & 117(13) & \textrm{---} & \textrm{---} & \textrm{---} & \textrm{---} & \textrm{---} & \textrm{---} & \textrm{---}  & \textrm{---}\\
\textrm{B}0021\text{$--$}72\textrm{J} & 198(23) & 130(15) & \textrm{---} & \textrm{---} & \textrm{---} & \textrm{---} & \textrm{---} & \textrm{---} & \textrm{---}  & \textrm{---}\\
\textrm{B}0021\text{$--$}72\textrm{L} & \textrm{---} & 110(12)  & \textrm{---} & \textrm{---} & \textrm{---} & \textrm{---} & \textrm{---} & \textrm{---} & \textrm{---}  & \textrm{---}\\
\enddata
\tablecomments{Cyclic merits and upper and lower observing frequency limits on the full deconvolution regime for LOFAR, NenuFAR, CHIME, and uGMRT. Note that some values that may pass either the optimistic or conservative cyclic merit threshold may not have additional markers in cases where we could not find reported $W_{10}$ values, and so their full deconvolution regime maximum frequencies could not be estimated. Estimated cyclic merits for NANOGrav pulsars may be listed with asymmetric errors due to their flux density uncertainties originating from 16$^{\rm th}$ and 84$^{\rm th}$ percentile measurements of flux density.
\\ $^\$$Cyclic that pass the only the optimistic threshold for cyclic deconvolution within error but are not within the full deconvolution regime. \\ $^{*}$Cyclic merits that pass only the optimistic threshold for cyclic deconvolution within error and are within the full deconvolution regime. \\ $^{\&}$Cyclic merits that pass the conservative cyclic merit threshold within error but are not within the full deconvolution regime. \\ $^!$Cyclic merits that pass the conservative threshold for cyclic deconvolution within error and are within the full deconvolution regime.}
\end{deluxetable*}

\bibliography{turner_1937_cs.bib}{}
\bibliographystyle{aasjournal}
\end{document}